\documentclass[lettersize,journal]{IEEEtranTCOM}

\normalsize
\usepackage{amsmath,amssymb,amsfonts}
\usepackage{cite}
\usepackage{algorithmic}
\usepackage{algorithm}
\usepackage{lipsum,afterpage}
\usepackage{xcolor}
\usepackage{colortbl}
\definecolor{LightCyan}{rgb}{0.88,1,1}
\definecolor{Gray}{gray}{0.9}
\usepackage{graphicx}
\usepackage{float}
\usepackage{multirow}
\usepackage{stfloats}
\usepackage{textcomp}
\usepackage{multirow}
\usepackage{epstopdf}
\usepackage{alphalph}
\usepackage{newtxmath}
\usepackage{caption}
\usepackage{acronym}
\usepackage{subcaption}
\usepackage{tikz}
\usepackage{flushend}
\begin{document}
\bstctlcite{IEEEexample:BSTcontrol}
\title{RIS-Assisted Visible Light Communication Systems: A Tutorial}

\author{Sylvester Aboagye,~\IEEEmembership{Graduate Student Member,~IEEE,}
        Alain R. Ndjiongue,~\IEEEmembership{Senior Member,~IEEE,}
        Telex~M.~N.~Ngatched,~\IEEEmembership{Senior Member,~IEEE,}
        Octavia~A.~ Dobre,~\IEEEmembership{Fellow,~IEEE,}%
         ~and H.~Vincent~ Poor,~\IEEEmembership{Life~Fellow,~IEEE}
\IEEEcompsocitemizethanks{\IEEEcompsocthanksitem This work has been submitted to the IEEE for possible publication.
Copyright may be transferred without notice, after which this version may
no longer be accessible.
}
}

\maketitle

\begin{abstract}
Recent intensive and extensive development of the fifth-generation (5G) of cellular networks has led to their deployment throughout much of the world. As part of this implementation, one of the challenges that must be addressed is the skip-zone problem, which occurs when objects such as trees, people, animals, and vehicles obstruct the transmission of signals. In free-space optical (FSO) and radio frequency (RF) systems, dead zones are most often caused by buildings and trees, while in visible light communications (VLC), obstructions are caused by individuals moving around a room or objects placed in the room. A signal obstruction can significantly reduce the signal-to-noise ratio in RF and indoor VLC systems, whereas in FSO systems, where the transmitted signals are directional, the obstruction can completely disrupt data transmission. Therefore, the skip-zone dilemma must be resolved to ensure the smooth and efficient operation of  5G and beyond networks. By placing a relay between a transmitter and a receiver, the effects of obstacles can be mitigated. As a result, the signal from the transmitter will reach the receiver. In recent years, reconfigurable intelligent surfaces (RISs) that are more efficient than relays have become widely accepted as a method of mitigating skip-zones and providing reconfigurable radio environments. However, there have been limited studies on RISs for optical wireless communication (OWC) systems. Through the RIS technology, OWC and RF communication channels can be reconfigured. This paper aims to provide a comprehensive tutorial on indoor VLC systems utilizing RIS technology. The article discusses the basics of VLC and RISs and reintroduces RISs for OWC systems, focusing on RIS-assisted indoor VLC systems. We also provide a comprehensive overview of optical RISs and examine the differences between optical RISs, RF-RISs, and optical relays. Furthermore, we discuss in detail how RISs can be used to overcome line-of-sight blockages and the device orientation issue in VLC systems while revealing key challenges such as RIS element orientation design, RIS elements to access point/user assignment design, and RIS array positioning design problems that need to be studied. Moreover, we discuss and propose several research problems on integrating optical RISs with other emerging technologies, including non-orthogonal multiple access, multiple-input multiple-output systems, physical layer security, and simultaneous lightwave and power transfer in VLC systems. Finally, we highlight other important research directions that can further improve the performance of RIS-assisted VLC systems.
\end{abstract}

\section*{LIST OF ACRONYMS}
\begin{acronym}[ NY]  
\acro {5G} 5th generation cellular network
\acro {ADT} Angle diversity transmitter
\acro {AF} Amplify and forward 
\acro {AP} Access point
\acro {A-QL} asynchronous quick link
\acro {B5G} Beyond 5G
\acro {BER} Bit error rate
\acro {CSI} Channel state information
\acro {CM-FSK} Camera M-ary frequency-shift-keying
\acro {C-OOK} Camera-based on-off keying
\acro {CSK} Color shift keying
\acro {DC} Direct current
\acro {DD} Direct detection
\acro {DF} Decode and forward
\acro {FoV} Field-of-view
\acro {FSO} Free space optical 
\acro {HA-QL} Hidden asynchronous quick link
\acro {HS-PSK} Hybrid spatial phase-shift-keying
\acro {IEEE} Institute of Electrical and Electronics Engineers
\acro {IR} Infrared
\acro {IM} Intensity modulation
\acro {IMA} Intelligent mirror array
\acro {IMR} Intelligent metasurface reflector 
\acro {IoT} Internet-of-things
\acro {ITO} Indium tin oxide
\acro {LB} Lower bound
\acro {LC} Liquid crystal
\acro {LD} Laser diode
\acro {LED} Light emitting diode
\acro {LiFi} Light fidelity
\acro {LoS} Line-of-sight
\acro {MEMS} Micro-electro-mechanical systems
\acro {mmWave} Millimeter-wave
\acro {MIMO} Multiple-input multiple-output
\acro {NLoS} Non-line-of-sight
\acro {NOMA} Non-orthogonal multiple access
\acro {OFDM} Orthogonal frequency division multiplexing
\acro {OFDMA} Orthogonal frequency division multiple access
\acro {OOK} On-off keying
\acro {OWC} Optical wireless communication
\acro {PD} Photodetetctor
\acro {PHY} Physical
\acro {PIN} Positive-intrinsic-negative
\acro {PLS} Physical layer security
\acro {PSK} phase-shift-keying 
\acro {QSM} Quadrature spatial modulation
\acro {QCM} Quad-LED complex modulation
\acro {QoS} Quality of service
\acro {RF} Radio frequency
\acro {RIS} Reconfigurable intelligent surfaces
\acro {SISO} Single-input single-output
\acro {SNR} Signal-to-noise ratio
\acro {SS2DC} Sequential scalable 2D code
\acro {SSK} Space shift keying
\acro {SM} Spatial modulation
\acro {THz} Terahertz
\acro {TDMA} Time division multiple access
\acro {TIA}  Transconductance amplifier
\acro {UB} Upper bound
\acro {UFSOOK} Undersampled frequency shift-OOK
\acro {VLC} Visible light communication
\acro {VPPM} Variable pulse position modulation
\acro {VTASC} Variable transparent amplitude-shape-color
\acro {WiFi} Wireless fidelity
\end{acronym}
\section{Introduction} \label{intro}
\IEEEPARstart{B}{eyond} fifth-generation cellular networks (B5G) are expected to deliver data rates of up to gigabits per second, provide massive connectivity and enhanced reliability while reducing deployment costs and power consumption \cite{Bariah102020}. In order to achieve these unique objectives, B5G networks are envisioned to rely on a number of wireless technologies, including millimeter-wave (mmWave), terahertz (THz) communications, and optical wireless communication (OWC). The main reason for this is that the mmWave, THz, and optical bands allow larger bandwidths, which results in higher data rates. Among the problems that should be addressed for the efficient operation of B5G mobile networks, which will utilize such high frequency bands, is the loss of signals due to obstructions by buildings, trees, walls, thick concrete, and people \cite{Islam102016,Ding102017,Dai052018,Yin122016}. More specifically, the mmWave, the THz, and the optical bands typically suffer higher penetration loss in non-line-of-sight (NLoS) scenarios and are more sensitive to shadowing and blockages than lower frequency bands due to their short wavelength. This blockage engenders skip zones which significantly affect network coverage and impact the system's performance and quality of service (QoS). Therefore, the occurrences and the impact of blockages need to be reduced by implementing strategies that establish alternative line-of-sight (LoS) paths in these communication systems.

In data transmission systems, the dead zone problem has been addressed in various ways, including relays and cooperative communications for both radio frequency (RF) systems using mmWave and THz \cite{Bhardwaj102021,Mamaghani022022,Xia122021,Chen012021,Ruiz022021,Yalcin062020,Kim062020,Zhang022021} and optical communication systems \cite{Bayaki122012,Liu102020,Najafi072017,Kamga052021,Wang072019,Kizilirmak102015,Aboagye102021,Pan112020,Guzman112015,Guzman072020,Aboagye062021}. Recently, reconfigurable intelligent surfaces (RISs) have emerged as one of the most effective solutions for solving skip-zone situations, in which an obstacle exists between a transmitter and receiver, preventing the LoS path of the transmitted signal from reaching the receiver \cite{basar022020,wu012020,liu112020,Yang082020,Shtaiwi062021, He092021,Abumarshoud042021,ndjiongue112021,Gao072021}. Unlike relays and cooperative communication techniques that cannot alter the behavior of the channel, RISs are able to proactively reconfigure the wireless propagation channel to enhance the communication performance. Over the past few years, research on the application of RISs in RF communication systems has increased significantly (e.g., \cite{Gong062020,Ning022021,Pei122021,Zhang082020,Zeng062021,Hao082021} and references therein). However, only few works have successfully studied and incorporated RISs into visible light communication (VLC) systems \cite{GLYBOVSKI052016,Abumarshoud042021, ndjiongue052021, Aboagye122021,ndjiongue112021,Cao022020,Qian062021,Abdelhady142022,Sun112021,Ssun2022}. As mentioned in \cite{Abumarshoud042021}, the interplay between RISs and VLC can lead to innovative and progressive applications in B5G networks. Since different mechanisms are involved in transmitting RF and optical signals, RF RIS techniques are significantly different from that of VLC systems and, hence, cannot be straightforwardly adopted in the latter. As a result, the progress that has been made in combining these two fields of telecommunication engineering warrants a tutorial to be prepared and presented to the research community and industry. This is the motivation of this tutorial. It focuses on the application of RIS technology to VLC systems. For the first time, this article provides a comprehensive tutorial that emphasizes the integration of RIS technology into VLC systems, while addressing their most strategic aspects.

Similar to the free-space optical (FSO), the infrared (IR) link, and other communication links that utilize frequencies around the THz region, VLC is an OWC technology that employs the visible spectrum to transmit data. When compared to its RF counterpart, VLC offers several advantages, including higher bandwidth and the use of light sources as transmitting antennas. In view of the nature of light, the VLC channel does not function as a bidirectional transmission medium since the transmitted data is encrusted within the light utilized for illumination. In addition to this disadvantage, the short transmission distance of visible light signals as well as their high susceptibility to blockages prevent the VLC technology from being a true competitor to the RF technology. VLC is therefore exploited in conjunction with RF to alleviate the saturation of the RF spectrum. Light Fidelity (LiFi), a VLC related technology, is a means of streaming internet content within an environment. LiFi may be considered as a real competitor to wireless fidelity (WiFi) in situations such as indoor internet broadcasting. This will however only occur if a carefully selected return path is used for the VLC system. As mentioned earlier, VLC systems suffer from signal loss in both indoor and outdoor environments due to obstructions, specifically in the indoor environment where human obstructions are prominent. Optical relays have been proposed as a solution to this problem. However, in light of the advent of optical RISs, it is reasonable to consider integrating the technology into the VLC as it also creates a reconfigurable optical transmission environment. 

{{\color{black}{On a more general level, the application of RISs in VLC systems is similar to RISs in mm-Wave/THz/sub-6GHz (i.e., RF communication systems). This is because RISs are known to perform roles such as signal reflection and, until recently, signal transmission \cite{9437234} in RF systems. However, on a high-level, the application of RISs in VLC systems is completely different (e.g., VLC signal characteristics, RIS materials, and functionalities) and presents novel challenges when compared to the application of RISs in RF communication systems. Such unique features/challenges, which serve as the motivation for this tutorial paper, can be summarized as follows:
\begin{enumerate}
\item Visible light signal characteristics: The transmission of VLC signals is significantly different from signal transmission in RF communication systems due to their unique characteristics. Firstly, intensity modulation  (IM) and direct detection (DD)  are the most commonly used modulation and demodulation techniques, respectively. Secondly, the dual role of illumination and communication are performed simultaneously. Hence, RISs in VLC must optimize both communication and illumination performance.  Thirdly, the transmitted signals should be real and non-negative. Moreover, signal transmission and reception in VLC systems is highly influenced by the fields-of-view (FoVs) of the transmitter\footnote{{\color{black}{The FoV of an LED is defined as the angle between the points on the radiation pattern at which the directivity is reduced to $50\%$ \cite{svilen}. Generally, it is specified by the semi-angle at half power.}}} and the receiver. Finally, light emitting diodes (LEDs) and photodetectors (PDs) are used as the transmitter and receiver, respectively. Since the signal features mentioned above are not common to RF signals, the proposed methods in RIS-aided RF systems cannot be directly employed in VLC systems.

\item Typical functionalities: RISs are typically deployed and optimized in RF systems (and VLC systems) to \textit{overcome link blockages and provide coverage extension}. However, the unique characteristics of VLC signals enable other communication and illumination performance-enhancement functionalities in VLC systems. As mentioned, signal transmission at the transmitter and reception at the receiver are directly influenced by their FoVs. While a large FoV at the transmitter side permits wider beam angle for more uniform illumination coverage, it reduces the illumination intensity at the receiver. On the other hand, a small FoV at the transmitter side can provide high illumination intensity at the receiver but would require an unobstructed line-of-sight alignment. At the receiver side, large FoV for the PD enables significant amount of light beams to be detected. However, PDs with large FoVs can result in performance degrading factors such as increased manufacturing cost, increased receiver noise, and decreased receiver bandwidth, which can render VLC systems unsuitable for high data-rate applications. RISs can be deployed at the transmitter and receiver side to \textit{dynamically control the FoV} of the transmitter and the receiver to perform the role of beam focusing and steering. Depending on the distribution of access points (APs) and receivers, RISs can be used to configure small or large FoVs for the transmitters and receivers such that the system's performance is maximized. Moreover, optical concentrators made up of convex lenses are typically placed in front of the PD to focus impinging light beams on the center of the PD. However, the use of such convex lenses can result in up to 30$\%$ losses in the incident light power due to reflections at the lens' upper surface \cite{ndjiongue052021}. RISs can be used inside the receiver to overcome such losses by \textit{amplifying the incident light beam} \cite{ndjiongue052021,slysub}. Furthermore, RISs can be used for \textit{wavelength filtering and interference suppression} \cite{9609592,9860058,8839844}. Finally, RISs can be used to offer differentiated services satisfying different illumination requirements (i.e., \textit{illumination relaxation}).

\item {Place of deployment and propagation model: RISs are typically deployed in the medium between the transmitter and the receiver in RF communication systems. This is, however, not the case in VLC systems where RISs can be deployed at the transmitter side \cite{GLYBOVSKI052016, Abumarshoud042021}, the receiver side \cite{ndjiongue052021,slysub}, or in the medium between the transmitter and the receiver \cite{Aboagye122021}, depending on the desired functionality. Novel channel models are required to characterize signal propagation in RISs when placed at the transmitter or the receiver side. Each of these different configurations provides the system with unique advantages.  The type of RIS materials suitable for deployment at the transmitter or receiver side, the optimal operating conditions, and performance optimization and analysis have not been investigated previously.}

\item{Hardware and performance optimization:} Typical RIS hardware in VLC includes metasurfaces, mirror arrays, and liquid crystals (LCs) (another type of metasurface). The optimization of RISs in RF systems mainly focuses on optimizing the phase shifts of the RIS elements for communication related performance improvements. Since VLC serves both illumination and communication purposes, the deployment of RISs must seek to optimize both. Specifically, the formulation of optimization problems for RIS-aided VLC systems must consider illumination in either the objective function or as part of the constraint set. Since the position of the RIS when deployed in the channel affects the illumination and communication (e.g., data rate) performance, novel RIS placement optimization problems need to be studied.  In addition to the phase shift for optical metasurfaces, RIS parameters such as mirror orientation angles and  refractive index of LCs are other decision variables that cannot be neglected in the design of RIS-aided VLC systems. As a result, novel channel capacity, horizontal illumination, and channel gain expressions are required to effectively analyze RIS-aided VLC systems. Since the problem structure for such systems may completely differ from those in RIS-aided RF systems, novel solution techniques are required.
\item{Propagation model:} Novel irradiance and channel gain expressions are required for metasurface, mirror arrays, and LCs-based RISs in VLC systems as the proposed models for RIS-enabled RF communication systems cannot be applied to VLC systems.
\end{enumerate}}}}

\begin{figure}
	\centering
	\includegraphics[width=0.5\textwidth]{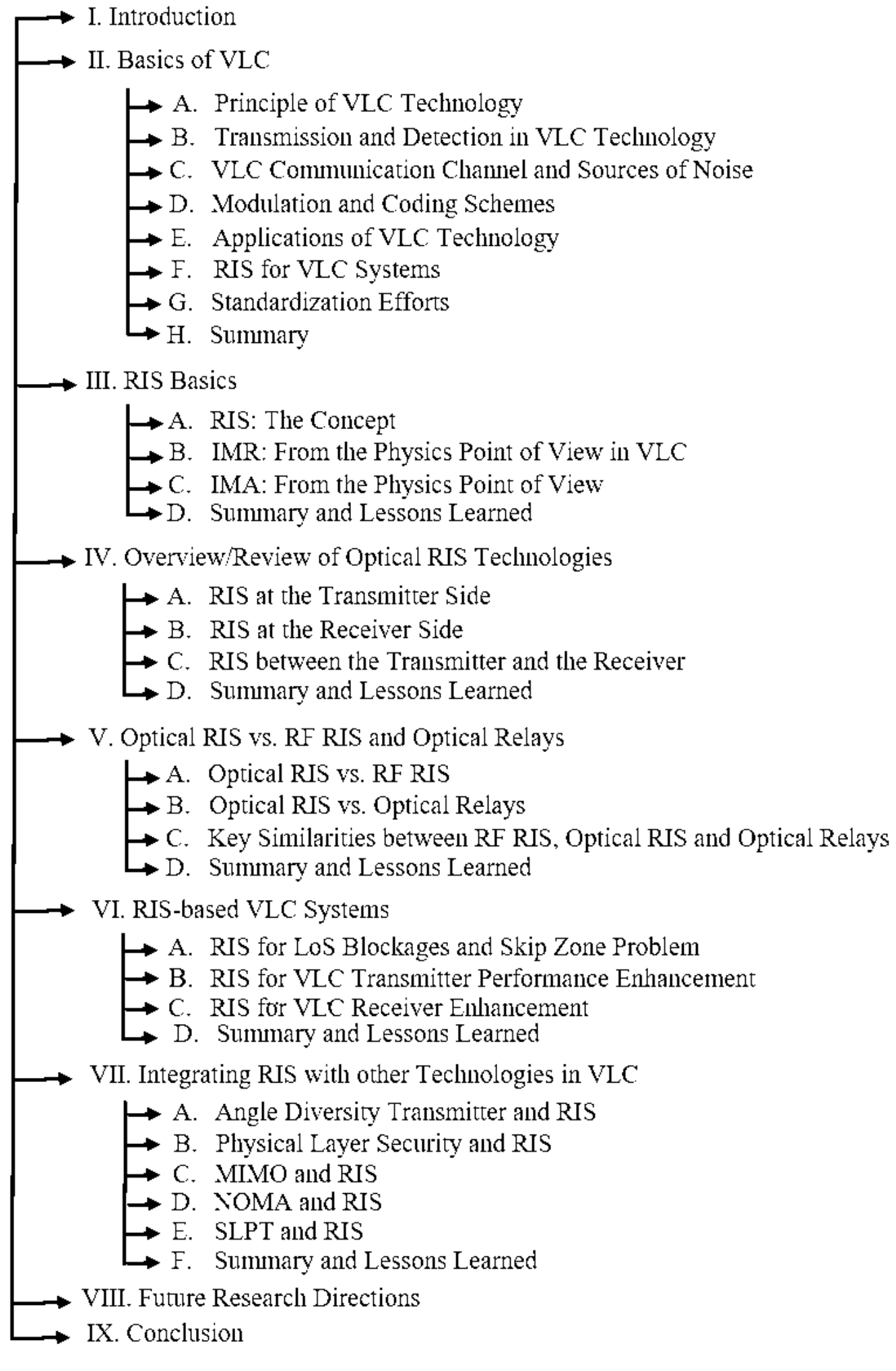}
	\caption{\small{Organization of this paper.}}
	\label{Fig:organa}
\end{figure}

To achieve the main objectives of this tutorial, we elaborate on the use of RIS technology in OWC systems, with an emphasis on the VLC system. We first review the VLC technology, its principle, transmitters and receivers, and signal detection. We also discuss the channel and noise in VLC systems. We review the modulation schemes, the applications of the VLC technology, and finally, we touch on the standardization effort for the VLC technology. We then discuss RIS basics and fundamental concepts. We also highlight and explain the main types of RISs used in VLC from the physics point of view, including intelligent metasurface reflector (IMR), intelligent mirror array (IMA), and LC-based RIS. This is followed by an overview of the optical RIS technology where we describe different RISs configurations for the VLC technology with an emphasis on RISs located inside the transmitter, receiver, and over the channel. Next, we compare the optical RIS to the RF-RIS, and to optical relays. We examine the particular case of RIS-based VLC systems and discuss the various use-cases and performance optimization approaches. Later, we discuss the use of an RIS when placed inside the receiver to improve the VLC receiver's FoV and when placed inside the transmitter to improve beamforming. Furthermore, we look at the integration of RISs with other technologies in VLC systems. To this end, we discuss transmitter and receiver diversity with RISs, physical layer security (PLS), multiple-input multiple-output (MIMO), and non-orthogonal multiple access (NOMA) with RISs. Moreover, we explore simultaneous lightwave and power transfer and RISs. Finally, possible extensions to the integration of RISs in VLC systems are outlined and conclusions are drawn. To facilitate the reading flow of the paper, we provide its flowchart, showing its organization and the relations among different sections in Fig.~\ref{Fig:organa}. 

\section{Basics of Visible Light Communications} \label{vlc}
This section provides the basics of the VLC technology while presenting a comprehensive overview of its communication principle, transmission and detection schemes, communication channel and sources of noise, modulation and coding schemes, its typical applications, the adoption of RISs in VLC, as well as standardization efforts. Note that several other survey and tutorial papers focusing on related issues on its early development, channel modeling, design principles (including user and network-centric), networking techniques, noise and its reduction techniques, performance optimization techniques, and recent emerging
applications of VLC can be found in \cite{Obeed032019,9241073,7932857,9351549,8015106,6497926,8308722,8698841,7072557,kumarapr2010,tsonevfeb2014,wunovdec2014,pathak4q2015,sahaoct2015,qiuoct2016,sindhubala042016,do052016,7096279}.

\subsection{Principle of VLC Technology}
\begin{figure*}[t]
	\centering
	\includegraphics[width=0.99\textwidth]{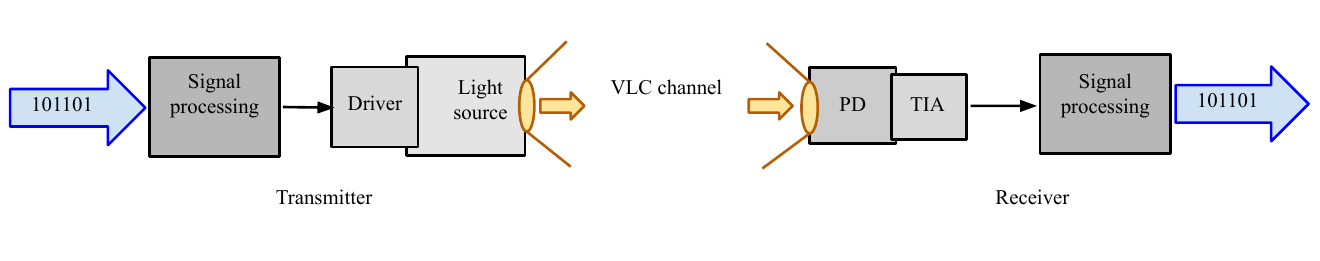}
	\caption{\small{Block diagram of data transmission using VLC technology.}}
	\label{Fig:PrincipleVLC}
\end{figure*}

As with most OWC technologies, the VLC technology utilizes high switching rate LEDs to incrust the incoming data into the generated light as depicted in Fig.~\ref{Fig:PrincipleVLC}. The generated light carries the message signal towards the PD. 
Its intensity is modulated to accommodate the transmitted data within a frequency bandwidth. At the receiver, with the help of a transconductance amplifier (TIA), the PD converts the detected light intensity into a voltage,  readily understandable by the signal processing unit. One of the main dilemmas of the VLC technology is the limited modulation bandwidth since LEDs switch into saturation mode as the modulation frequency increases. In addition to this, the transmitted signal is required to be of a real and positive value. 

\begin{figure}[t]
	\centering
	\includegraphics[width=0.5\textwidth]{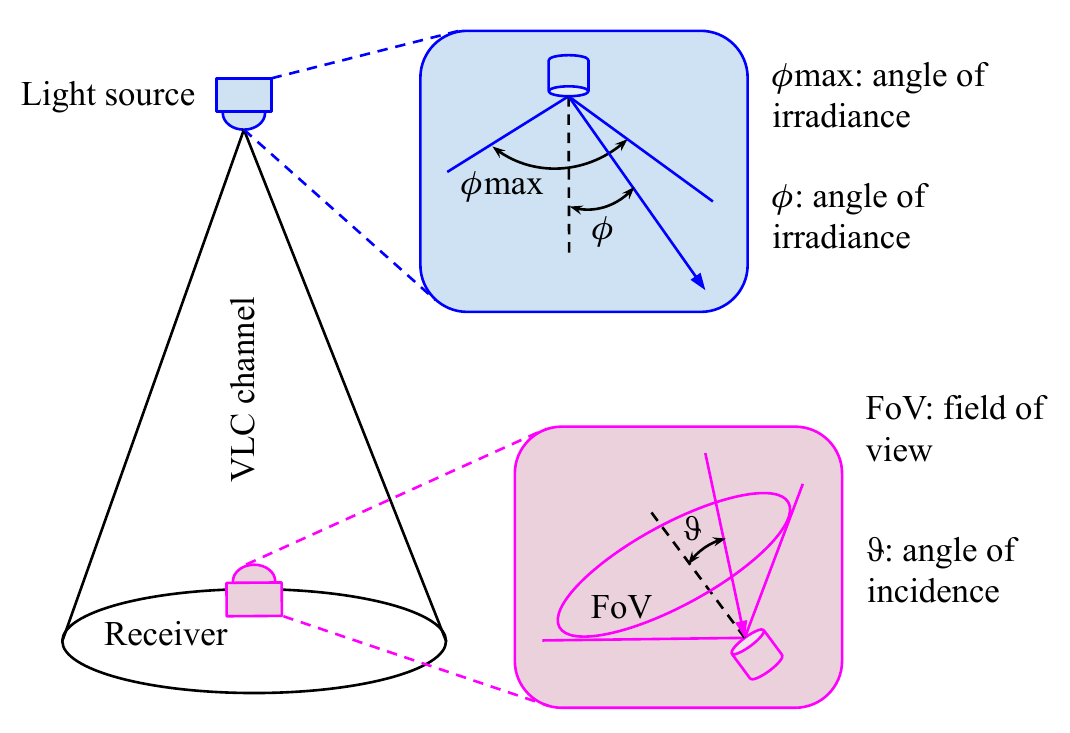}
	\caption{\small{Principles of data transmission using VLC technology.}}
	\label{Fig:VLC-principle}
\end{figure}

Figure~\ref{Fig:VLC-principle} depicts the typical principle of indoor VLC systems, which illustrates the double utilization of the light source. It broadcasts both light and a message. A span of both the light source and PD displays geometrical structures of transmitted and received beams, as well as the detection criterion, which is based on the receiver’s FoV. At the light source, Fig.~\ref{Fig:VLC-principle} shows the Lambertian scattering, the parameters of which include the incidence and irradiance angles, the Lambertian order, and define the channel parameters. However, the end-to-end channel gain also integrates the distance between a light source and the PD, the receiver’s FoV, the receiving angle, and the responsiveness of the PD. 

\subsection{Transmission and Detection in VLC Technology}

\subsubsection{VLC signal transmission}
A closer look at the transmitter depicted in Figs.~\ref{Fig:PrincipleVLC} and \ref{Fig:VLC-principle} shows that the VLC transmitter differs from the conventional one by the constitution of the different blocks. For example, an RF transmitter does not need a LED’s driver, which cannot be omitted in VLC. The VLC encoder is similar to the RF encoder in its functionality. The modulator may be different from those used in RF by the fact that transmitted signals in VLC are purely positive and real. In Section III-D, we discuss the process related to the asymmetric transmitted VLC signal. In the VLC technology, two main types of diodes are used in the light source package, namely laser diode (LD) and LED. Both are IM-based. In this technique, the incoming bits modulate the current intensity flowing through the LD/LED. Typically, the latter transfers its waveform to the resulting light beam. 

Several types of luminescent diodes are available. They differ by wavelength or by the process through which light is produced. Thus, we have phosphor-LEDs, red-green-blue-LEDs, high-power LEDs, IR-LEDs, ultraviolet-LEDs, LDs, matrices of LEDs, organic LEDs, and Quantum dot LEDs. Originally, light-emitting semiconductors were manufactured in several colors (or not perceptible colors) and wavelengths, such as yellow for 570 nm $\leq \lambda \leq$ 590 nm; red for 610 nm $ \leq \lambda \leq  $ 760 nm, blue for 450 nm $ \leq \lambda \leq  $ 500 nm, and green for 500 nm $ \leq \lambda \leq  $ 570 nm. For LDs, 630 nm $ \leq \lambda \leq  $ 950 nm. The white color can be constructed from two different processes: (\textit{i}) by a combination of red, green, and blue, or (\textit{ii}) by phosphor conversion. In the latter case, the phosphor is incorporated in the body of a blue-LED with a peak wavelength of around 450 to 470 nm. Part of the blue light is converted to yellow light by the phosphor, and the combination of the obtained yellow color and the remaining blue produces a white color. The former offers an opportunity to apply specific modulation techniques for data transmission such as color shift keying (CSK), MIMO, or diversity techniques. Note that most power LEDs are white-colored, and that ultraviolet-LEDs are part of visible light sources. Most of the electromagnetic semiconductors are low-cost and contribute to the complexity aspects of VLC systems by the ease of their current modulation. All the above-mentioned light sources represent only the antenna, which physically corresponds to the bridge between the modem and the transmission channel. After signal processing, the current sent through the light source should allow adequate lighting, while performing data transfer.
\subsubsection{VLC signal detection}
At the receiver, detection may be based on DD or heterodyne modes. However, the IM/DD combination provides advantages in cost and complexity. The key elements in VLC detectors, which make its receiver different from the RF receiver, are the PD and TIA. However, they present many similarities. In the following paragraphs, we comment on PDs and TIAs. 

\begin{figure}
	\centering
	\includegraphics[width=0.5\textwidth]{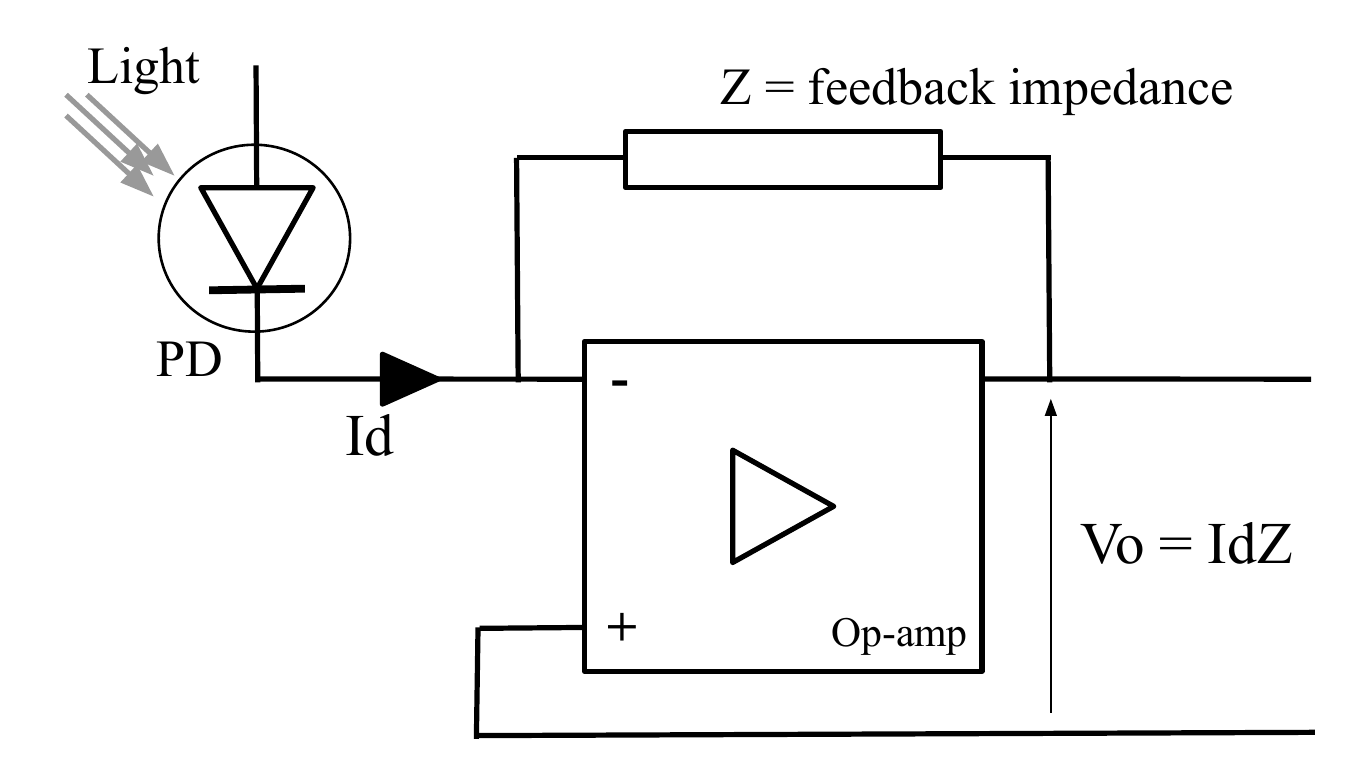}
	\caption{\small{Trans-conductance amplifiers.}}
	\label{fig:4}
\end{figure}

In general, PDs have the same doping structure as illuminating semiconductors. For a PD to detect a specific waveform, it must naturally be prepared to detect the corresponding frequency range, i.e., it must be sensitive to that specific wavelength. Thus, an IR-PD detects light from an IR-LED, a laser PD is sensitive to a signal from an LD, and so forth. Significantly, PDs, as with LEDs and LDs, are cost-efficient and low-power components, which make the entire receiver a cost-effective device. A TIA is a current-to-voltage converter made of operational amplifiers. The VLC processing modules which include the analog-to-digital converter, de-modulator, and decoder, are voltage-oriented components, i.e., their input requires a signal in voltage form. The TIA converts the PDs' output current to a voltage, which is acceptable by these blocks. Both can then process signals based on their functions. Figure~\ref{fig:4} shows a principle diagram in which the PD is combined with an operational amplifier circuit to form a TIA.

\subsection{VLC Communication Channel and Sources of Noise}
{\color{black}{\textbf{VLC channel:}
The communication channel in VLC, as in any other telecommunication technology, is the medium between transmitting and receiving antennas, i.e. bounded by the light source and PD. It represents the optical part of the VLC system. The VLC channel suffers from optical path loss and multi-path induced dispersion. However, the configuration of the VLC system typically determines how the channel impacts the transmitted signal. For LoS configurations, the reflected light components do not need to be taken into consideration, and consequently, the VLC channel is impacted by path loss which can be easily calculated from the knowledge of the transmitter beam divergence, receiver size, and separation distance between the transmitter and receiver. The LoS channel  gain is given by \cite{komine5012004}

	\begin{equation}\label{cg11a}
	{G_{\rm LoS}} = \left\{ {\begin{array}{*{20}{l}}  
	{\frac{{\left( {m + 1} \right)}{A_{\rm{PD}}}}{{2\pi}d^2}{\cos ^m}\left( {{\Phi }} \right) T\left( {{\vartheta }} \right)G\left( {{\vartheta }} \right) {\cos }\left( {{\vartheta }} \right), {0\le\vartheta\le {\vartheta_{\rm {FoV}}}}
	}
	\\
	{0,\,\,{\rm{otherwise,}}}
	\end{array}} \right.
	\end{equation} 
where $m$ is the Lambertian index which is calculated by $m=-1/\log_2{\left(\cos\left(\Phi_{1/2}\right)\right)}$, with $\Phi_{1/2}$ the half-intensity radiation angle, $A_{\rm{PD}}$ is the physical area of the PD, $d$ denotes the distance between the AP and the user, $\Phi$ is the angle of 
irradiance, $\vartheta$ is the angle of incidence, $T\left( {{\vartheta }} \right)$ and $G\left( {{\vartheta}} \right)$ are the gains of the optical filter and the non-imaging concentrator, respectively, and ${\vartheta_{\rm FoV}}$ is the FoV of the PD. The gain of the concentrator can be expressed as $G\left( {{\vartheta }} \right)~=~f^2/\sin^2{\vartheta_{\rm FoV}},\,0\le \vartheta \le {\vartheta_{\rm FoV}},$ where $f$ is the refractive index. 

With regard to NLoS configurations (which occur mainly in indoor deployments), reflections from wall surfaces and furniture need to be considered. According to   \cite{komine5012004,Tang022021} the optical power received from signals reflected more than once is negligible. As a result, only the signals from the LoS path and those from the first reflected links are typically considered. By focusing on the effect of reflective light by any wall surface $k$, the channel gain of the first reflection is given as \cite{komine5012004}

	\begin{equation}\label{cg11}
	{G_{\rm NLoS}^{{\rm wall}_k}} = \left\{ {\begin{array}{*{20}{l}} \rho_{\rm{wall}} \frac{{\left( {m + 1} \right)}{A_{\rm{PD}}}}  {{2\pi^2} \left(d_{k}^{a}\right)^2 \left(d_{k}^{u}\right)^2} dA_k  \cos^{m}\left(\Phi_k^a\right){\cos }\left( {{\vartheta_k^a}} \right){\cos }\left( {{\Phi_u^k}} \right)\\
	\times \,{\cos }\left( {{\vartheta_u^k}} \right)T\left( {{\vartheta}} \right)G\left( {{\vartheta}} \right),\,\, 0\le \vartheta_u^k\le {\vartheta_{\rm {FoV}}}\\\\
	
	{0,\,\,{\rm{otherwise,}}}
    \end{array}} \right.
	\end{equation} 
where $\rho_{\rm{wall}}$ denotes the reflection coefficient of the wall surface, $d_k^a$ is the distance between the AP and reflective surface $k$, $d_k^u$ is the distance between reflective surface $k$ and the user, $\Phi_k^a$ is the angle of irradiance from the AP to reflective surface $k$, $\vartheta_k^a$ is the angle of incidence on the reflective surface $k$, $\Phi_u^k$ is the angle of irradiance from the reflective surface $k$ towards the user, and $\vartheta_u^k$ is the angle of incidence of the reflected signal from surface $k$.

Unlike RF communication systems, VLC links do not suffer from the effects of multi-path fading since the receivers use detectors with a surface area typically of magnitude bigger than the transmission wavelength. Another unique feature of the VLC channel is its susceptibility to blockages and shadowing as well as impact of the device's orientation. As a result of the short wavelength of VLC signals, specific shadows are formed when the light signals encounter any opaque obstacle such as a human body. As a consequence, a receiver in the shadowed area will be in communication outage. With regard to the impact of device orientation, PDs have limited FoVs. This restricts the angle at which a PD can  receive the optical signals as the angle of the incident light significantly affects the intensity of the received optical signal. While the angle of irradiance is not affected by the random orientation of the user's device, the angle of incidence is highly influenced by the orientation of the device. It is shown in \cite{Soltani032019} that the cosine of the angle of incidence $\vartheta$ can be expressed in terms of the device's polar angle $\alpha$ and the azimuth angle $\beta$ as

\begin{equation}\label{mr1}
\begin{array}{*{20}{l}}
{\cos}\left( {{\vartheta }} \right)=\left(\frac{x_a-x_u}{d}\right){\sin}\left( {{\alpha}} \right){\cos}\left( {{\beta}} \right) + \left(\frac{y_a-y_u}{d}\right){\sin}\left( {{\alpha}} \right)\times\\\,\,\,\,\,\,\,\,\,\,\,\,\,\,\,\,\,\,\,\,\,\,\,\,{\sin}\left( {{\beta}} \right) + \left(\frac{z_a-z_u}{d}\right){\cos}\left( {{\alpha}} \right),
\end{array}
\end{equation}
where $\left(x_a,y_a,z_a\right)$ and $\left(x_u,y_u,z_u\right)$ denote the position vectors specifying the locations of the AP and the user, respectively. According to  \cite{Soltani032019}, the polar angle can be modeled using the truncated Laplace distribution with a mean and standard deviation of $41^{\circ}$ and $9^{\circ}$, respectively, and its value lies in the range $[0,\frac{\pi}{2}]$. The azimuth angle follows a uniform distribution: $\beta \sim \mathcal{U}[-\pi,\pi]$ \cite{Soltani032019}.}}

\textbf{Noise over the VLC channel: }Several noise sources are identified over the VLC channel. They occur in both the optical and electrical domains, and are present in both indoor and outdoor environments. Among these, shot and thermal noises are the most prominent. Shot noise, also called Poisson or quantum noise, is an optical noise and is related to the particle nature of light. This noise defines the variation of the number of electrons generated after the photons hit the PD and may originate from coherent or thermal lights. When due to the former, it follows the Einstein statistics, and follows the Poisson statistics when resulting from the latter \cite{Ndjiongue012020}. Shot noise bears a normal distribution for a high number of photons falling on the PD's area \cite{Ndjiongue012020}. The electronic circuitries of the transmitter and receiver generate thermal noise, which is also called Johnson or Nyquist noise, and follows the normal distribution since it is modeled by the central limit theorem. Other noises such as background and Fano noises are present in the VLC environment, but their amplitude is small enough to be neglected.

\textbf{Interference in the VLC channel: }Signal deterioration in VLC is also due to other light sources which interfere with the message signal.They are mainly two groups: (\textit{i}) natural sources such as the sun and moon. Sun and moon rays may disturb the message encrusted in the light beam. In general, they increase the number of photons which land on the effective area of the PD and force it to work in the saturation region; and  (\textit{ii}) interference from artificial light sources such as other LEDs, fluorescent bulbs and other light sources in the environment.
\subsection{Modulation and Coding Schemes}
\subsubsection{Type of transmitted signal}
The signal transmitted over the VLC channel is optical since the message signals are carried over by a light beam. Because the light is positive by nature, the transmitted signal is positive. It is worth emphasizing that the component which produces the light, i.e., LD or LED, are diodes and do not allow a negative current to flow.
\subsubsection{Modulation schemes}
Most modulation schemes proposed for VLC systems relate to the asymmetric and positive aspects of the VLC signal. An analysis of the VLC technology considers two main groups of modulation schemes, namely, standardized and non-standardized techniques. IEEE 802.15.7 D3a proposes most of the standardized modulation schemes which are all associated with a specific physical (PHY) layer \cite{Ndjiongue012020}. Here, except for those that use phase shift keying (PSK) for example, most schemes naturally produce the required positive signal. Besides this constraint, the modulation technique should also satisfy dimming and flickering requirements of VLC, and efficiently convey information. Most of these schemes produce the required real and positive-valued signal after one or sometimes several operations, such as direct current (DC) offset-orthogonal frequency division multiplexing (OFDM) and asymmetrically clipped optical OFDM, amongst others.

\textbf{Standardized modulation schemes:} Among these, we underline on-off keying (OOK) and variable pulse position modulation (VPPM), which are used with PHY I and II. CSK proposed for PHY III which can be used in combination with OOK. Optical variances of OOK and VPPM such as undersampled frequency-shift-OOK (UFSOOK), twinkle VPPM, and offset VPPM for PHY IV, camera-based OOK (C-OOK) for PHY V, and hidden asynchronous quick  link (HA-QL) for PHY VI, are also proposed in IEEE 802.15.7 D3a. A complete description of these modulation techniques, their corresponding data rate and coding schemes used to generate both inner and outer codes, are provided in \cite{Ndjiongue012020}.

\textbf{Non-standardized modulation schemes: }OFDM cannot be applied directly in VLC due to the restrictions of IM/DD schemes (real and positive values of transmitted signals). Therefore, different variations of OFDM have been proposed, such as DC-biased optical OFDM \cite{hu022019}, asymmetrically clipped DC-biased optical OFDMs \cite{na032018}, asymmetrically clipped optical OFDM \cite{wang012019}, fast-OFDM, and polar-OFDM. Among the OWC versions of OFDM, optical OFDM techniques were proposed with an aim of applying schemes such as quadrature amplitude modulation to VLC systems. All these schemes try to provide a modulated signal which meets the asymmetric aspect of VLC, while keeping the system cost-effective and efficient. Note that all of these versions of OFDM suffer from a high peak-to-average power ratio. The optical version of MIMO (index modulation) has been investigated in order to improve VLC transmission. Other schemes, such as space shift keying (SSK), generalized SSK, spatial modulation (SM), and multiple active SM, have also been proposed. There is also evidence in the literature that other higher-order schemes have been developed for VLC systems. This includes quad-LED and dual-LED complex modulation, as well as quad-LED complex modulation (QCM), which are used in MIMO VLC systems, quadrature spatial modulation (QSM), and dual-mode index modulation.
\subsubsection{Coding schemes}
Coding schemes such as the Reed Solomon, Manchester, and convolutional codes, to mention only three, are used to correct errors in VLC transmission systems. Most are proposed by IEEE and are related to the six physical layers \cite{Ndjiongue012020}. In PHY I, for example, Reed Solomon is indicated to be used with OOK as outer code while convolutional codes is for inner code, and Manchester is to be used as line code. 
\subsection{Applications of VLC Technology}
In general, two sets of applications are considered for the VLC technology: indoor and outdoor applications. Since the sun lights up the environment during daylight, the illumination feature of VLC is necessary outdoors only at night \cite{Ndjiongue062018}. The most prominent applications of the VLC technology are Internet broadcasting and LiFi \cite{Haas032016, Alshaer092018}, which can be deployed indoors and outdoors, and indoor positioning.
\subsubsection{Light fidelity}

\begin{figure}
	\centering
	\includegraphics[width=0.5\textwidth]{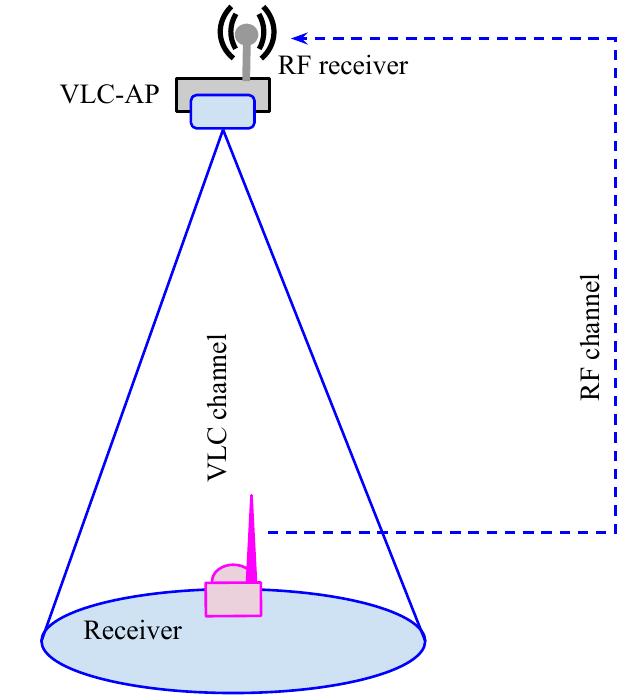}
	\caption{\small{A duplex VLC transmission system.}}
	\label{fig:5}
\end{figure}
In LiFi \cite{Haas032016, Alshaer092018}, the VLC AP broadcasts internet and the light source acts as a portal for devices to connect to a local area network. The message to be transmitted is moved to the AP. Figure~\ref{fig:5} depicts the configuration of LiFi which is a duplex system in which VLC is in charge of the downlink traffic (lighting plus data transfer), while a different technology is used in the uplink \cite{Ndjiongue012020}. Thus, it belongs to the category of hybrid networks, where the uplink employs RF because using VLC in both downlink and uplink produces much interference and leads to poor communication performance. Other technologies, such as IR or laser, can also be used. However, RF offers the possibility of a fully mobile receiving node. In this situation, the mobile terminal integrates a VLC receiver and an RF transmitter (see Fig.~\ref{fig:5}). In this system, VLC is used to supplement the WiFi downlink \cite{figueiredo102017, papanikolaou032018}. It is worthwhile emphasizing that, although LiFi is seen as an application of VLC, it is also considered as an emerging technology that involves VLC and RF. Besides the utilization of LiFi discussed above, it can also be used in multiple other applications, including Internet-of-things (IoT) \cite{albraheem072018}, intelligent transportation systems, road security, and massive data transfer. It is also used in combination with existing systems and technologies, such as Bluetooth and indoor positioning, smart communication and lighting, with hybrid systems \cite{wu122017}, or in combination with an universal serial bus dongle. In most of these cases, the technology may face challenges related to the handover mechanism \cite{soltani032017} or interference \cite{surampudi082018}.

\begin{figure}
	\centering
	\includegraphics[width=0.45\textwidth]{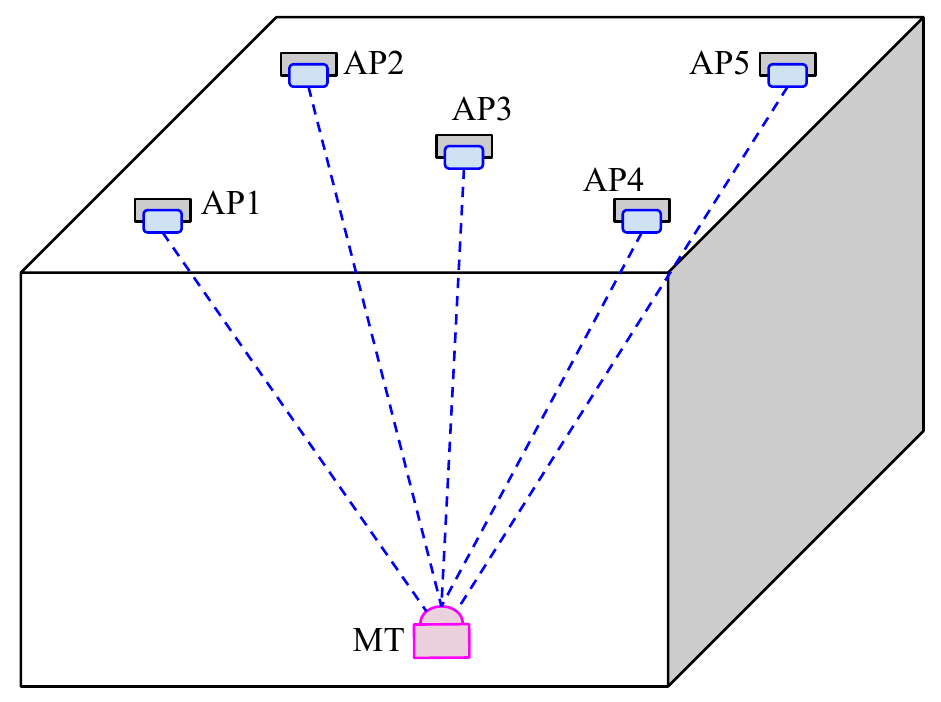}
	\caption{\small{Principles of indoor positioning using VLC.}}
	\label{fig:6}
\end{figure}
\subsubsection{Indoor positioning}
An indoor positioning system is a useful application of VLC which allows users to move inside an illuminated building and be notified of their physical position should they request this. It is one of the most frequently investigated topics in VLC technology \cite{pittolo042016, artale032018, prasad112020, belhassen032020, lin042017, li072018, zhu012018, park042017, gu052016, alam072019, feng052016, hou122016, qin012019, du082018, guo012019, yu122018}. The same AP used to deploy LiFi is exploited to give an accurate address to the node on its current position in the building. As stated earlier, in indoor positioning, localization and navigation systems, the light bulbs included in the AP serve as illumination devices. The indoor positioning system is appended on these light sources to create a network of APs to locate users. Figure~\ref{fig:6} depicts an illustration of an indoor positioning system in which the mobile terminal is moving in an indoor environment. As it moves from one point to another, its PD detects several signals. The receiver included in the mobile terminal selects the signal with the best signal-to-noise ratio (SNR), which in general, corresponds to the closest AP. The demonstration of VLC-based indoor positioning systems \cite{lin042017, alam072019, du082018} shows high-accuracy positioning. Most studies on this topic are theoretical, but touch on different aspects of the system. For example, they may deal with a single PD, multiple PDs \cite{yu122018}, optical cameras \cite{li072018}, off-the-shelf components \cite{hou122016}, as well as in relation with artificial intelligence \cite{guo012019}, amongst others.

{\color{black}\subsubsection{Underwater Wireless Communication}
Underwater wireless communication typically relies on RF or acoustic signal. However, acoustic communication links are characterized by large propagation delays and low bandwidth (tens of hertz and hundreds of kilohertz), which limits the amount of information that can be transmitted. Although RF links outperform acoustic waves in terms of data rate and high tolerance to turbulence and turbidity \cite{9664274}, their high energy consumption and deployment cost coupled with the short transmission distances necessitates the search for alternative underwater communication technologies to meet the extremely high data rate requirements of beyond 5G. VLC has recently attracted considerable research attention for underwater wireless communication due to its ability to provide the highest transmission bandwidth and data rate as well as the lowest link delay and implementation cost \cite{7593257}. Moreover, underwater VLC systems are more secure because the transmitted signal is highly directional and energy efficient since low cost laser diodes and PDs are used as the transmitter and receiver, respectively. To that end, recent works have focused on path loss channel modeling \cite{8449320}, turbulence channel modeling \cite{8370053,9140399}, multiple access schemes \cite{9664274,9585115}, modulation scheme \cite{9590553,9324916}, transceiver design \cite{8610103,8935430}, and security \cite{9760160}.}  

\subsection{Re-configurable Intelligent Surfaces for VLC Systems}
The use of RISs  in VLC systems is very recent and is attracting a great deal of research interests \cite{di042020, wu012020, hu032018, liang062019, basar091019, han062019, huang062019, basar022020}. This is due to its potential to revolutionize the design of future VLC-based wireless networks. The technology was firstly introduced by Berry in 1963, where he proposed the reflectarray antenna \cite{liang062019}. Since then, the research community has studied absorbing boards, selective windows and walls, and frequency selective surfaces. Nowadays, it bears several other names. The concept is called: (\textit{i}) large intelligent surface to indicate the area exploited to contain the RIS units; (\textit{ii}) large intelligent meta-surface or reconfigurable meta-surface because the surface accommodating the RIS elements is characterized by a complex and artificial electromagnetic structure; (\textit{iii}) intelligent reflecting surface or smart reflect-arrays because the incident signal on the RIS components can be reflected; (\textit{iv}) passive intelligent surface or passive intelligent mirrors, when the RIS elements do not amplify the received signal.  

\subsection{Standardization Efforts for VLC Technology}
The three main factors governing the evolution of a technology are the market, the technology, and regulation. Regulation refers to rules, requirements, and guidelines for VLC products, process, and services \cite{Ndjiongue012020}. The effective implementation of the VLC technology and the integration of the RIS technology must follow a specific guideline. The VLC technology is regulated by standards on short-range optical wireless communications. Up to date, a few drafts of these standards have been proposed, including from IEEE and the VLC consortium. In IEEE, the IEEE 802.15.7 Task Group specifies wireless personal area network standards and deals with rules and regulations for the VLC technology. They have successively lunched IEEE Std 802.15.7-2011, IEEE Standard for Local and Metropolitan Area Networks--Part 15.7: Short-Range Wireless Optical Communication Using Visible Light \cite{IEEE092011}. 
\begin{table*}[t]
	\centering
	\caption{IEEE 802.15.7/D3a: Summary of the operating modes for PHY I to VI \cite{Ndjiongue012020}.}
	\label{tab:table1}
	\begin{tabular}{*7c}
		\hline
		\textbf{Mod. scheme} & \textbf{Line Code} & \textbf{Clock Rate} & \multicolumn{2}{c}{\textbf{Forward Error Correction}} & \textbf{Data Rate}\\
		{}   & {}   & {}  & \textbf{Outer Code}   & \textbf{Inner Code} & {}\\
		\hline
		\multicolumn{6}{c}{|\textbf{PHY I}|} \\
		\hline
		OOK   &  Manchester & 200 kHz   & RS  & CC & 11.67 kbps to 100 kbps \\
		VPPM   &  4B6B & 400 kHz   & RS  & CC& 35.56 kbps to 266.6 kbps \\
		\hline
		\multicolumn{6}{c}{|\textbf{PHY II}|} \\
		\hline
		VPPM   &  4B6B & 3.75 MHz/7.5 MHz  & RS &RS & 1.25 Mbps to 5 Mbps \\
		OOK   &  8B10B & 15 MHz to 120 MHz  & RS&RS & 6 Mbps to 96 Mbps \\
		\hline
		\multicolumn{6}{c}{|\textbf{PHY III}|} \\
		\hline
		CSK   &  -.- & 12 MHz/24 MHz  & RS&RS & 1.25 Mbps to 5 Mbps \\
		OOK   &  8B10B & 15 MHz to 120 MHz  & RS&RS & 12 Mbps to 96 Mbps \\
		\hline
		\multicolumn{6}{c}{|\textbf{PHY IV}|} \\
	\hline
		UFSOOK   &  -.- & Multiframe rate  & \multicolumn{2}{c}{MIMO path dependent} & 10 bps\\
		Twinkle VPPM   &  -.- & 4x bit rate  &RS&RS & 4 kbps \\
		S2-PSK   &  Half-rate code & 10 Hz  & \multicolumn{2}{c}{Temporal error correction} & 5 kbps \\
		HS-PSK   &  Half-rate code & 10 kHz  & RS&RS & 22 kbps \\
		Offset VPPM   &  -.- & 25 Hz  & RS&RS & 18 bps \\
		\hline
		\multicolumn{6}{c}{|\textbf{PHY V}|} \\
	\hline
		RS-FSK   &  -.- & 30 Hz  & XOR FEC&XOR FEC & 120 bps \\
		C-OOK   & Manchester/4B6B & 2.2 kHz/4.4 kHz  & Hamming code & Optional/RS & 400 bps \\
		CM-FSK   & -.- & 10 Hz  & -.- & Optional & 60 bps \\
		MPM   &  -.- & 12.5 kHz  & \multicolumn{2}{c}{Temporal error correction} & 7.51 bps \\
		\hline
		\multicolumn{6}{c}{|\textbf{PHY V}|} \\
		\hline
		A-QL   &  -.- & 10 Hz  & RS & CC & 5.54 kbps \\
		HA-QL   & Half-rate code & 10 Hz & RS & CC & 140 bps \\
		VTASC   & -.- & 30 Hz  &  RS&RS & 512 kbps \\
		SS2DC   &  -.- & 30 Hz  &  RS&RS & 368 kbps \\
		IDE-MPSK blend   & -.- & 30 Hz  & RS&RS & 32 kbps \\
		IDE-WM   &  -.- & 30 Hz  &  RS&RS & 256 kbps \\
			\hline \\
	\end{tabular}\par
	{{\bf OOK}: on-off keying; {\bf VPPM}: variable pulse position modulation; {\bf CSK}: color shift keying; {\bf UFSOOK}: undersampled frequency shift on-off keying;\\
	{\bf S2-PSK}: spatial 2-phase-shift-keying; {\bf HS-PSK}: hybrid spatial phase-shift-keying; {\bf C-OOK}: camera on-off keying; {\bf RS}: Reed-Solomon;\\ {\bf CM-FSK}: camera M-ary frequency-shift-keying; {\bf MPM}: mirror pulse modulation; {\bf A-QL}: asynchronous quick link; {\bf HA-QL}: hidden asynchronous quick link; \\{\bf VTASC}: variable transparent amplitude-shape-color; {\bf SS2DC}: sequential scalable 2D code; {\bf IDE-MPSK}: invisible data embedding M-ary phase-shift-keying;\\ {\bf IDE-WM}: invisible data embedding watermark; {\bf CC}: convolutional coding; {\bf -.-}: no line code or forward error correction code has been proposed.}
\end{table*}
This standard was successively revised several times. Thus, in 2018, the IEEE 802.15.7 Task Group proposed a new draft, IEEE P802.15.7/D2a, IEEE Draft Standard for Local and metropolitan area networks - Part 15.7: Short-Range Optical Wireless Communications \cite{IEEE072018} with a slight difference that the focus is not only on the VLC technology, but all technologies with similar characteristics such as infrared. The IEEE P802.15.7/D2a draft was improved to IEEE P802.15.7/D3, which led to an approved draft, P802.15.7/D3a, in August 2018 \cite{IEEE122018}. Finally, in 2019, the IEEE task group released revised version of the standard for VLC technology, IEEE 802.15.7-2018, in 2019 \cite{IEEE042019}. The main focus of all these versions of the IEEE 802.15.7 standards are the modulation schemes, the forward error correction and line codes, and data rates over short range optical channels in local and metropolitan networks. 

In Table~\ref{tab:table1}, we summarize the activities of IEEE 802.15.7 and classify the related important parameters. The PHY layer described in IEEE 802.15.7 is divided into 6, namely PHY I, II, III, IV, V, and VI. Each of these has specific modulation scheme, coding techniques, and data rate for any communication. They also have different clock rates, which vary with the different schemes used.
\subsection{Summary}
In this section, we have revisited the VLC technology and highlighted the multiple-use of its light source, channel, and receiver. We have discussed its principle, the modulation schemes, and coding techniques, and highlighted its applications. Finally, standardization efforts have been provided. It turns out that the VLC technology is a good candidate to overcome the spectrum shortage of RF systems, especially in the indoor environment.

\section{RIS Basics}
\subsection{RIS: The Concept}
An RIS can be defined as a metasurface or a mirror consisting of an array of low-cost nearly passive reflecting elements for reconfiguring incident signals and manipulating (e.g., reflecting, refracting, focusing, etc.) them in an intelligent way to improve communication performances. Specifically, each of the RIS elements can be configured  individually, and in real-time, to induce controllable manipulation of some characteristics (e.g., amplitude, phase, polarization, etc.) of the incident signal. For instance, the use of RISs enable the direction of any reflected wave to be controlled such that all the waves converge to a point (i.e., anomalous reflection) rather than having specular reflection. For that to happen, the electromagnetic response of each of the reflecting elements is first adjusted by tuning the surface impedance through electrical voltage stimulation. This causes each element of the RIS array to induce a phase shift to the incoming signals, and as a result, controlling the main direction of the reflected signals. In general, the control mechanism in RISs can be realized by using ultra-fast switching elements such as varactors, positive-intrinsic-negative (PIN) diodes, or micro-electro-mechanical systems (MEMS) switches that communicate with a central controller. As opposed to requiring human subjective judgement and recognition in controlling the operation of traditional metasurfaces, an RIS controller has the capability to sense the environment \cite{Renzo052019}, and make use of intelligent algorithms to actively identify and judge environmental changes and make autonomous decisions on its operations \cite{Luo082021,Ma102019}. As a result, a dense deployment of RISs in any wireless communication network will allow full manipulation of transmitted and reflected waves to enable an intelligent control of the communication channel and signal propagation to enhance the end user's quality of experience.

The RIS technology has recently gained significant research attention in wireless communications due to the numerous benefits it offers including: (\textit{i}) metasurfaces that are used in RISs are easy to fabricate using traditional nanofabrication techniques such as photolithography and electron-beam lithography due to the rapid advancement in the semiconductor industry; (\textit{ii}) their ease of deployment since RISs can be deployed on existing infrastructure like  the exterior and interior of buildings, roadside billboards, t-shirts, etc.; (\textit{iii}) their low energy consumption and carbon footprint; (\textit{iv}) key performance metrics (spectral efficiency, throughput, energy efficiency, and coverage) enhancements especially in the absence of a LoS path between the transmitter and the receiver; and (\textit{v}) compatibility with the standards and hardware of existing wireless networks. {\color{black}{There has been extensive research on its application in RF communications. However, the RIS technology and its application in RF communication systems cannot be directly adapted to VLC systems due to the reasons summarized in Table~\ref{vlc_rf_ris}.}}

\begin{table*}[t]
	\centering
\caption{A comparison of VLC and RF RISs.}
	\label{vlc_rf_ris}
 \begin{tabular}{|l|l|l|}
\hline
\textbf{Feature}         & \textbf{VLC}                                                                                                                                                                                                                                                               & \textbf{RF}                                                                                                                                                                                                                     \\ \hline
Signal characteristics   & \begin{tabular}[c]{@{}l@{}}Wavelength ranges from 350 nm to 800 nm\\ Intensity modulation and direct detection\\ Real- and positive-valued signals \\ Intended for communication and illumination\\ Emitted from LEDs and received by \\ PDs\end{tabular} & \begin{tabular}[c]{@{}l@{}}Wavelength ranges from 1 mm to 10 m\\ Coherent modulation and demodulation\\ Complex-valued signals\\ Intended for communication\\ Emitted from and received by electromagnetic \\ transceivers\end{tabular} \\ \hline
Typical functionalities  & \begin{tabular}[c]{@{}l@{}}Dynamic FoV control\\ Light amplification\\ Wavelength filtering and interference \\ suppression\\ Coverage expansion and beam focusing\\ Illumination relaxation\end{tabular}                                                        & \begin{tabular}[c]{@{}l@{}}Coverage expansion and beam focusing\\ Interference nulling\end{tabular}                                                                                                                             \\ \hline
Hardware                 & \begin{tabular}[c]{@{}l@{}}Metasurfaces\\ Mirror arrays\\ Liquid crystals\end{tabular}                                                                                                                                                                                     & Metasurfaces                                                                                                                                                                                                                    \\ \hline
Place of deployment      & \begin{tabular}[c]{@{}l@{}}At the transmitter side (e.g., in front of the LED)\\ At the receiver side (e.g., in front of the the PD)\\ In the channel between transmitter and receiver\end{tabular}                                                                        & In the channel between transmitter and receiver                                                                                                                                                                                 \\ \hline
Performance optimization & \begin{tabular}[c]{@{}l@{}}Decision variables include roll and yaw \\ orientation angles of mirror arrays, phase shift\\ for metasurfaces, and refractive index for liquid\\ crystals\\ Communication and illumination constraints\end{tabular}                            & \begin{tabular}[c]{@{}l@{}}Phase shift as the decision variable\\ Communication related constraints\end{tabular}                                                                                                                \\ \hline
Propagation model        & \begin{tabular}[c]{@{}l@{}}Novel channel models required for metasurface,\\  mirror arrays, and liquid crystals-based RISs\end{tabular}                                                                                                                                          & \begin{tabular}[c]{@{}l@{}}Novel channel model required for metasurface-based\\  RISs\end{tabular}                                                                                                                               \\ \hline 

Technological maturity        & Moderate                                                                                                                                        & High                                                                                                                               \\ \hline 
Cost        & Low                                                                                                                                        & Moderate                                                                                                                               \\ \hline 

\end{tabular}
\end{table*}

In the subsections that follow, the two different setups for any RISs in VLC systems, namely, intelligent metasurface reflector (IMR) and intelligent mirror array (IMA) are briefly introduced. Then, detailed discussions on their operating principles and functions, in the context of communication, are also provided.

\subsection{Intelligent Metasurface Reflector (IMR): From the Physics Point of View}
A typical IMR consists of three main layers, namely, a metasurface for the outermost layer, a conducting back plane that prevents energy leakage as the second layer, and a control circuit that connects to a micro-controller as the third layer. It is important to note that this third layer distinguishes an RIS from classical reconfigurable reflectarrays and array lenses \cite{Hum012014}. In this subsection, a description of the typical structure of a metasurface is provided. Then, the various tuning mechanisms that enable the reconfigurable properties of metasurface reflectors are discussed in terms of the tuning material, the operating frequency range, and the typical application.

\subsubsection{Structure and tuning mechanisms}
A metasurface is a two-dimensional artificially nanostructured interface that is composed of spatially arranged meta-atoms of a subwavelength size on a flat substrate. These meta-atoms typically consist of dielectric \cite{Saman012016, nicolas072015} or plasmonic \cite{Boltasseva012011} nanoantennas that can directly reconfigure properties such as the phase, the amplitude, and the polarization, of any incident signal by manipulating the outgoing photons. The types of substrates used in metasurfaces include silicon, gallium arsenide, sapphire, germanium, quartz, polymide, and parylene. Metasurfaces in general have been widely investigated in the past decades because of their unique abilities for blocking, absorbing, focusing, reflecting, or guiding incident waves ranging from the microwave band through the optical frequency bands \cite{Li062018}. Such unique abilities result from their strong interaction with electric and/or magnetic fields, which is typically provided by resonant effects controlled by the geometry of the meta-atoms. In the early development stages of metasurfaces, they were mostly designed for specific functions. For instance, a  metasurface absorber composed of a reflective backplane and a microwave absorption layer sandwiched between two dielectric substrates, only works for a certain or a narrow range of frequencies. As such, complete redesign and re-fabrication were required for the metasurface to be able to absorb signals of different frequency range. 

Recently, real-time re-configurable (or programmable or tunable) metasurfaces have received enormous research attention due to their ability to offer multiple unique functionalities without any re-fabrication processes \cite{liu052018, chang072018}. Such RISs generally consist of a metastructure and a tuning mechanism, and both components communicate through a control circuit. Several tuning mechanisms for realizing real-time re-configurable metasurfaces have been proposed in the literature. Popular tuning mechanisms and their corresponding practical applications in the design of IMR are presented below: 

\begin{itemize}
    \item \textit{Liquid crystals:} Infusing a metasurface with LCs and varying its permittivity or refractive index with external stimuli (e.g., electric field, magnetic field, temperature, etc.) is a known tuning mechanism for re-configurable metasurfaces \cite{Lininger082020}. A typical structure of an operational re-configurable LC-based metasurface for VLC receivers is shown in Fig.~\ref{figssep13}. In this figure, the polarizer filters any incoming light and allows only the wavelengths carrying the transmitted information signals. The glass substrates generate the preferred direction of orientation for the LC molecules when an external electric field (e.g., source voltage) is applied to the electrode. The indium tin oxide (ITO) material assists with heat generation and control for the LC cell, and finally, the photoalignment layer guides light-rays through the preferred direction in the LC cell. By infiltrating the metasurface with LCs, the resonance condition of the meta-atoms (i.e., nanodisks) become dependent on the refractive index of the LC medium. By varying the source voltage at the request of the RIS controller, the refractive index of the LCs can be easily adjusted by undergoing molecular reorientation and, in return, can modify the phase and amplitude of any incident signal. 

\begin{figure}%
    \centering
    {{\includegraphics[width=0.5\textwidth]{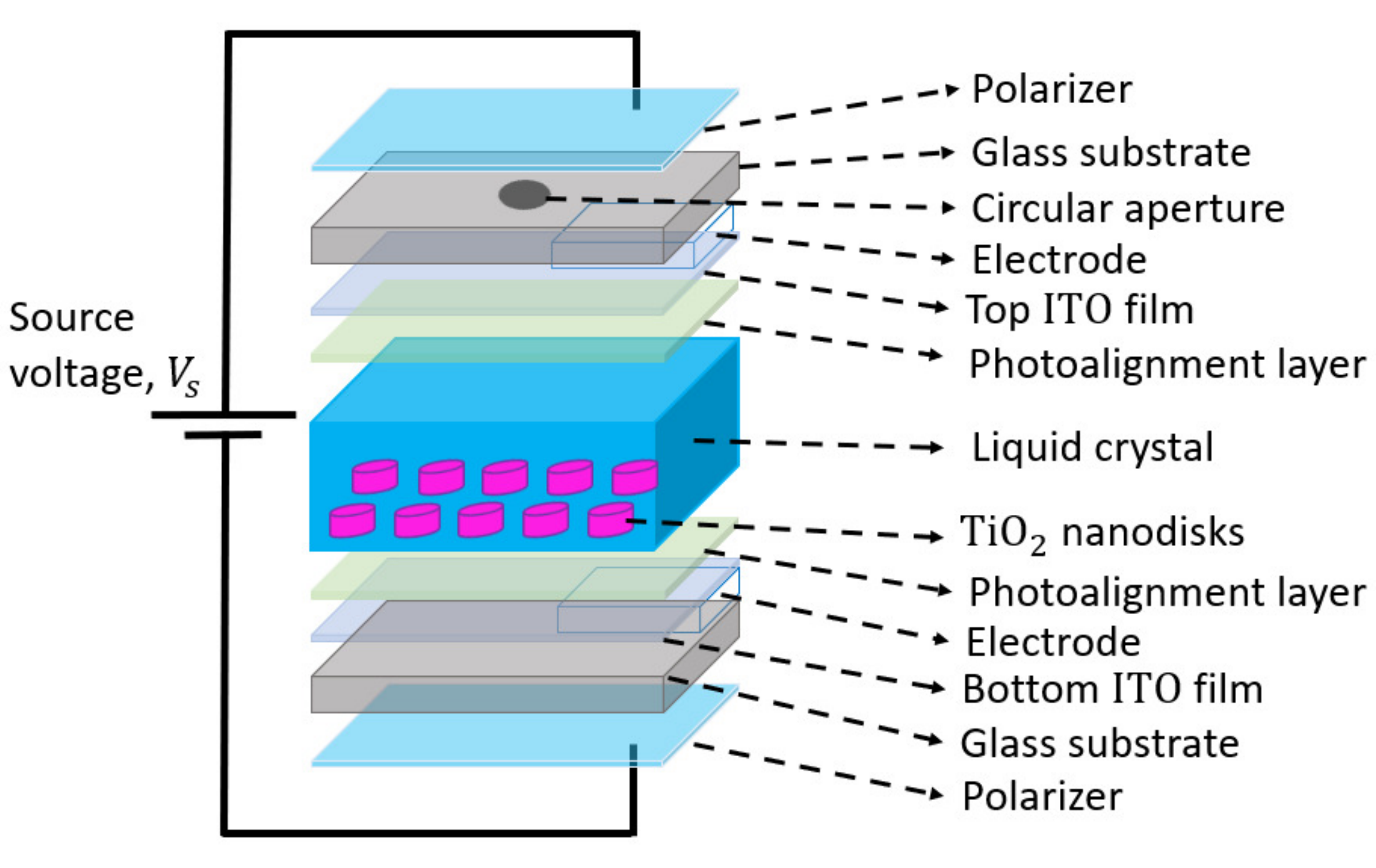}}}
    \caption{\small{An electronic tunable tin oxide $\rm{\left(TiO_2\right)}$ metasurface with LC infiltration sandwiched between two transparent plate electrodes deposited upon  $\rm{\left(ITO\right)}$-coated glass substrates \cite{ndjiongue052021, Sun062019}.}}%
  \label{figssep13}
  \vspace*{-1mm}
\end{figure}

       \begin{figure}%
    \centering
    {{\includegraphics[width=0.5\textwidth]{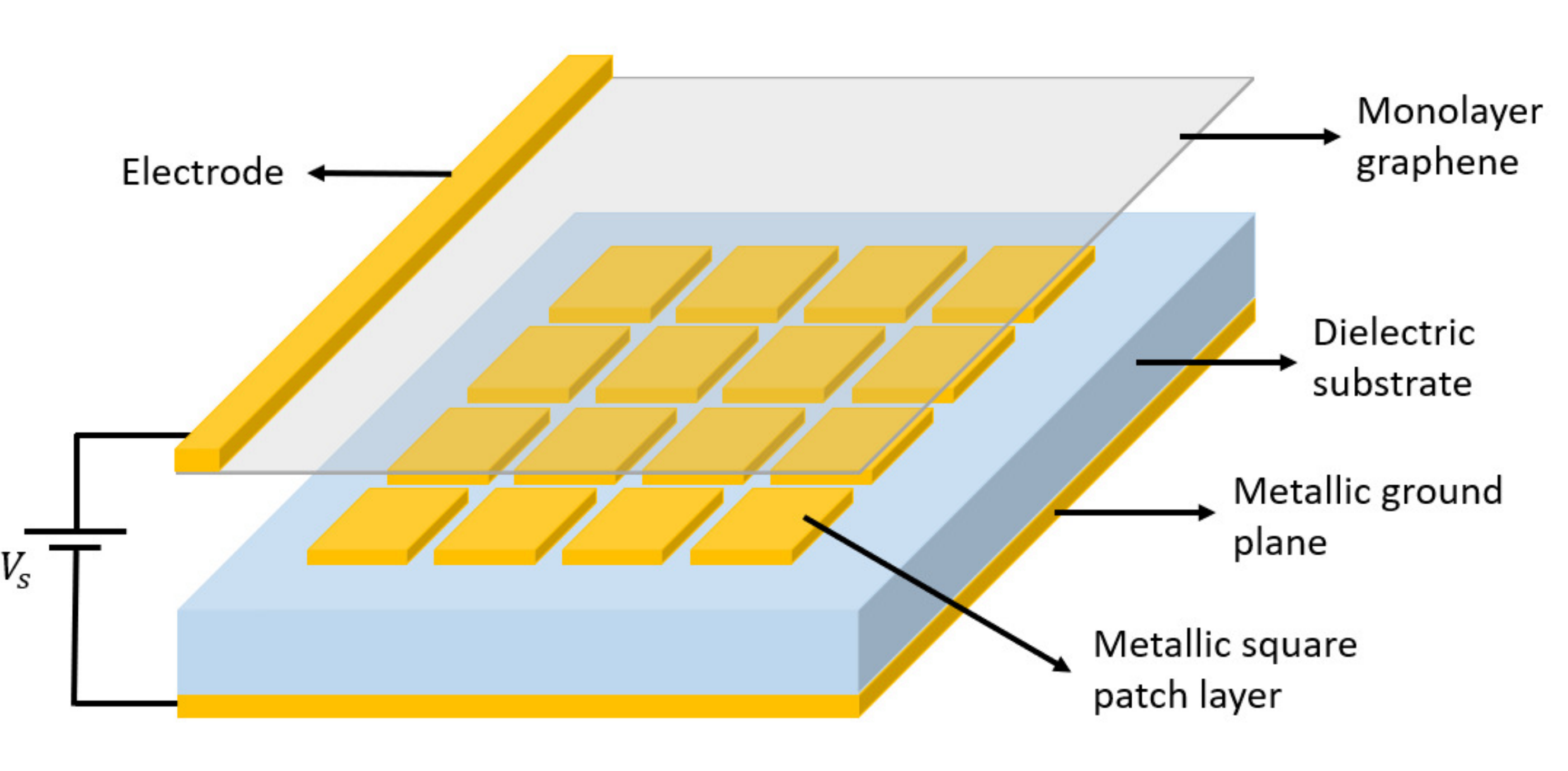}}}
    \caption{\small{An electronically tunable metasurface composed of a graphene sheet directly placed on the array of metal patches \cite{Wang122017}.}}%
  \label{figsep15b}
  \vspace*{-2mm}
\end{figure}
    
    \item \textit{Graphene:} Graphene, one of the most preferred RIS materials, is a flat monolayer of carbon atoms tighly packed into a two-dimensional honeycomb lattice whose surface conductivity and, consequently, impedance can be easily tuned by applying a proper gate voltage \cite{Aldrigo102014}. Specifically, embedding a metasurface (i.e., array of meta-atoms) with graphene and adjusting its Fermi level by modifying the Fermi energy through a bias voltage enables the realization of a tunable metasurface \cite{Lee092012, Wang122017}. Graphene possesses a chemical-potential-dependent complex conductivity which can be tuned under electrostatic bias to enable the dynamic control of the reflection phase and amplitude characteristics of any incident beam \cite{Giddens042018}. Figure~\ref{figsep15b} shows a typical graphene-based metasurface, which consists of a continuous graphene sheet (i.e., tunining material) transfer-printed on an open dielectric surface with an array of metal patches to form a patterned surface, and backed by a metallic ground plane. In this figure, the metal layer on the backside of the substrate acts as a reflector and an electrode, and the applied voltage is between the top and the back electrodes. By varying the applied voltage to adjust the Fermi level, the chemical potential of the graphene sheet can be dynamically tuned  to enable different functionalities like wave absorption and the control of wave characteristics like amplitude, phase, and polarization. Thus, by utilizing the tunable conductivity properties of a graphene  material in a metasurface connected to a control unit, intelligent metasurfaces with dynamic reconfiguration  can be efficiently designed.

    \item \textit{Photoconductive semiconductor:} Introducing light-sensitive materials (i.e., photoconductive semiconductor materials) in the metasurface composition and varying its property (e.g., conductivity) through carrier photo-excitation with an IR pump beam is another tuning mechanism for re-configurable metasurfaces\cite{liu052018}. Photoconductivity can be defined as an optical and electrical phenomenon in which a material becomes more electrically conductive when exposed to light (e.g., visible light, IR light, or ultraviolet light) of sufficient energy. This phenomenon occurs when light -- with energy higher than the bandgap energy -- strikes a photoconductive material and causes some electrons to move from the valence band into the conduction band (i.e., move across the band gap), and as a result, changes the electrical conductivity of that material. Typical examples of such materials include silicon, gallium arsenide, silicon carbide, and lead sulfide. Typically, a photoconductive metasurface consists of a pair of metal contacts placed on a photosensitive semiconductor substrate (photoconductor) \cite{Jiang092021}, as depicted in  Fig.~\ref{figsep24}. The tunable operation of this metasurface is explained as follows \cite{Bhattacharya102019}: An optical beam, with photon energy exceeding the bandgap of the semiconductor substrate, is used to irradiate the photoconductive gap between the electrodes. This causes photocarrier generation in the semiconductor substrate. At the same time, the bias voltage applied across the electrodes accelerates the photogenerated carriers towards the anode, resulting in an instantaneous drift current density. The rate of change of the movement of the charged carriers in the semiconductor substrate (i.e., change in its conductivity) enables the dynamic manipulation of the amplitude characteristics of the incident wave and the operating  frequency of the metasurface. In addition to the references above, other design ideas of such metasurfaces can be found in \cite{Shen012011, MAJDAZDANCEWICZ201892, Shen042009}.
    
    \begin{figure}%
    \centering
    {{\includegraphics[width=0.5\textwidth]{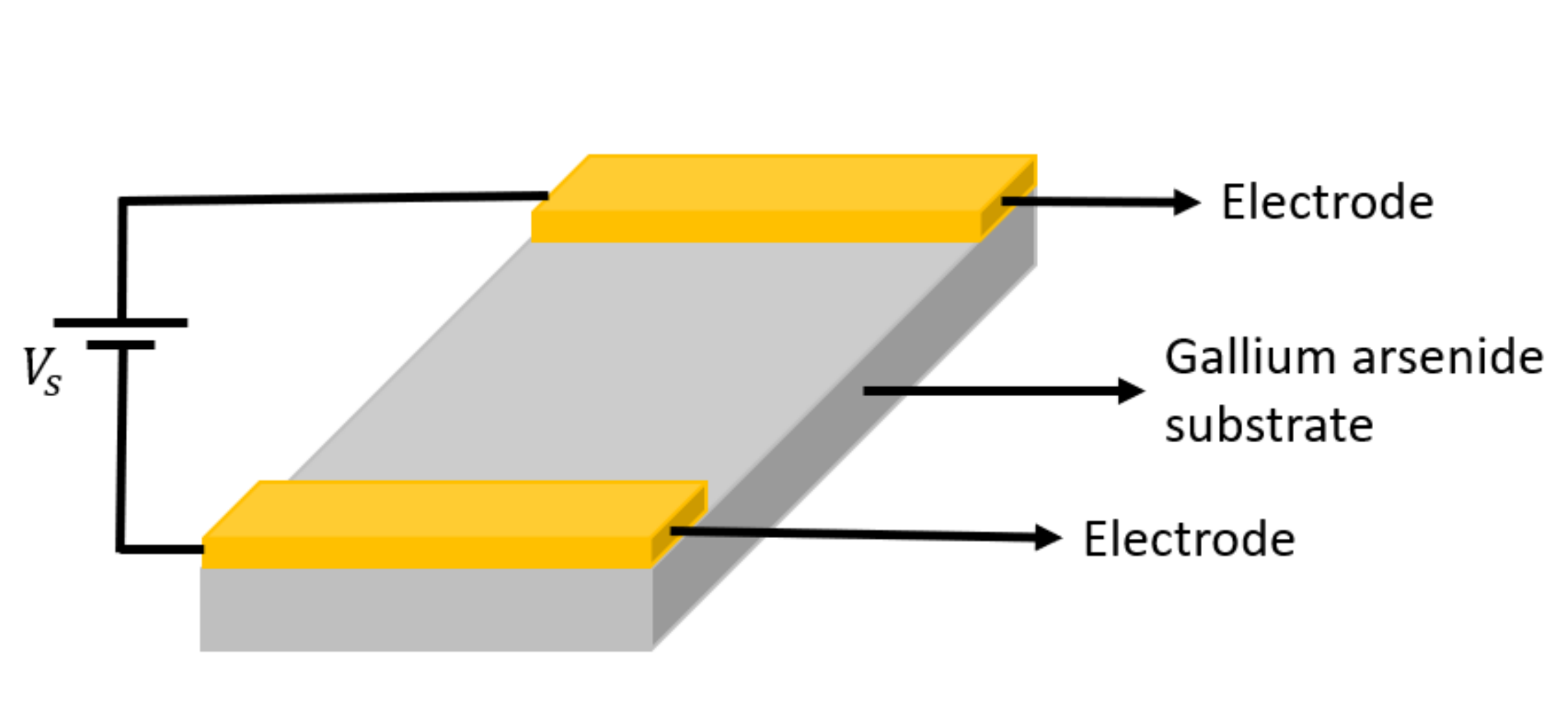}}}
    \caption{\small{A tunable photoconductive metasurface.}}%
  \label{figsep24}
\end{figure}
    
\end{itemize}

\subsubsection{Typical tunable functionalities and applications in communication systems}
\begin{figure*}[t]%
    \centering
    {{\includegraphics[width=0.6\textwidth]{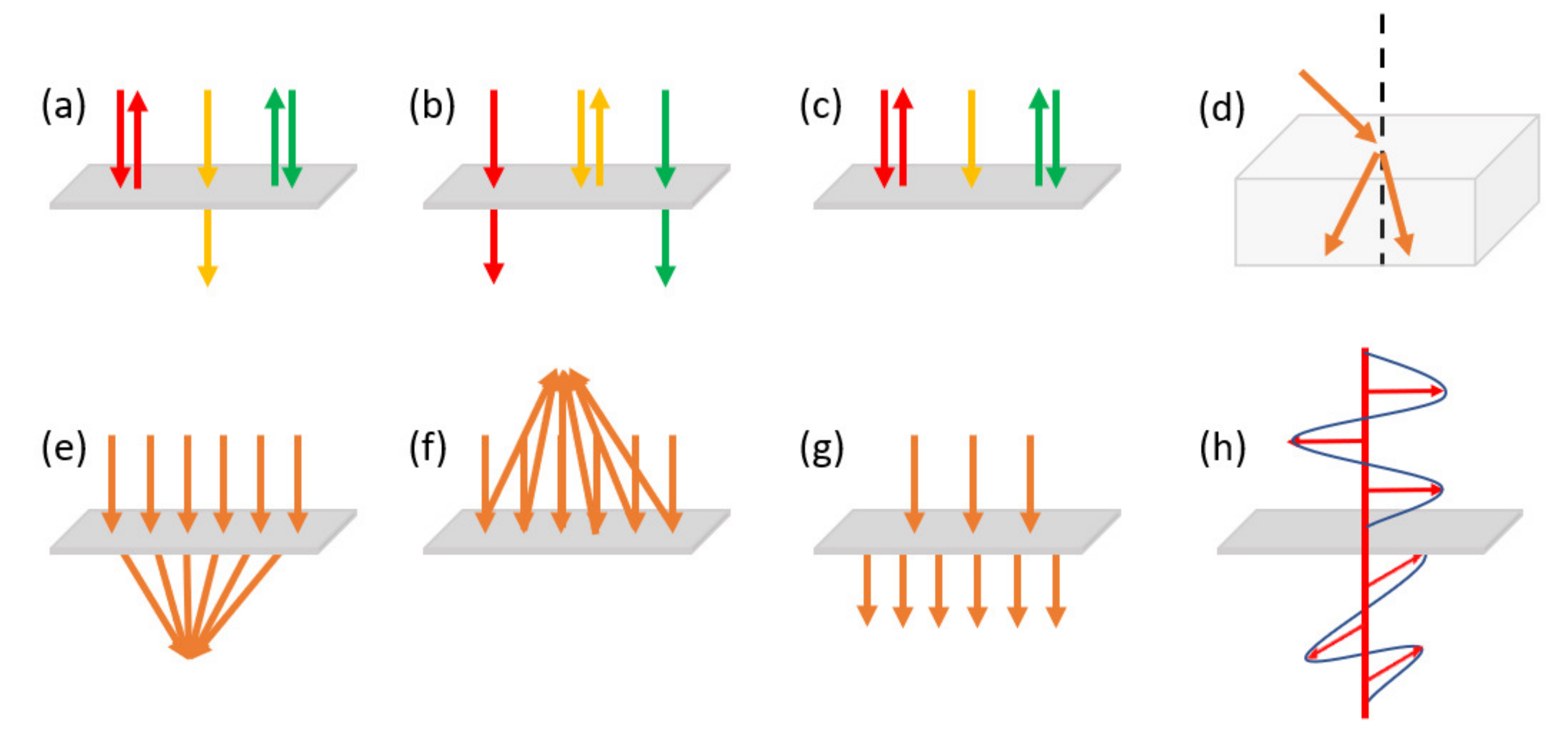}}}
    \caption{\small{Selected functionalities of metasurfaces \cite{GLYBOVSKI052016, Abumarshoud042021}: (a) bandpass frequency selective surface; (b) bandstop frequency selective surface; (c) narrowband perfect absorber; (d) refractive index tuning; (e) beam steering transmitarray; (f) beam steering reflectarray; (g) beam amplification; (h) polarization transformation.}}%
  \label{figssep27}
\end{figure*}

Figure~\ref{figssep27} shows some of the typical functionalities performed by metasurfaces in wireless communication systems. Particularly, Figs.~\ref{figssep27}(a) and (b) demonstrate the use of a metasurface to perform spectral filtering. This functionality has several signal processing applications  in communication systems. Notable among them is the design of a tunable optical filter (\textit{i}) for signal detection in multi-color VLC systems, (\textit{ii}) to block outdoor light noise in VLC systems, and (\textit{iii})  to function as a low-cost, nearly passive optical identifier in VLC-based indoor positioning systems. Figure~\ref{figssep27}(c) shows a scenario whereby a metasurface has been used as a perfect absorber. A material is said to be a perfect absorber when it absorbs 100$\%$ of the incident wave power under a specified angle of incidence at a single frequency \cite{GLYBOVSKI052016}. Perfect absorbers, including narrowband absorbers, have useful applications in many areas including  
interference management in RF radars where they are used to suppress backscattering by large metal targets. Figure~\ref{figssep27}(d) depicts a scenario where the incident signal is refracted towards the opposite side of the impinging signal. This particular functionality of RISs is crucial to the development of intelligent omni-surfaces that are capable of reflecting and refracting impinging signals towards both sides of the metasurface \cite{9491943}. Figures~\ref{figssep27}(e), (f), and (g) depict scenarios of wavefront shaping with metasurfaces. Specifically, Fig.~\ref{figssep27}(e) shows the use of a metasurface to steer the beam from the transmitter in a particular direction. This is useful in coverage extension for wireless communication systems. Figure~\ref{figssep27}(f) shows the use of a metasurface to control the main direction of a reflected signal (i.e., achieve anomalous reflection). For instance in optical transmission, when light waves leave a source, they spread out in all directions and upon striking any smooth, finite-sized flat surface, they get reflected away from the surface at the same angle as they arrived and the intensity of the reflected light is not always equal to that of the incident light as some of the light get absorbed by the surface (i.e., specular reflection). However, anomalous reflection can be achieved if each element of the metasurface induces a certain phase shift to the incoming signal and the overall joint effect of all phase shifts is a reflected beam in a specified direction. Figure~\ref{figssep27}(g) and Fig.~\ref{figssep27}(h) depict the scenario whereby a metasurface is used for signal amplification and polarization transformation, respectively.  

A summary of the various tuning materials and their typical frequency range of application is provided in Table~\ref{tab_meta}. In this table, typical functionalities of the tuning materials and their maturity are also provided.

\begin{table*}[t]
	\centering
\caption{A comparison of the various tuning mechanisms for optical RIS and their corresponding typical functionalities.}
	\label{tab_meta}
\begin{tabular}{|l|l|l|l|}
\hline
\textbf{Tuning Material}  & \textbf{Frequency Range of Application}                                                                                                                 & \textbf{Typical functionality}                     &    Maturity                                                               \\ \hline\hline
LCs            & \begin{tabular}[c]{@{}l@{}}Millimeter wave band (30  -- 300 GHz)\\ Terahertz band (300 GHz -- 3 THz)\\ Near-infrared band (214 -- 400 THz)\\ Visible light band (430 -- 730 THz)\end{tabular} & \begin{tabular}[c]{@{}l@{}} Wave absorption\\Amplitude and phase manipulation   \end{tabular}                                                                                               & Medium \\ \hline
Graphene                   & \begin{tabular}[c]{@{}l@{}}Terahertz  band\\ Far-infrared band (0.3 -- 20 THz)\\ Mid-infrared band (37 -- 100 THz)\\ Near-infrared band\\ Visible light band \end{tabular}                                                              & \begin{tabular}[c]{@{}l@{}}Wave absorption\\ Amplitude, phase, and polarization\\ state manipulation\end{tabular}  & Low \\\hline
\begin{tabular}[c]{@{}l@{}} Photoconductive\\ semiconductor \end{tabular}& \begin{tabular}[c]{@{}l@{}}Terahertz  band\\ Far-infrared band\\ Mid-infrared band\\ Near infrared band\\ Visible light band\\ Millimeter wave band\\ Microwave band \end{tabular}                                                              & \begin{tabular}[c]{@{}l@{}}Wave absorption\\ Amplitude, phase, and polarization\\ state manipulation\end{tabular}  & Low \\\hline
\end{tabular}
\end{table*}

\begin{figure}%
    \centering
    {{\includegraphics[width=0.5\textwidth]{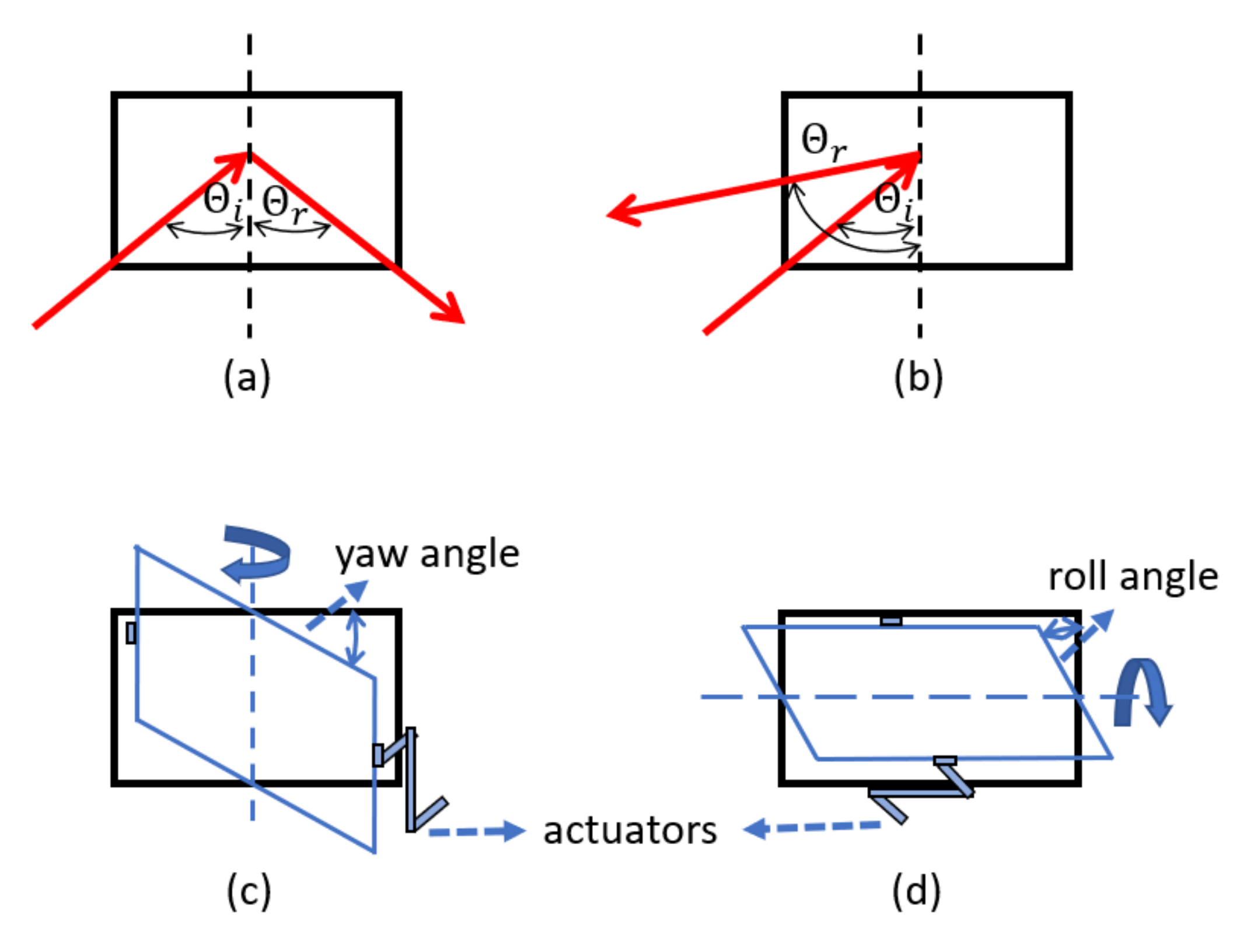}}} 
    \caption{\small{Controllable mirror as RIS: (a) specular reflection $(\Theta_i = \Theta_r)$; (b) anomalous reflection $(\Theta_i \ne \Theta_r)$; (c) mirror array orientation according to the yaw angle; (d) mirror array orientation according to the roll angle.}}%
  \label{figsoct14}
\end{figure}
\subsection{Intelligent Mirror Array (IMA): From the Physics Point of View}
Mirrors offer another approach to realize RISs, especially for optical communication systems. As depicted in Fig.~\ref{figsoct14}(a), the relationship between the angles of incidence, $\Theta_i$, and reflection, $\Theta_r$, of any plane mirror is governed by Snell's law. According to this law, on reflection from a smooth surface, the angle of the reflected ray is equal to the angle of the incident ray (i.e., $\Theta_i = \Theta_r$) and the reflected ray is always in the same plane defined by the incident ray and the normal to the surface. However, recent advancements in MEMS technology has enabled the development of reconfigurable mirrors that can guide and control light dynamically. MEMS are particularly well suited for optical applications because they are well matched to optical wavelengths, and can be manufactured in high volume and high density arrays in semiconductor manufacturing processes \cite{Chen122004}. As shown in Fig.~\ref{figsoct14}(b), the use of an electro-mechanical mirror array  allows the arbitrary control of the direction of the reflected ray and, as a result, the angles of incidence and reflection are no longer necessarily the same (which is in accordance to the generalized Snell's law). As shown in Figs.~\ref{figsoct14}(c) and (d), the micro-mirror uses a compactly folded actuator design to tune its yaw and roll angles via electrostatic actuation, respectively, enabling it to perform a wide range of operations including wave-front shaping and beam steering. Typical design of such controllable mirror array and its operating principle have been reported in \cite{Lei122010,David082016}. In comparison to metasurfaces, the MEMS-based mirror arrays offer numerous advantages such as (\textit{i}) relatively lower power consumption, (\textit{ii}) all of the incident light is always reflected with the same intensity (i.e., there is no absorption), (\textit{iii}) they work at very low temperature, and (\textit{iv}) despite using mechanically mobile parts, their lifetime is long due to miniaturization \cite{Hillmer072018}.

\subsection{Summary and Lessons Learned}
This section has revealed that RISs allow  real-time and autonomous intelligent reconfiguration of any impinging signal in contrast to traditional metasurfaces that require human subjective judgement and recognition in controlling any characteristic of the incident signal. The design structure and the tuning mechanisms utilized in the two main types of RIS (i.e., IMR and IMA) have been presented. In particular, the typical tuning mechanisms for the IMR include LCs, tunable chips, graphene, ferrite, and photoconductive semiconductor. With regards to the design and fabrication of the IMA, MEMS are widely used as the tuning mechanism, where actuators are used to configure the individual micro-mirrors via electrostatic actuation. In addition, the typical functionalities and the frequency range of operations for both IMR and IMA have been provided. The discussions in this section can aid engineers and network designers in selecting an appropriate RIS (and tuning mechanism) for a communication system by considering the frequency range of application, the desired required functionality, as well as the maturity of the technology.

\section{Overview/Review of Optical RIS Technologies} 
In this section, a thorough discussion on the various RISs technologies for OWC systems and their deployment scenarios is presented. This is carried out according to the place of deployment (i.e., at the transmitter side, receiver side, or in the channel) of  RISs in any OWC network. As mentioned earlier in Section~III, LCs, graphene, mirror arrays, and photoconductive semiconductor tuning materials can be used to design RISs for OWC systems. 
\begin{figure}%
    \centering
    \includegraphics[width=0.5\textwidth]{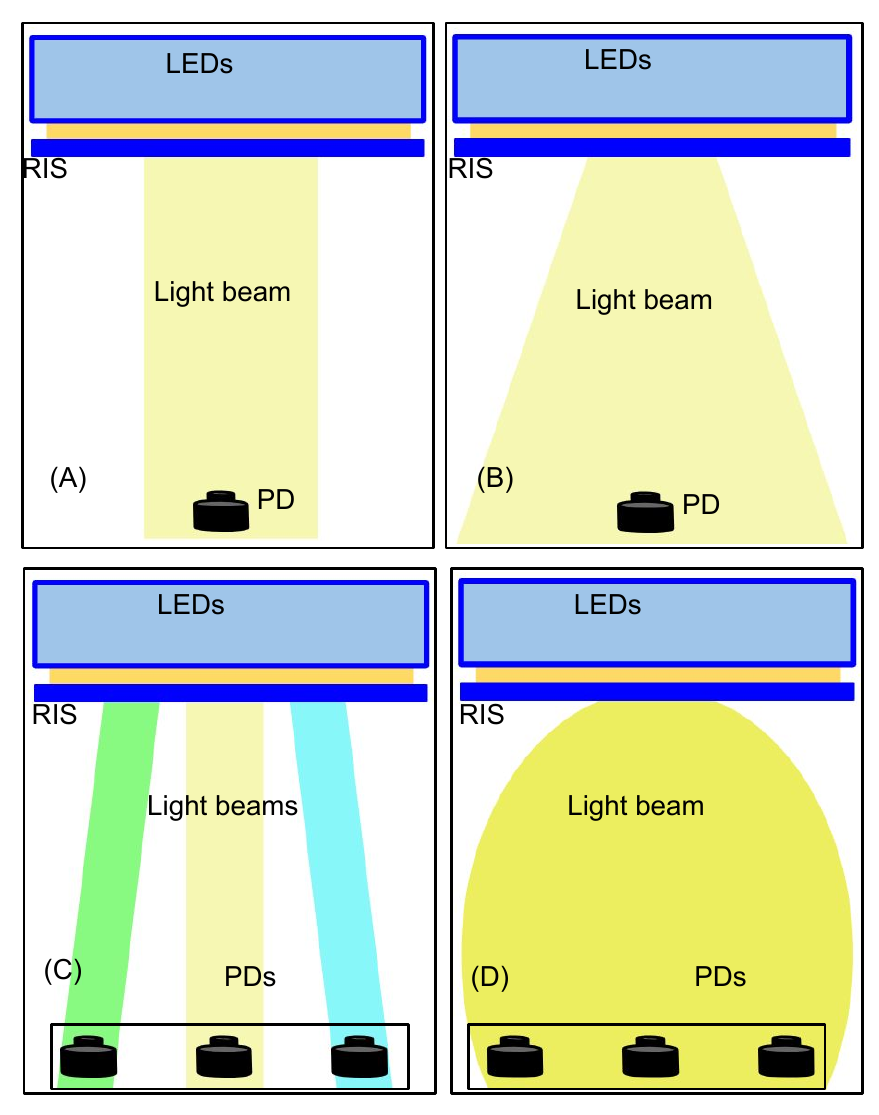} 
    \caption{\small{Envisioned deployment scenarios and functionalities of RIS at the transmitter side.}}%
  \label{figsNov25a}
  \vspace*{-2mm}
\end{figure}   

\subsection{RIS at the Transmitter Side}
The most widely used transmitters in OWC are LEDs and LDs. LEDs are multimode and incoherent light sources that typically emit light from an extended surface area (i.e., large FoV) in almost all directions (i.e., non-directional) and over a wide range of wavelengths. As a result, LEDs are able to provide larger coverage areas and more uniform illumination, but with varying light intensities across different areas. In contrast, LDs are single mode, coherent light sources that emit light from a small area in one direction (i.e., narrow and highly directional beams) and at one wavelength (or a few closely spaced wavelengths). Based on the above-mentioned properties of LEDs and LDs, and the fact that RISs are typically employed at the transmitter side to perform the roles of beam steering and amplification, it is strongly recommended to use RISs at a transmitter equipped with LEDs. {\color{black}Fig.~\ref{figsNov25a} illustrates the many ways that RISs can be exploited at the transmitter side in any VLC system to enable performance improvement factors such as coverage extension, beam concentration, and beam directivity through the principle of light refraction.} In this figure, an RIS is located inside the light source and it is configured to provide different shapes of the light beam for various transmission specifications and applications as demonstrated in Figs.~\ref{figsNov25a} (A) - (D). Specifically,  Fig.~\ref{figsNov25a}~(A) demonstrates the use of an RIS to control the FoV of the VLC transmitter and focus the emitted light beam to the user. Whenever the location of the user changes, the directivity of this beam can be adjusted accordingly with the help of the RIS controller. Figure~\ref{figsNov25a}~(B) depicts the placement of an RIS in front of the transmitter to increase the FoV as well as the illumination coverage of the LEDs and ensure seamless communication for all users in the indoor environment. Specifically, the use of an RIS enables the generation of a Gaussian shaped light beam. Figure~\ref{figsNov25a}~(C) shows the use of RIS to perform configurable wavelength division multiplexing to enable the use of the different colors that combine to form white light (e.g., trichromatic LED-based white light source) or in multichromatic light source to transmit different data streams. In this same figure, it can be noticed that the RIS can be used to design reconfigurable angle diversity transmitters (ADTs) to allow dynamic beam steering and focus to multiple users such that there is no overlap of the optical beams. More particularly, when several single color LEDs are deployed as a point source transmitter, an RIS can be used to focus different information-carrying beams towards the location of the multiple users such that there is no overlap and, as a result, less interference in the communication system. Finally, a spherical type of light beam is generated by the light source in Fig.~\ref{figsNov25a}~(D). In this figure, the use of the RIS offers the ability to change the  shape of the emitted light beams for different design specifications. LC-based RIS has been identified as an appropriate RIS technology for deployment inside the transmitter.     
 
\subsection{RIS at the Receiver Side}
\begin{figure}%
    \centering
    {{\includegraphics[width=0.5\textwidth]{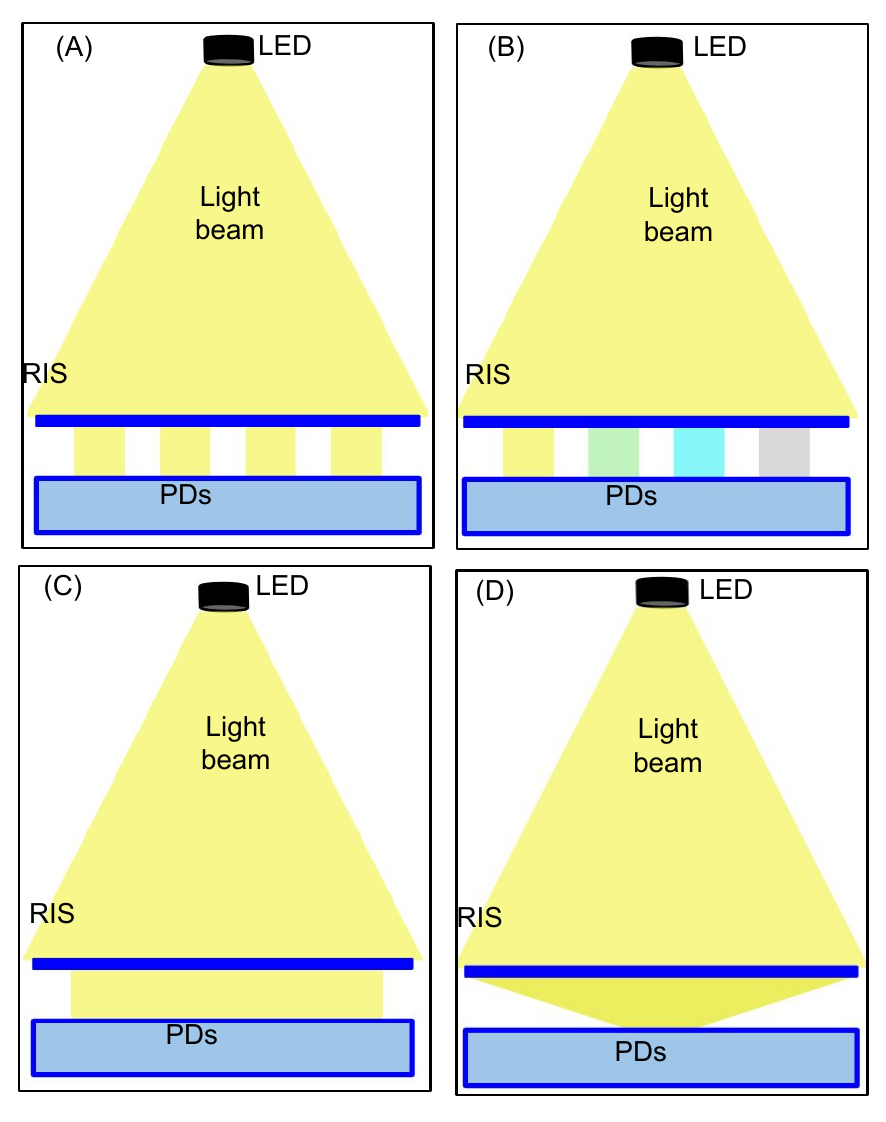}}} 
    \caption{\small{Envisioned deployment scenarios and functionalities of RIS at the receiver side.}}%
  \label{figsNov25b}
  \vspace*{-2mm}
\end{figure} 
{On the other hand, Fig.~\ref{figsNov25b} depicts a RIS-based indoor VLC system where an RIS is placed at the receiver. In this scenario, the RIS helps to steer the detected light and focus its beam on the PD. The RIS can generate multiple parallel identical beams as shown in Fig.~\ref{figsNov25b} (A). It can also generate a multiple parallel beams with  multiple frequencies, as depicted in Fig.~\ref{figsNov25b} (B). Specifically, it can act as a filter to allow light signals of specific wavelength to pass while blocking any unwanted light and, thereby, reducing interference.  The RIS can also generate a parallel beam as shown in Fig.~\ref{figsNov25b} (C), or focus all lights towards the detector surface as illustrated in Fig.~\ref{figsNov25b} (D)}.

\begin{figure}%
    \centering
    {{\includegraphics[width=0.5\textwidth]{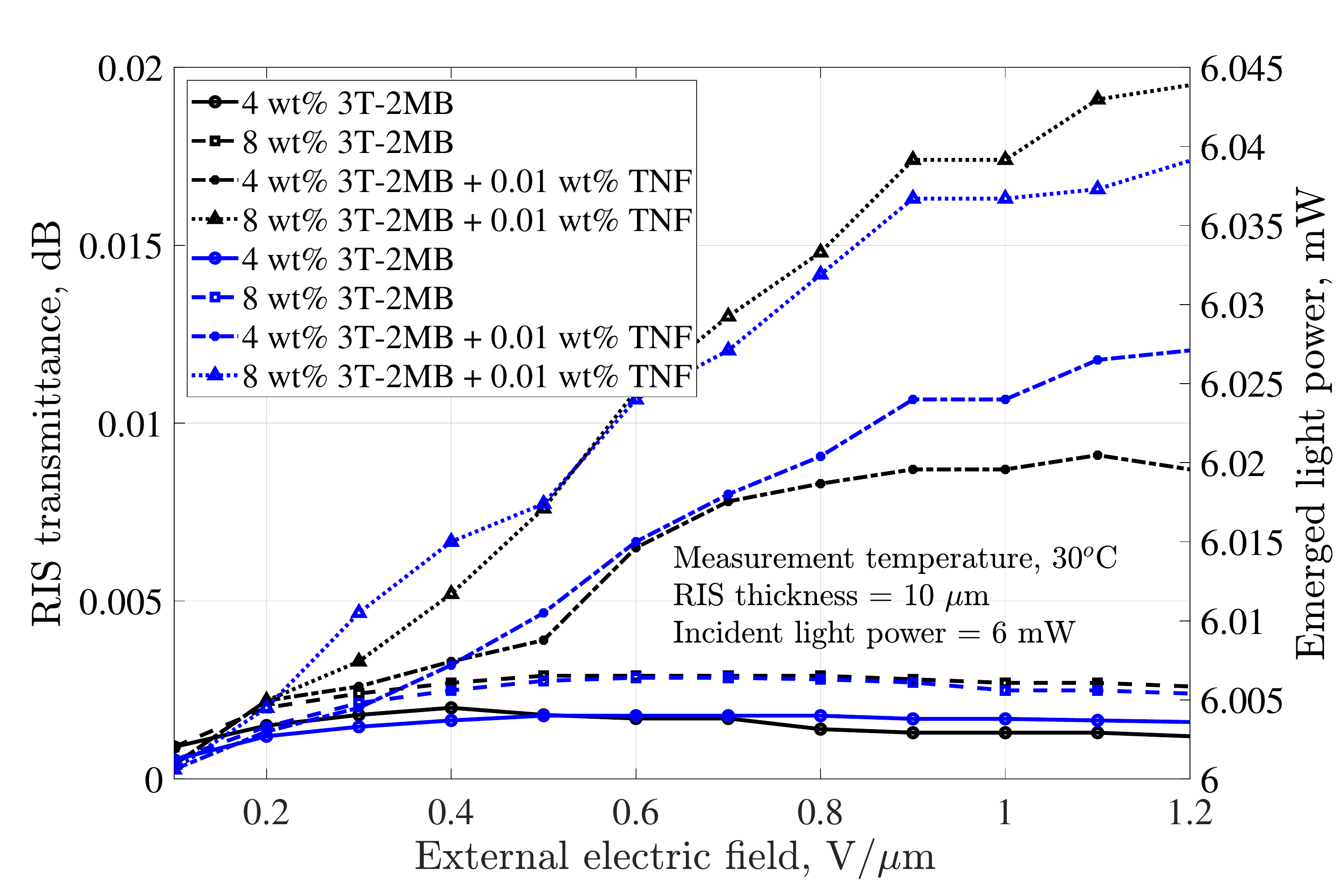}}} 
    \caption{\small{LC-based RIS transmittance and emerged light power for selected LC samples \cite{ndjiongue052021}.}}%
  \label{figRIS_Receiver}
\end{figure} 

\begin{figure}%
    \centering
    {{\includegraphics[width=0.5\textwidth]{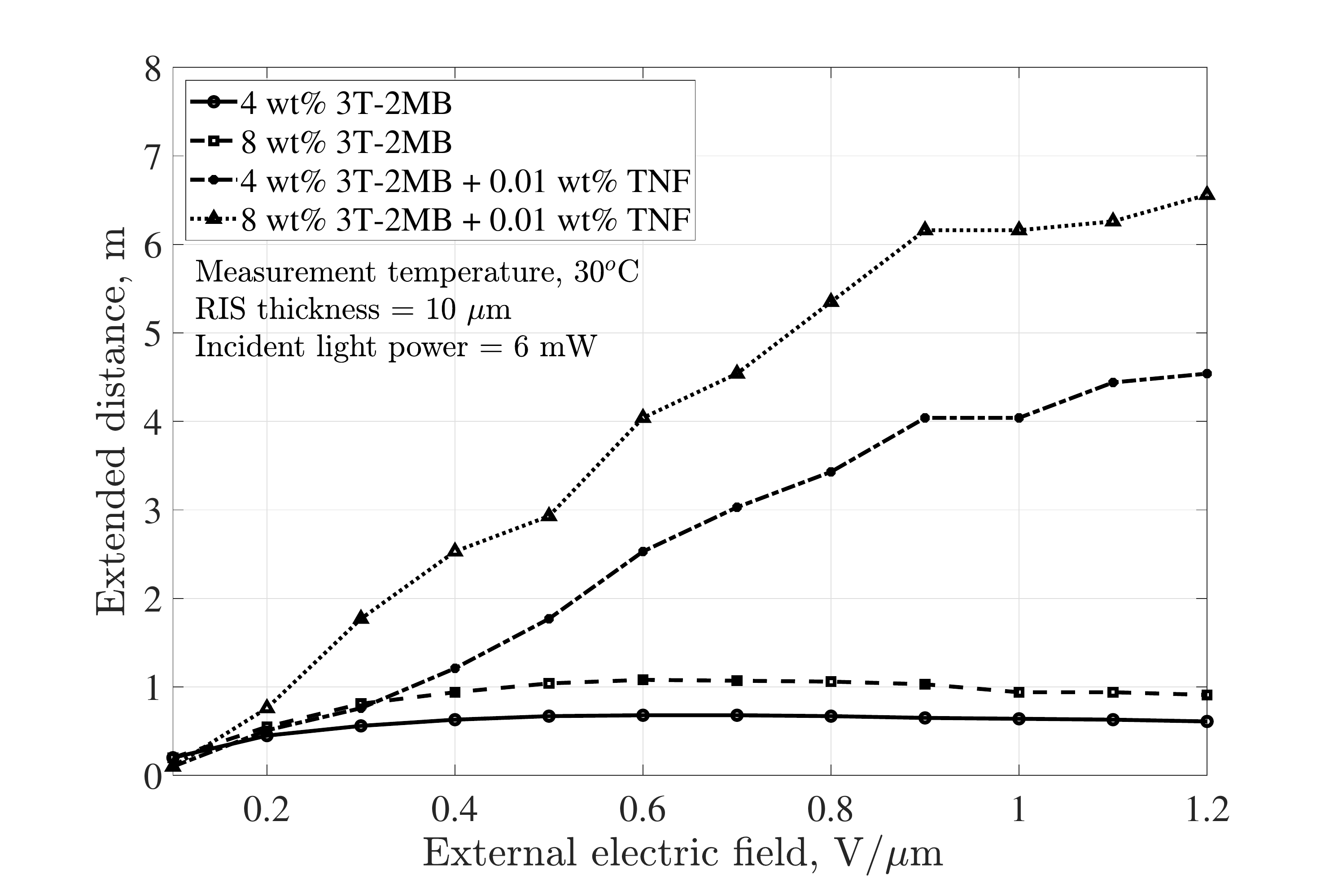}}} 
    \caption{\small{Extended transmission range of a VLC system due to LC-based RIS light amplification, for selected LC samples  \cite{ndjiongue052021}.}}%
  \label{figRIS_Receiver_Distance}
  \vspace*{-2mm}
\end{figure} 

{The most common type of RIS at the receiver is the LC-enabled RIS, which offers several interesting advantages, such as removing unwanted colors from the incoming light, improving the SNR at the PD, extending the transmission range, and most importantly, increasing the receiver's FoV. LC-based RIS, as outlined in \cite{ndjiongue052021,ndjiongue062021, Abumarshoud042021}, are capable of filtering the incoming light to remove all unnecessary colors, including noise, and allow only monochromatic light containing the transmitted data to reach the PD. Meanwhile, LCs are also efficient in amplifying the filtered light and improving the signal power at the PD, thereby improving the SNR \cite{ndjiongue052021}. Finally, LC-based RIS can improve the receiver FoV to an angle of approximately 90$^{^{\circ}}$. However, despite the fact that the transmitted intensity is significantly reduced near 90$^{^{\circ}}$, the amplification provided by the LC compensates for the power loss incurred by the high FoV.}

{These three main advantages are related to the application of an external electric field in the RIS. The RIS transmittance and emerged light power depend upon the externally applied voltage. The transmission coefficient of LC-based RIS increases when a dye is added to the LC's substance. An experimental example of transmittance and emerged light power, and extended range profiles are provided in Figs.~\ref{figRIS_Receiver} and \ref{figRIS_Receiver_Distance}, respectively. In this experiment, we used an LC of 10 $/mu$m thickness at 30 $^{\circ}$C with a variable external electric field of 0 to 1.2 V/$\mu$m. These figures show that any increment of the electric field yields an increase of the RIS transmittance. The resulting light power, and the transmission distance are improved accordingly. These parameters' values are enhanced if a dye is added to the LC substance. Moreover, the figures show that a suitable combination of dyes leads to an improved RIS transmittance, resulting in a greater power allocation to the emerged light.}  

\subsection{RIS between the Transmitter and the Receiver}
\begin{figure*}[t]%
    \centering
    {{\includegraphics[width=0.8\textwidth]{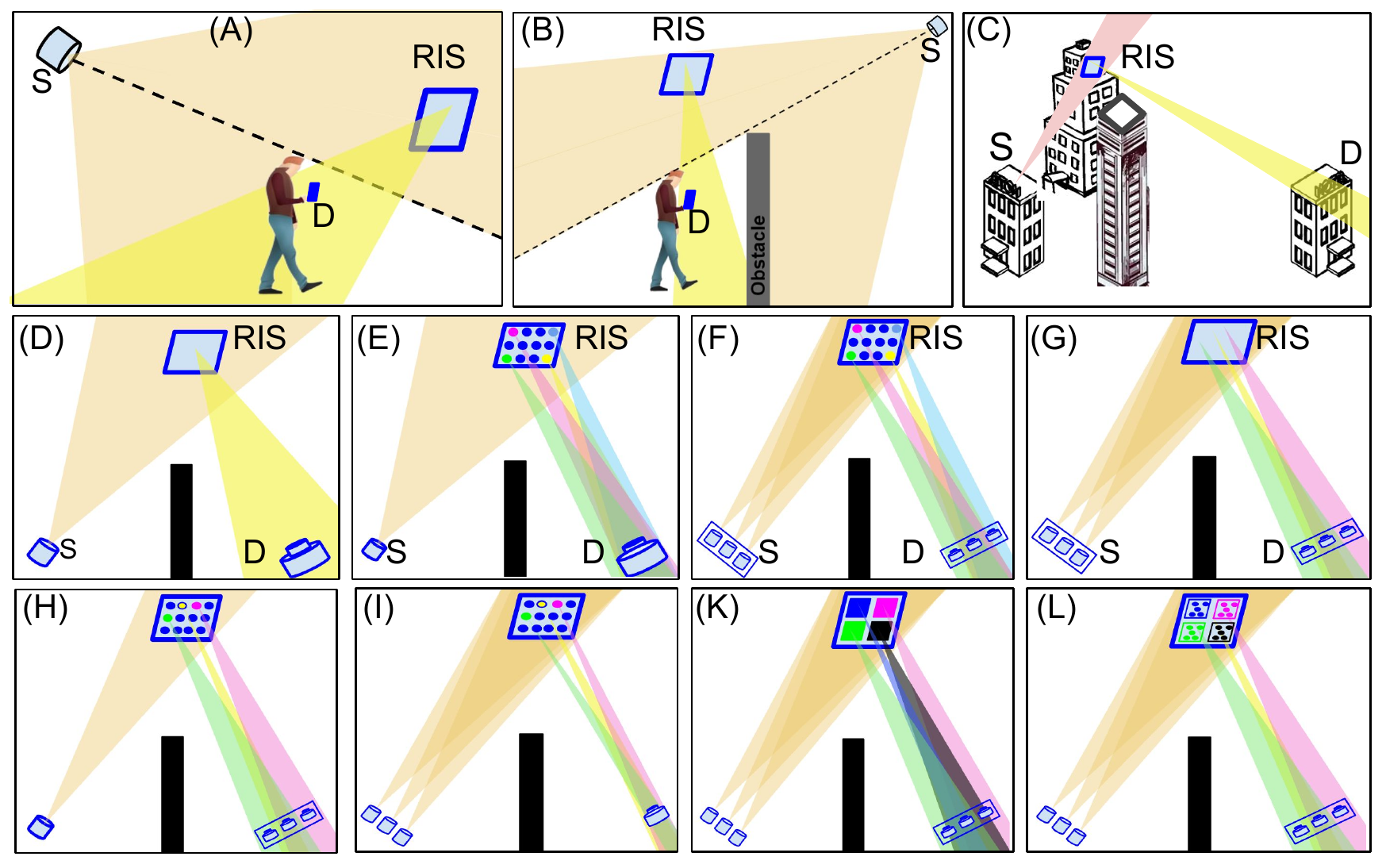}}} 
    \caption{\small{Envisioned deployment scenarios and functionalities of RIS between the transmitter and the receiver.}}%
  \label{figsNov9b}
\end{figure*}

The typical performance impairment in OWC systems is the obstruction of the direct LoS path between the transmitter and the receiver. As such, it is important to investigate the different ways that RISs can be utilized to create virtual LoS paths to ensure successful data transmission. Figures~\ref{figsNov9b} (A) and (B) present the two types of blockage that can be found in any indoor OWC environment. The human and mobile blockages are depicted in Fig.~\ref{figsNov9b} (A) while fix blockage is shown in Fig.~\ref{figsNov9b} (B). The main difference between these two types of blockages is that in the former, the blockage, which is probably human, can move within the environment, while in the latter the blockage is stationary. Fix blockages are walls, tables, or any other object disposed between the light source and the receiver. Figure~\ref{figsNov9b} (C) depicts the use of RISs in outdoor OWC systems such as outdoor VLC and terrestrial FSO. Between the transmitter and the receiver, as an example, a recently constructed new building, obstructs the light from the light source to reach the PD, which is located on another building. As the figures show, the signal is re-directed towards the receiver by appropriately deploying an RIS on a neighboring building. Note that, as in the case of the indoor, the outdoor environment is also composed of mobile blockages. 

Figure~\ref{figsNov9b} (D) to (L) details the different indoor scenarios of RIS-based indoor OWC and highlight the number of light sources, RIS elements, and PDs without any regard to the type of blockage. In Figs.~\ref{figsNov9b} (D) and E, the transmitter and receiver are made of a single light source and PD, respectively. In Fig.~\ref{figsNov9b} (D), the RIS is made of one panel and a single element, while in Fig.~\ref{figsNov9b} (E), the RIS is made of one panel of multiple elements. Note that with multiple elements, the RIS structure helps to reduce geometric losses at the receiver for a more efficient data transmission.

Figures~\ref{figsNov9b} (F) and (G) respectively depict scenarios where the transmitter and receiver are made of multiple light sources and PDs, while the RISs are made of panel with multiple elements for Fig.~\ref{figsNov9b} (F) and a panel of single element for Fig.~\ref{figsNov9b} (G). In Figs.~\ref{figsNov9b} (H) and (I), the RIS is made of a panel of multiple elements. In Fig.~\ref{figsNov9b} (H), the transmitter is made of a single light source and the receiver contains multiple PDs, while in Fig.~\ref{figsNov9b} (I), we have a transmitter with multiple light sources and a receiver with multiple PDs. Figures~\ref{figsNov9b} (K) and (L) depict scenarios with a more sophisticated RIS structure. In both sub-figures, the RIS is respectively made of multiple panels of a single element and multiple elements, respectively, while transmitter and receiver are made of multiple light sources and PDs. Typical optical RISs technologies that can be deployed between the transmitter and the receiver can be an IMR (as discussed in Table ~\ref{tab_meta}) or an IMA. 

\subsection{Summary and Lessons Learned}
OWC systems are able to offer secured high data-rate communication links needed for B5G communication systems. However, the requirement of the existence of LoS path between the transmitter and the receiver remains a major limiting factor in their deployment. This section has discussed an overview of the potential role of RISs in relaxing the LoS requirement as well as improving the performance of OWC systems. Specifically, the different optical RIS technologies have been presented based on their location of deployment in an OWC system. Then, the various ways RISs can be deployed at the transmitter side, receiver side, or in the  channel for performance improvement functions have been studied. This section has revealed the following insights: (\textit{i}) LC-based metasurface can be deployed at the transmitter side to perform various roles such as beam steering and amplification, coverage extension, as well as configurable wavelength division multiplexing; (\textit{ii}) LC-based metasurface can be deployed at the receiving end to act as a filter and/or to focus the impinging optical signal to improve the SNR and/or to adjust the receiver's FoV; and (\textit{iii}) both IMR and IMA are optical RISs that can be deployed between the transmitter and the receiver to enable virtual LoS paths and thus improve the communication performance.

\section{Optical RISs  and Optical Relays}
Optical RIS- and relay-assisted technologies can be used to relax the LoS requirements and improve the SNR at the receiver in OWC systems. However, the two technologies are different in many aspects. Optical RISs normally consists of nearly-passive elements (which could be in the form of intelligent metasurfaces, mirror arrays, or LCs) that control incident lights using principles such as generalized Snell's law of reflection and refraction, EM wave manipulation, and the orientation of the molecules in LCs \cite{9662064}. On the other hand, optical relays are active devices that typically receive, amplify, and re-transmit incident signals (i.e., require a full transceiver architecture) via a duplexing protocol. Thus, the two technologies use different approaches to achieve mostly the same functionalities. In this section, we discuss the key differences and similarities of optical RISs compared with optical relays from their implementation point of view.

\subsection{Comparison of Optical RISs and Optical Relays}
\begin{itemize}
    \item Solving skip-zones and SNR improvement
\end{itemize}
Both optical RISs and relay can solve the lost of the transmitted signal problem. In an outdoor VLC environment and over the FSO channel, optical RISs and relays can forward the obstructed transmitted light to the receiver. Generally, the optical signal outdoor is a monochromatic laser light, which, when obstructed, does not reach the receiver at all since there are no reflected components (i.e., NLoS paths to the receiver). The use of an RIS or relay enables the signal to reach the receiver. Thus, RISs and relays can solve the skip-zone problem in outdoor VLC environment. In addition to the skip-zone problem, RISs and relays can be used to improve the SNR at the receiver even when there is no direct light path from the transmitter in the indoor VLC environment. Unlike the outdoor environment that no signal reaches the receiver when the laser light is obstructed, reflected light signals in the indoor environment do reach the receiver but with a reduced SNR. RISs and relays can be used to focus any reflected light component towards the receiver and, thus, improve the SNR. 

\begin{figure}
    \centering
    {\includegraphics[width=0.5\textwidth]{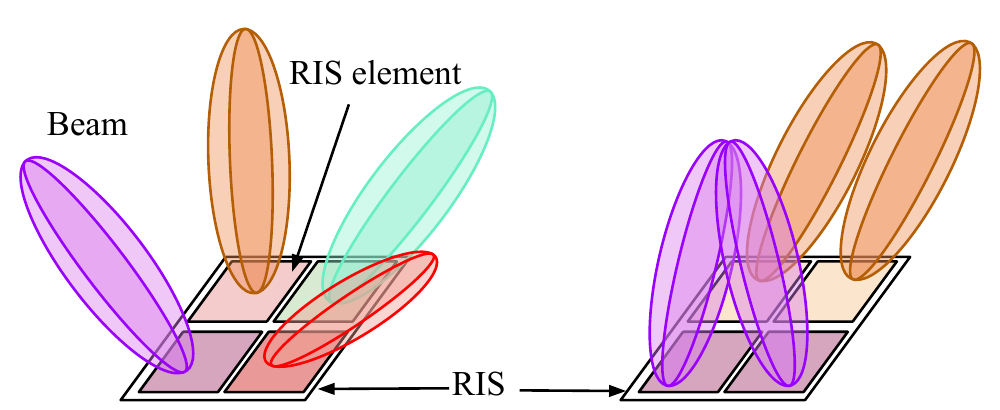}}
    \caption{\small{The principle of beamforming with RIS.}}
  \label{fig:BeamForming}
\end{figure}

Relaying can be classified in two main scenarios, namely amplify and forward (AF) and decode and forward (DF). In the former, the signal is just multiplied by a coefficient and re-transmitted, while in the latter, a few processing is required as the signal is first decoded, re-encoded, then re-transmitted. Unlike relays, the operation of RISs in DF mode has not yet been demonstrated. However, mirror-based RISs are typical types of relays with an amplification coefficient equal to unity. Recently, it was demonstrated in \cite{ndjiongue112021} that optical RISs can be exploited actively to solve the double fading occurring over the two subchannels created by the introduction of RISs. This shows that an RIS can be used as an AF module.
\begin{itemize}
    \item Beamforming
\end{itemize}
Apart from solving skip zones, beamforming is one of the target applications of the RIS technology. It has been demonstrated that RISs modules are reputed in creating beams as each cell can be manipulated individually to create a different beam \cite{liu112020}. This represents one of the main advantages of the RIS technology over relays that typically have one antenna (due to size and complexity constraints) and can only create one beam. Several materials and substances have been proved to be efficient in creating individual beams in OWC systems. Two examples among many others are LCs and elsatomers. LCs are characterized by their refractive index, which can be controlled to provide full tunability of both reflection and refraction, leading to a tunable RIS phase shift \cite{ndjiongue062021, ndjiongue052021}. Since RIS elements are independently controlled, the emerged beam can be individually controlled. The same is true for elastomers, where either compression or extension allows full control of incident light as it enters the material \cite{ndjiongue062021}. Fig.~\ref{fig:BeamForming} illustrates beamforming using the RIS technology. It clearly shows how each RIS element redirects the emerging light towards a specific direction. Note that, as in the case of relays in RF systems \cite{lee032010, gu062020, mahboobi082015, louie062019, nakai022019, liu052021, liu082012}, optical relays have also been exploited to control beams in OWC systems \cite{oliveira062021, liu08092010, liu052011}.

\begin{figure}%
    \centering
    {{\includegraphics[width=0.48\textwidth]{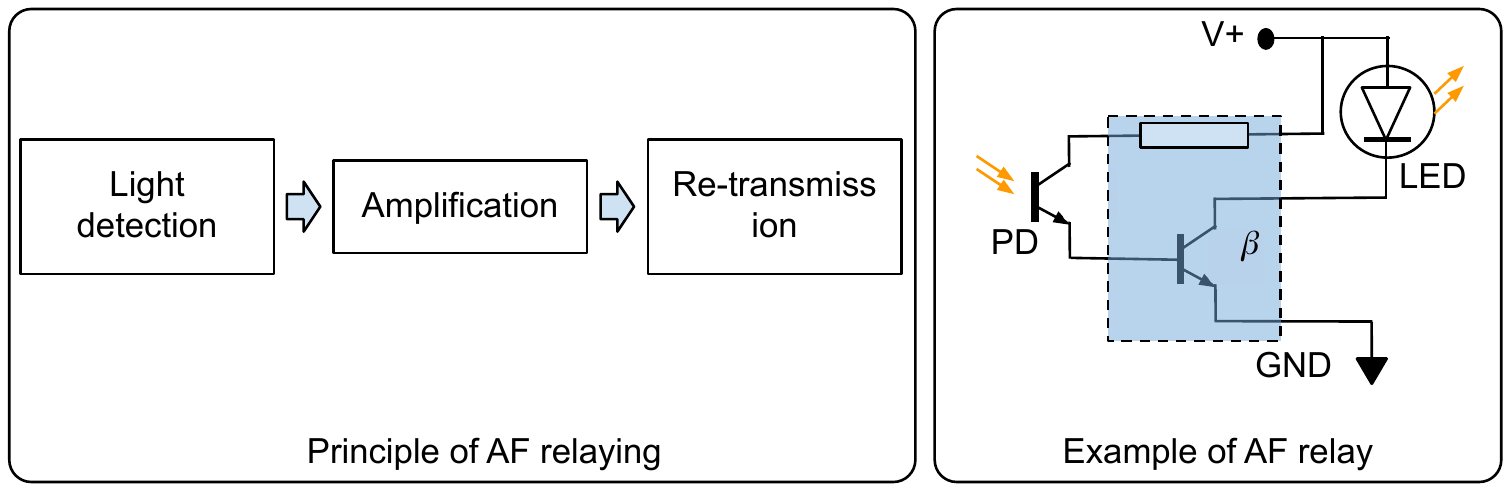}}} 
    \caption{\small{The principle of AF optical relays.}}%
  \label{fig:AF_OR}
\end{figure}

\begin{itemize}
    \item {New frontier of optical receivers}
\end{itemize}
As an amplifier, RISs can be used to enhance the OWC signal at the receiver. During this operation, the RIS also performs beam steering and filtering. When used at the receiver, the RIS is able to improve the FoV of the VLC receiver. According to \cite{ndjiongue052021}, VLC receivers equipped with convex lenses have a FoV of approximately 36.2$^{\circ}$, whereas those equipped with gradient index lenses and compound parabolic concentrators can achieve a FoV of 40$^{\circ}$. Spherical lenses provide VLC receivers with a FoV that is close to 45$^{\circ}$. On the other hand, a catadioptric monolithic bi-flat bi-parabolic lens can detect light with a FoV of 85$^{\circ}$. Lastly, it is demonstrated in \cite{ndjiongue052021} that lenses with adjustable elements, meta-lenses with artificial muscles, and LC-based RIS enable optical receivers with FoV that approach 90$^{\circ}$. The LCs-based RIS also provides greater flexibility, as the refractive index changes under a variable external electric field, influencing both the direction and the intensity of the light. However, optical relays cannot perform the roles mentioned above.\\

\begin{figure}%
    \centering
    {{\includegraphics[width=0.48\textwidth]{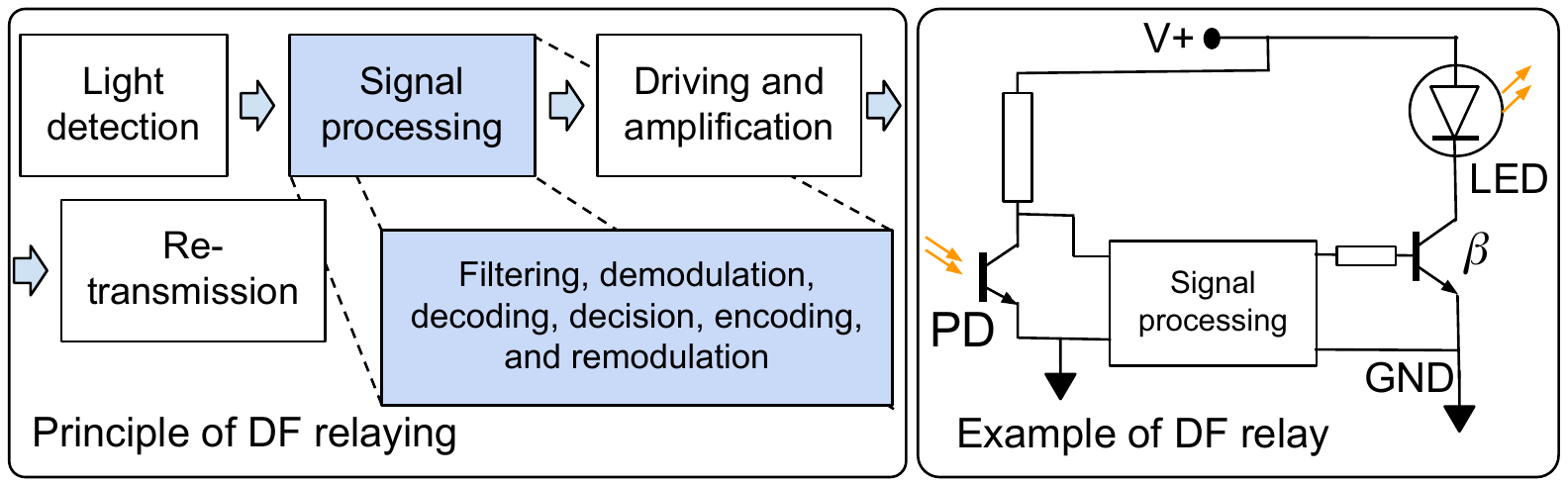}}}
    \caption{\small{The principle of DF optical relays.}}%
  \label{fig:DF_OR}
\end{figure}

\begin{itemize}
    \item Implementation considerations
\end{itemize}
The implementation of optical relays considers their operating mode, i.e., AF or DF. On this basis, an optical relay designed for the AF strategy will not operate in the DF scenario and vice-versa. The optical relays principles of implementation are illustrated in Figs.~\ref{fig:AF_OR} and \ref{fig:DF_OR}. They are based on solid-state systems. The relay does not operate as a true mirror since it always detects the originally transmitted signal, and then re-transmit it. Figure~\ref{fig:AF_OR} depicts an example of solid state system that is used in optical relays for an efficient relaying. The light signal is detected, amplified, either in the optical form or after conversion into an electrical current, and re-transmitted. The descriptive block diagram is given on the left-hand side of the figure, while a practical implementation is on the right-hand side. This example is proposed to show how this detection and amplification can be practically implemented. The incident light is detected by a PD, the obtained electric current is amplified and transmitted to an LED, which generates a light for re-transmission. On the other hand, Fig.~\ref{fig:DF_OR} shows an example of optical DF implementation. In this case, the detected light is strictly converted into an electrical current, which is transferred to the processing module. All signal processing including demodulation, filtering, decoding, decision, encoding, and re-modulation, required in optical receivers occur here, as shown on the left-hand side of the figure. After all the processing, the generated signal is amplified and used to power an LED for a current-to-light conversion, to end the re-transmission process. Note that all DF relaying strategies such as incremental relaying and incremental selective relaying, are implemented within the processing block. Also note that, unlike the DF strategy, the AF does not include a processing block. 

\begin{figure}%
    \centering
    {{\includegraphics[width=0.47\textwidth]{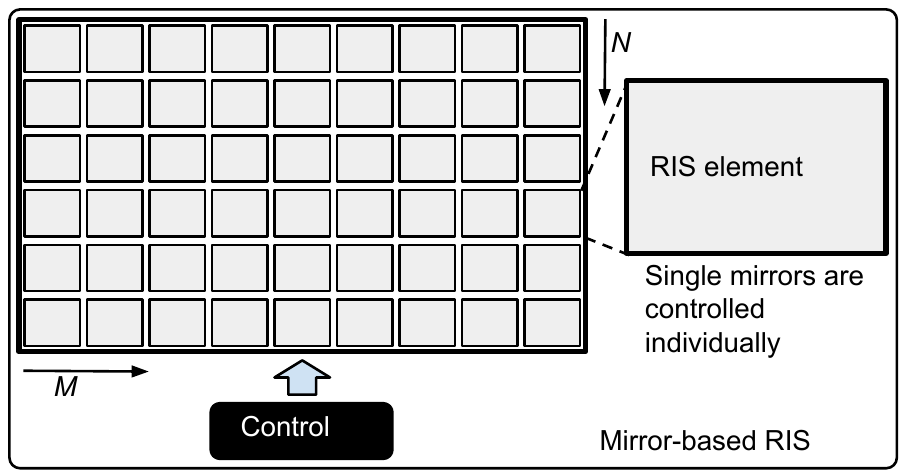}}} 
    \caption{\small{The principle of mirror-based RIS.}}%
  \label{fig:Mirror_RIS}
\end{figure}

\begin{figure}%
    \centering
    {{\includegraphics[width=0.47\textwidth]{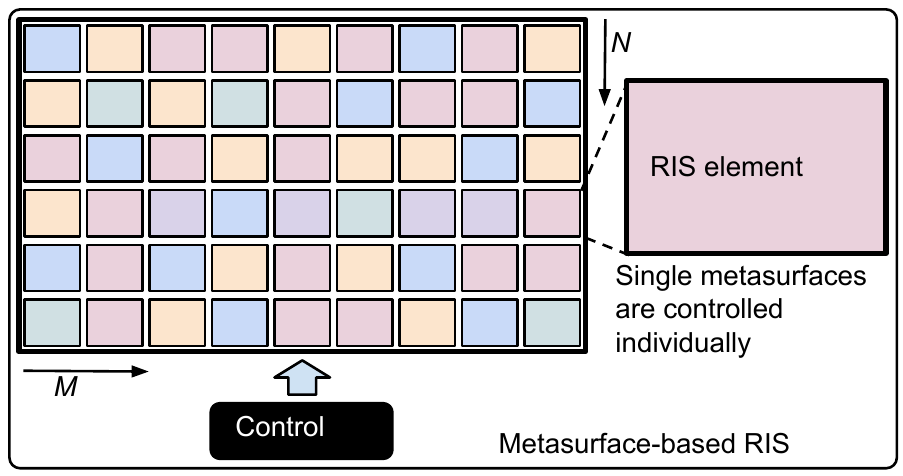}}} 
    \caption{\small{The principle of meta-based RIS.}}%
  \label{fig:Meta_RIS}
\end{figure}

The implementation of optical RISs is totally different from optical relays. Optical RISs are based on the light reflection or refraction principle. The RIS is a 2 dimensional structural set of meta-surfaces or mirrors, while the optical relay is a system made of antennas and electronic circuits. Figures~\ref{fig:Mirror_RIS} and \ref{fig:Meta_RIS} illustrate the two main types of optical RISs, namely the mirror array and the meta-surface array, respectively. All mirror arrays have the same intrinsic characteristics. They are typically made of glass with a flat or curvy surface, and have a reflective coat covering it. Each individual mirror is equipped with a mechanical control system, implying that the mirror arrays contain mechanical structures to dynamically orient the different mirrors. When there is no mechanical structure to control the physical orientation of the mirror, some mirrors would have a curvy surface to assure a perfect reflection of the light\footnote{Note that by opposition to metasurface arrays, mirror arrays do not operate as refractive devices since they only reflect the incident light.}. 

The control signal of these mirrors are generated from the control unit depending on the feedback from the target receiver in relation with specific requirements of the transmission system. On the other hand, metasurface arrays are RISs made of meta-materials. They are more complex and more expensive when compared to mirror arrays and are able to produce all phenomena related to the impact of photon on a surface. These include reflection, refraction, scattering, and absorption. Several ranges of materials and meta-materials are available for the implementation of metasurface arrays, including and not limited to metallic nanospheres in a dielectric medium, thin metallic rods isotropically distributed in a dielectric medium, splitring metallic elements in a dielectric medium, negative index meta-materials, and hyperbolic meta-materials. Among these materials, liquid based metasurfaces are recognized to provide the RIS with the ability to generate negative indices. Due to the variability of their refractive indices, LCs are suitable candidate to be used in manufacturing metasurface arrays for OWC applications. Figure~\ref{fig:Meta_RIS} shows a structure of a RIS made of metasurfaces. During control, the individual cells are uniquely controlled and each or several elements may have the same characteristics depending on the required resulting light orientation. No mechanical structure is required to remotely change the emerged light orientation. The two types of optical RISs described above, namely mirror- and metasurface-based, clearly illustrate the difference between optical RISs and optical relays.

\begin{itemize}
    \item Applications
\end{itemize}
From the application perspective, optical relays and optical RISs differ on a few points. Due to the high attenuation of light, mirror arrays are mostly well indicated for indoor VLC, while metasurface arrays can also be used outdoor, in which case the material used in the RIS must be selected accordingly. For example, a highly doped LC based RIS can be utilized in FSO systems to solve skip-zone problems, especially for a monochromatic laser light, while simultaneously amplifying the incident light to remedy the power loss due to the double fading generated by the two LoS links obtained after the introduction of the RIS \cite{alain122021}. Based on this, metasurface arrays can also be exploited as a relay in the AF strategy. On the other hand, optical relays are used to solve dead-zones in OWC systems, and serve as relays to extend the system coverage for both indoor and outdoor OWC environments.

\begin{table*}[t]
\centering
	\caption{Key differences between optical RISs and optical relays.}
\label{risrelcomp}
{\color{black}{
\begin{tabular}{|l|l|}
\hline
\textbf{Optical RISs}                                                                                                                                                                     & \textbf{Optical relays}                                                                                                     \\ \hline
Near passive reception and transmission                                                                                                                                                   & Active reception and transmission                                                                                           \\ \hline
Operate in full duplex mode                                                                                                                                                              & Operate in either full or half duplex mode                                                                                 \\ \hline
\begin{tabular}[c]{@{}l@{}}No need for analog-to-digital/digital-to-analog \\ converters, and power amplifiers\end{tabular}                                                            & \begin{tabular}[c]{@{}l@{}}Require analog-to-digital/digital-to-analog \\ converters, and power amplifiers\end{tabular} \\ \hline
Low energy consumption                                                                                                                                                                    & High energy consumption                                                                                              \\ \hline
\begin{tabular}[c]{@{}l@{}}Can be deployed at the transmitter, receiver, or in\\ the optical wireless channel\end{tabular}                                                                & \begin{tabular}[c]{@{}l@{}}Can only be deployed in the optical wireless \\ channel\end{tabular}                             \\ \hline
\begin{tabular}[c]{@{}l@{}}Can perform incident light amplification, control \\ the FoV, and wavelength filtering in addition to\\ signal enhancement and coverage extension\end{tabular} & \begin{tabular}[c]{@{}l@{}}Can perform only signal enhancement and \\ coverage extension\end{tabular}                       \\ \hline
Low deployment and maintenance cost                                                                                                                                                                    & High deployment and maintenance cost                                                                                               \\ \hline
\end{tabular}}}
\end{table*}

\begin{itemize}
    \item Hardware complexity
\end{itemize}
Optical relays and RISs have different hardware requirements and, as result, their hardware complexity are also different. As discussed in Section~IV, RISs are low-cost nearly passive reflecting elements that consist of metallic/dielectric patches printed on dielectric substrate (i.e., an IMR) or mirrors (i.e., IMA) that are connected to a central controller. The reconfigurability in IMR can be achieved by using tunable elements such as semiconductors and graphene. For IMA, their re-configurable properties are enable by the use of MEMS technology. In addition, RISs can amplify and forward any incident signal without the need of power amplifiers. In contrast, optical relays require a PD to receive the optical signal, an amplifier to enhance the signal power, and a transmitter  (e.g., LED) for re-transmission purpose as shown in Figs.~\ref{fig:AF_OR} and \ref{fig:DF_OR}. In the case of DF relaying, an additional circuitry for analog and digital signal processing is required. Hence, optical relays are viewed as active devices that need a dedicated power source and several electronic components for their operations. As a result, implementing optical relays may be more expensive in terms of the cost involved and the power consumption. Moreover, complicated interference management schemes are needed for self-interference cancellation for full duplex relays in OWC. On the other hand, optical RIS can operate in full duplex mode without the need of costly self-interference cancellation schemes. Finally, a dense deployment of optical relays may necessitate (i.e., depending on the frequency reuse scheme) the need of inter relay interference management schemes, while such schemes would not be needed in optical RISs. 

\begin{figure*}
    \centering
    {{\includegraphics[width=0.8\textwidth]{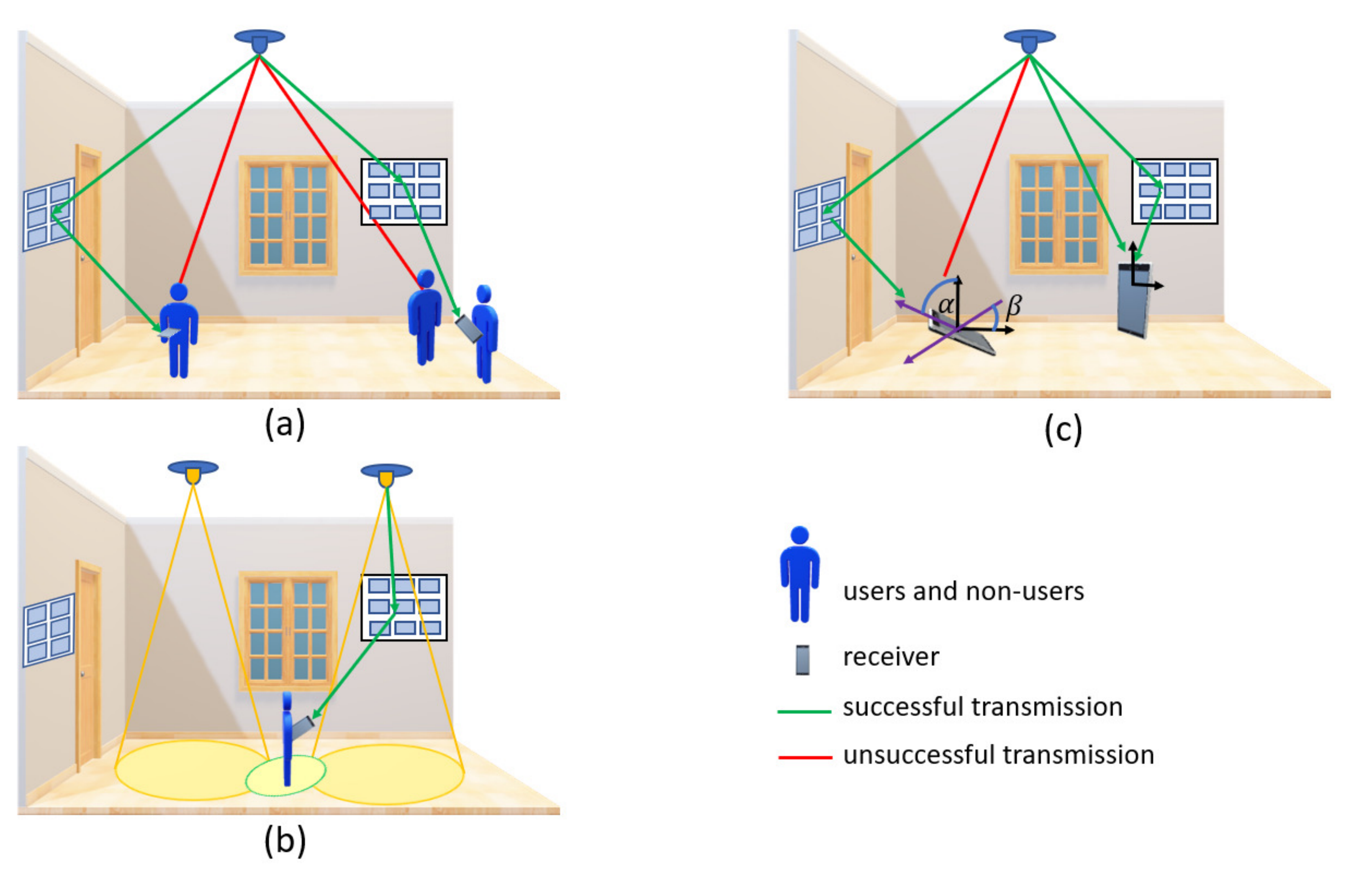}}}
    \caption{\small{An RIS-enhanced VLC system: (a) overcoming self-blockage and non-user blockage problems; (b) extending coverage to a user in dead-zone; (c) overcoming the impact of receiver's orientation on data transmission.}}%
  \label{bkge1}
\end{figure*}

\subsection{Summary and Lessons Learned} 
In this section, we discussed the key differences and similarities between optical RISs and optical relays. We learned that, while optical RISs and relays can be used to achieve coverage extension, SNR improvement, and beamforming, optical RISs offer several advantages such as lower energy consumption, operational complexity and  hardware cost, and less signal processing. The main differences between optical RISs and optical relays are summarized in Table~\ref{risrelcomp}. This comparison can serve as a useful guide for researchers and system designers to make informed decisions between optical RISs and optical relays for performance enhancement in VLC systems.

\section{RIS-based VLC Systems}
As already discussed, VLC has been identified as a promising solution to meet the high data requirements of the next-generation wireless systems. However, the requirement of the existence of direct LoS path between the transmitter and the receiver remains a key limiting factor for their successful deployment. In addition, the orientation of the VLC receiver also affects the existence and quality of the LoS channel from the transmitter. VLC systems therefore require the use of a different technology such as  RISs or optical relays to solve the signal obstruction and the random device orientation problems. As discussed in Section IV, RISs offer numerous advantages in terms of power consumption and hardware complexity. In addition to these important advantages, the integration of the RIS technology in VLC systems can help to enhance transmitter and receiver’s performances while a relay cannot. Hence, the use of RISs in VLC is motivated by two main reasons, namely solving the skip-zone and device orientation problems, and enhancing the performance of VLC transmitters and receivers. Although, there have been few studies on the integration of RISs and VLC \cite{Qian062021,Sun112021,Aboagye122021,Cao022020,Ssun2022,9756553,9784887,9799770}, the potential role of RISs in the context of VLC systems has not been well explored in the literature, and that is the focus of this section.

\subsection{RIS for LoS Blockages and Skip Zone Problem}
There are many reasons for which a receiver in a VLC system would not receive the transmitted data. This is due to the requirement of the existence of LoS path between the transmitter and the receiver and the fact that PDs have a limited FoV which restricts the angle at which any receiver can receive optical signals. In any indoor VLC environment, the unsuccessful transmission of data can result from the LoS blockage issue as shown in Figs.~\ref{bkge1} (a) and (c) or the user being in a skip-zone as demonstrated in Fig.~\ref{bkge1} (b). LoS signal from the VLC transmitter can be obstructed in the following ways:\\
\begin{itemize}
    \item The user carrying the receiver can obstruct the LoS signal by his own body (i.e., self-blockage), as shown in Fig.~\ref{bkge1} (a).
    \item A stationary or moving person (i.e., a non-user) or any object in the indoor environment can also block the LoS signal which is destined to the receiver held by a different person, as depicted in Fig.~\ref{bkge1} (a).
    \item The constant random changes in the direction and orientation of the receiver by the user as illustrated in Fig.~\ref{bkge1} (c).
\end{itemize}

\begin{figure}
    \centering
    {{\includegraphics[width=0.5\textwidth]{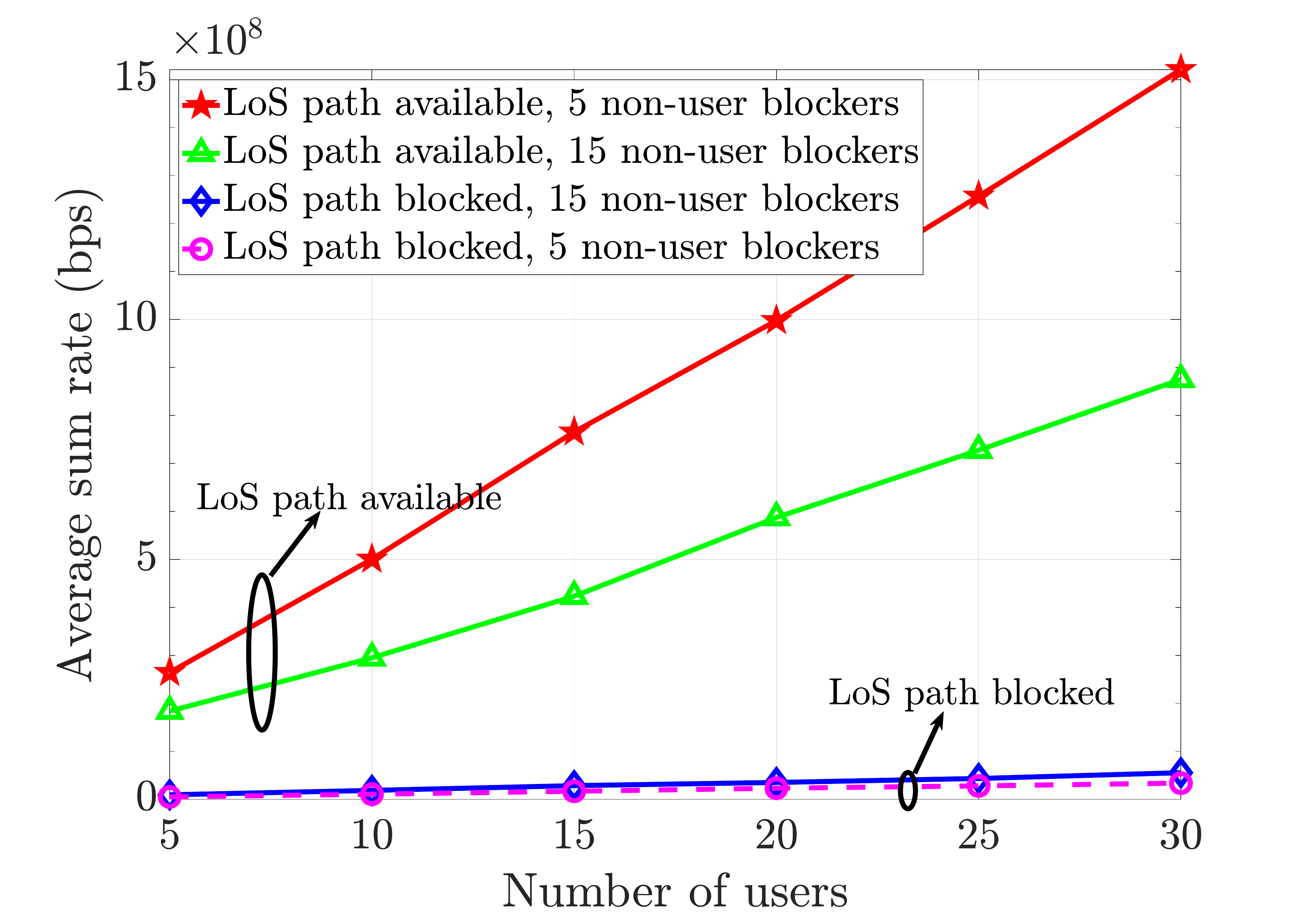}}}
    \caption{\small{Impact of blockages on sum-rate performance of VLC systems.}}%
  \label{fig2}
  \vspace*{-5mm}
\end{figure}
\begin{figure*}
    \centering
    {{\includegraphics[width=0.8\textwidth]{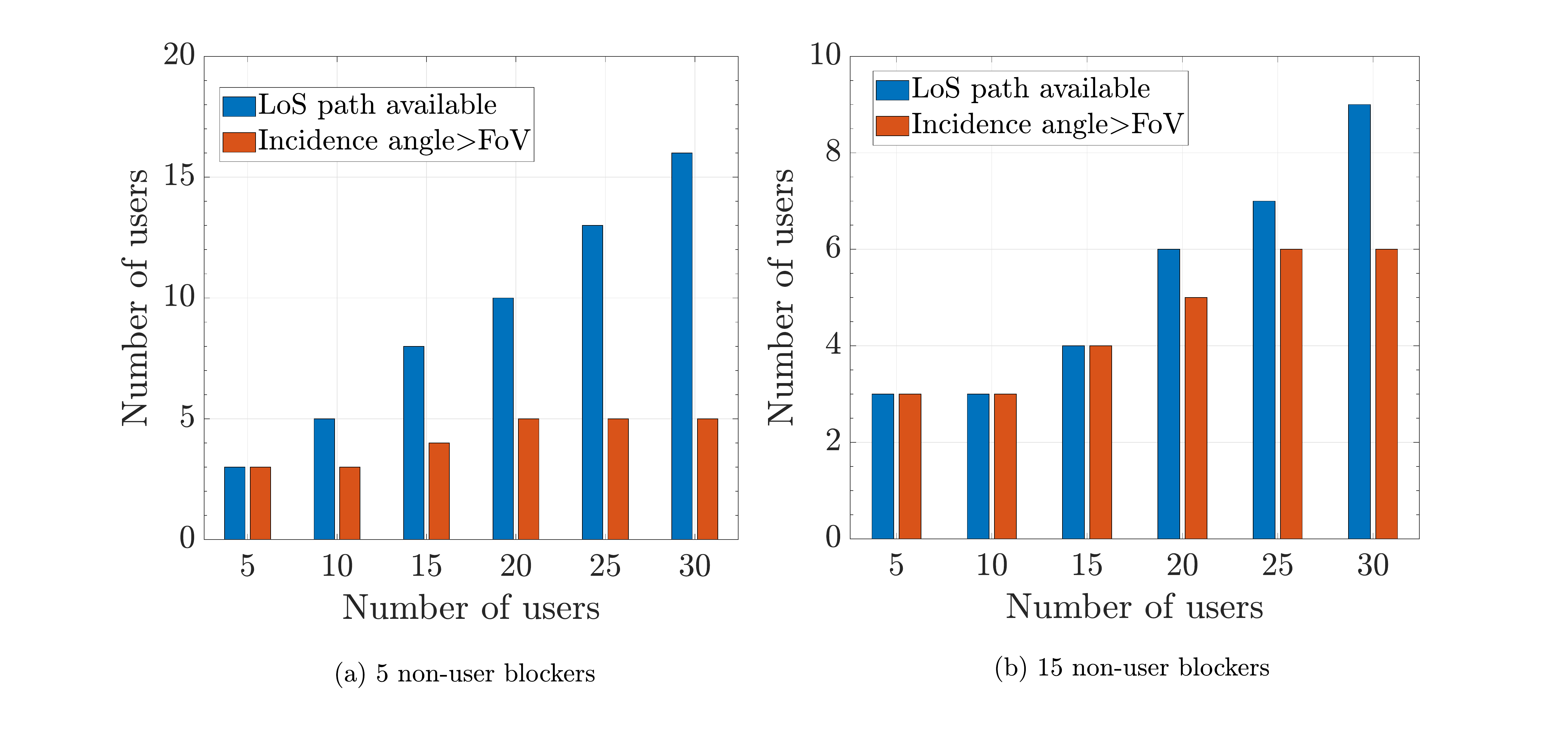}}}
    \caption{\small{Impact of receiver orientation on the availability of LoS path in VLC systems.}}%
  \label{fig3}
\end{figure*}

To illustrate the impact of LoS blockages (i.e., self and non-user blockages) on the performance of VLC systems, {\color{black}{we consider a setup of an indoor VLC environment of room size 5 m $\times$ 5 m $\times$ 3 m}} as shown in Fig.~\ref{bkge1} (a). In this setup, multiple users and non-users (i.e., blockers) are deployed randomly according to a uniform distribution. The users and non-users are modeled as cylinders with 0.30 m diameter and 1.65 m height. The receiver is held by the user at a distance of 0.75 m above ground and 0.36 m from the human body. The direction in which the user holds the receiver is random. The orientation of the receiver in any user's hand is defined by the polar angle $\alpha$ and the azimuth angle $\beta$, as illustrated in Fig.~\ref{bkge1} (c). The polar angle and azimuth angle are modeled according to a Laplace and uniform distribution, respectively, as discussed in \cite{Soltani032019}. The LoS channel gain and the channel gain of the first reflection (for the NLoS links) have been calculated using (\ref{cg11a}) and (\ref{cg11}), respectively.{\color{black}{The transmit optical power for each user is 2 W.}} Unless otherwise stated, all other simulated parameters for this section are chosen as in Table~I of \cite{Aboagye122021}. {\color{black}{It can be observed in Fig.~\ref{fig2} that there is a great disparity between the average sum rate of the VLC system when the LoS paths are available and  when they are blocked. This figure reveals that the diffused reflections from the walls have weak signal strength and, as a result, the achievable data rate in the absence of a direct path is very low. The sum-rate performance decreases when the number of non-user blockers increases from $5$ to $15$.}} 

Furthermore, due to the typical user behaviours in VLC system such as random receiver orientation, having an unobstructed direct LoS path from the transmitter does not necessarily guarantee successful data transmission since the quality of the received optical signal is highly dependent on the angle of the incident beam. This is because the transmitted data can be successfully received only when the incidence angle of the optical signal does not exceed the FoV of the receiver. Figure~\ref{fig3} shows a plot of the number of users with an unobstructed LoS path from the transmitter but who still cannot successfully receive data transmission due to the impact of random device orientation. It can be observed from this figure that about half of the total number of users with a direct LoS path from the transmitter has the incidence angle exceeding the value of the FoV of the PD (the value of the FoV is set as $85^{^{\circ}}$). Since future wireless networks are expected to support applications with diverse QoS requirements and high data rate demands, it is important to investigate solutions for the blockage and random device orientation issues for high-speed VLC systems. Many recent studies on VLC systems (e.g., \cite{Tang022021,Demir032021,Aboagye102021a,Guo102021,Aboagye102021,Obeed032021}) made a simplified assumption that the users' devices face upwards towards the ceilings and, hence, did not consider both the impact of device orientation and LoS path blockages on their performance analysis. Unlike those works, few studies have considered the effect of random device orientation and LoS blockages in VLC systems \cite{Eroglu022019,Soltani032019,Dehghani032019,Zeng082018,Purwita082019,Zeng052020} with \cite{Eroglu022019,Soltani032019,Zeng082018,Purwita082019} focusing on the modeling aspect and \cite{Dehghani032019,Zeng052020} on the performance analysis. 

The use of RISs to effectively overcome the LoS blockage and the device orientation issues is still an open research problem. In the following,  existing studies on the exploitation of RISs for performance improvement purposes in a system model similar to the RIS-enhanced VLC system in Fig.~\ref{bkge1}  are surveyed under the following design topics: (1) RIS elements orientation design, (2) RIS elements to AP/user assignment design, and (3) RIS array positioning design. The objective of these design topics is to properly utilize the elements (i.e., mirrors or metasurfaces) of the RIS array to improve key performance indicators such as achievable data rate \cite{Aboagye122021,Sun112021}, energy efficiency \cite{Cao022020}, spectral efficiecny \cite{Ssun2022}, and secrecy rate \cite{Qian062021}.  
\subsubsection{RIS elements orientation design} 
The RIS elements orientation  problem is very important in the design and deployment of RISs (i.e., a mirror array or a metasurface) in VLC systems. More specifically, finding the optimal orientation of the RIS mirror array elements\footnote{A mirror array is considered as it has been shown to outperform a metasurface reflector \cite{Abdelhady122021}.} is pivotal to harnessing  its full potential. The design problem of optimizing the orientation angles of the mirrors in an RIS array is generally formulated as:

\begin{equation}\label{dec8a}
    \begin{array}{l}
\mathop {\max }\limits_{{\boldsymbol{\gamma}},{\boldsymbol{\omega}}} {\mathcal R}\left({\boldsymbol{\gamma}},{\boldsymbol{\omega}} \right)\\
{\rm{s}}{\rm{.t}}{\rm{.}}\\
{C1: -\frac{\pi}{2}\le {\boldsymbol{\gamma}} \le \frac{\pi}{2},}\,\,\,\,\,\,
{C2: -\frac{\pi}{2}\le {\boldsymbol{\omega}} \le \frac{\pi}{2},}
\end{array}
\end{equation}

where ${\boldsymbol{\gamma}}=[\gamma_{ij}]$ and ${\boldsymbol{\omega}}=[\omega_{ij}]$ are the orientation angle matrices for the yaw and roll angles, respectively, with $i$ and $j$ being the row and column indicators of the mirrors. ${\mathcal R}$ represents the total network utility, which could be the achievable data rate, secrecy rate, or energy efficiency, and is a function of the orientation angles of the mirrors. Typically, the objective function in  (\ref{dec8a}) turns out to be a non-convex function. It is therefore hard to obtain the global optimal solution. Moreover, for any RIS array with $N$ mirrors, there will be $2\times N$ optimization variables, making (\ref{dec8a}) an intractable high dimensional optimization problem due to its extremely large search space. In comparison with an RIS metasurface array, there will be $N$ dimension of variables since each metasurface has a single phase shift variable. Performing an exhaustive search to obtain the global optimal solution for the problem of the form in (\ref{dec8a}) is computationally prohibitive even for an array with a small number of elements. Only a handful studies (i.e., \cite{Aboagye122021,Qian062021,Cao022020}) have attempted such a problem. Among those works, population based intelligent heuristic schemes (i.e., meta-heuristics) have been shown to be the appropriate method for overcoming the difficulties of solving (\ref{dec8a}). 

{\color{black}{The authors in \cite{Aboagye122021} proposed a sine-cosine based metaheuristic algorithm to obtain the optimal configuration angles  (i.e., ${\boldsymbol{\gamma}}$ and ${\boldsymbol{\omega}}$) of an RIS mirror array such that the achievable data rate from the NLoS VLC links is maximized. In that work, the authors considered all the various ways that humans and objects can impact VLC links (i.e., self blockages, non-user blockages, and random device orientation). To overcome the high-dimensional nature of the considered data rate optimization problem, the authors introduced the assumption that all the mirrors in the RIS array have identical yaw and roll angles. This significantly reduced the dimension of the optimization
variables. In the performance analysis, a comparison was made with (i) a design where each mirror of the RIS array assumes different roll and yaw angles and these
angle parameters are optimized using the sine-cosine based algorithm, and (ii) a baseline scheme that assigns random feasible values to the yaw and roll angles of each mirror of the RIS array. Simulation results showed that, in comparison with the data rate from wall reflections and the baseline scheme, the optimized RIS design attains at least a 400$\%$ gain in data rate. It was further revealed that, although optimizing RIS mirrors with different orientation angles  yields higher data rate performance than that of RIS mirrors with identical angles, such a design has a significantly higher computational complexity which grows rapidly with an increase in the number of RIS mirrors.}}

The authors in \cite{Qian062021} studied a secrecy rate optimization problem for an RIS-assisted VLC system. In this work, the authors focused on exploiting RIS mirror arrays in VLC to develop a secured communication system. To that end, the authors first derived a lower bound on the achievable secrecy rate, and then formulated an optimization problem to configure the orientation angles of the RIS array such that the difference between the channel gain of the legitimate user and the eavesdropper is enlarged.
To overcome the associated high dimensional nature and
the non-convexity of the resulting secrecy rate maximization problem, the authors first transformed the original problem into a reflected spot position finding problem. Then, a modified particle swarm optimization algorithm, which is a population based metaheuristic, was proposed to find the optimal solution.

An energy efficiency maximization problem was studied for a time division multiple access (TDMA) based RIS-assisted multi-user VLC system in \cite{Cao022020}. In this paper, the authors focused on the joint optimization of time allocation, power control, and phase shift matrix of the RIS metasurface array. Due to the non-convexity of this optimization problem, an iterative algorithm based on the interior point method \cite{boyd2004} and a one-dimensional search method was proposed to obtain a sub-optimal solution.

{\color{black}{A deep reinforcement learning solution based on a deep deterministic policy gradient algorithm has been proposed in \cite{9784887} to optimize the secrecy capacity of an RIS-assisted VLC system. 
 The algorithm maximizes the secrecy capacity by adjusting the beamforming weights of the LEDs and the mirror orientations of the RIS mirror array. It was shown in this paper that optimizing both the beamforming weights and the RIS mirror orientation provides the highest secrecy rate for the VLC system.}}

\subsubsection{RIS elements to AP/user assignment design}
RIS arrays can be deployed in multi-user, multi-AP VLC systems. In such a setting, it becomes important to determine which elements of the RIS array serve which particular APs and users such that a utility function is maximized. This is a typical RIS array and AP or RIS array and user assignment problem. This problem is important in the design of RIS-assisted VLC systems since the VLC channel model is highly sensitive to the geometric positions of APs and users. In this line of research, the authors in \cite{Sun112021} investigated the design problem
of associating RIS elements to the available LEDs in a cell free multi-user TDMA-based indoor VLC system. Under the assumption that the location of the RIS array, the LEDs, and the users are known, and with sum-rate maximization as the objective function, a binary programming problem to assign the RIS elements to the LEDs was formulated. By relaxing the requirement that association indicators should be binary, the authors proposed a solution based on the projected gradient descend algorithm. Then, a greedy algorithm was
introduced to recover the binary solution from the output of the projected gradient descend algorithm. Having obtained the RIS-LED association matrix, each RIS element configures its orientation properties, based on the available information on the geometrical positioning of the users and the LEDs, to focus any reflected signal to the target receiver. Simulation results showed that the proposed design outperforms benchmarks such as randomly assigning the LEDs to the RIS units and assigning LEDs to RIS elements based on the
minimum distance criterion.

In \cite{Ssun2022}, the authors studied the problems of user association, power allocation, and the RIS-AP assignment to maximize the achievable spectral efficiency and proposed solutions based on the frozen variable algorithm and the minorization-maximization algorithm. Simulation results showed signficant spectral efficiency performance when compared to a VLC system without RIS and other benchmarks. 

{\color{black}{The study in \cite{9756553} addressed the problem of assigning RIS elements to LEDs to optimize the secrecy rate performance of an RIS-aided VLC system. The authors formulated a secrecy rate maximization problem and then transformed it into a sequence of LED-RIS assignment subproblems via an objective function approximation. For this problem, an iterative Kuhn-Munkres algorithm was proposed and simulation results confirmed significant improvement in the secrecy rate performance of the VLC system.}}

It can be noted from the above discussion that there are only three studies on the RIS array and AP/user association problem and, hence, additional work needs to be done to fully harness the potential of RIS in VLC systems. Several assumptions were made in \cite{Sun112021} to aid the tractability of the considered optimization problem. Notable among them are (\textit{i}) the location of the APs, users, and the RIS array which were assumed to be known, and (\textit{ii}) the probability that the reflected light emitted from other LEDs  propagates to an arbitrary user which was ignored. However, such assumptions might not always hold in practice and, as a result, must be relaxed in future designs. Moreover, this design problem should be considered for different multiple access schemes to exploit their benefits. Furthermore, the RIS-user/AP association problem should be jointly considered with the RIS elements orientation design problem.
\subsubsection{RIS array positioning design}
The optimization of RIS array placement in indoor VLC systems is another research area that needs to be investigated since the positioning of the array greatly impacts the system's performance. For example, the RIS array should be positioned such that it is within the illumination region of the access point and it can receive a great amount of optical signals. Moreover, it is important to investigate if the RIS should be deployed closer to the receiver or the transmitter in a VLC system. However, there has been minimal research activity on the optimization of RIS positioning and performance analysis for VLC. A typical objective of such a design problem can be to determine the optimal position for the placement of the RIS array such that a utility function (e.g., achievable rate, energy efficiency, secrecy rate, etc.) is maximized while taking into account the distribution of users, the receiver's FoV, practical user behaviours (such as device orientation), interference effects, and the occurrences of LoS path blockages. The RIS array positioning design problem fits into the class of integer programming problem which is NP hard, since each RIS array can only be at one location every time. Although the exhaustive search method can be applied to such a complicating and difficult optimization problem to obtain the optimal placement solution, such a solution technique is highly intractable and cannot be realized in practice. Hence, novel solution methods as well as performance analysis are required. In \cite{9799770}, the authors investigated a novel approach of simultaneously enhancing  illumination uniformity and throughput of an indoor multi-element RIS-enabled VLC system. Specifically, the authors formulated throughput and illumination optimization problems to obtain the optimum mirror placement, LED power allocation, and LED-user association. Simulation results revealed a three-fold increase in the average illumination and a four-fold increase in the average throughput by optimizing the RIS mirror placement.

\begin{table*} 
\centering
{\color{black}{ 
	\caption{Summary of RIS-VLC Proposed Performance Enhancement Schemes.}
\label{aug4}
\begin{tabular}{|l|l|l|l|}
\hline
\textbf{Ref.} & \textbf{Main Contribution}                                                                            & \textbf{Key VLC Issues}                                                                 & \textbf{Proposed Technique(s)}                                                                                                                                                                                                                                     \\ \hline
\cite{Aboagye122021}     & \begin{tabular}[c]{@{}l@{}}Data rate maximization\\ for RIS-aided VLC systems\end{tabular}             & \begin{tabular}[c]{@{}l@{}}Link blockages\\ Device orientation\\ User orientation\end{tabular} & \begin{tabular}[c]{@{}l@{}}A sine-cosine based metaheuristic is used to obtain an\\ optimal configuration of the RIS mirror  arrays.\end{tabular}                                                                                                                        \\ \hline
\cite{Cao022020}      & \begin{tabular}[c]{@{}l@{}}Energy efficiency maximization \\ for RIS-enabled VLC systems\end{tabular}                                                            & Link blockages                                                                                 & \begin{tabular}[c]{@{}l@{}}A typical interior point method is used to obtain optimal \\time allocation, DC-offset distribution, and power control.\\ A one-dimensional search method is used to obtain a\\ sub-optimal phase shift solution.\end{tabular} \\ \hline
\cite{Qian062021}     & \begin{tabular}[c]{@{}l@{}}Secrecy rate maximization \\ for RIS-enabled VLC systems\end{tabular}       & Security                                                                                       & \begin{tabular}[c]{@{}l@{}}A modified particle swarm optimization algorithm is used \\to obtain an optimal mirror RIS  configuration.\end{tabular}                                                                                                                     \\ \hline
\cite{Sun112021}      & \begin{tabular}[c]{@{}l@{}}Sum rate maximization for\\  RIS-aided VLC systems\end{tabular}                   & Link blockages                                                                                 & \begin{tabular}[c]{@{}l@{}}A greedy algorithm is used to associate LEDs with RIS\\ elements.\end{tabular}                                                                                                                                                                  \\ \hline
\cite{Ssun2022}      & \begin{tabular}[c]{@{}l@{}}Spectral efficiency maximization\\  for RIS-aided VLC systems\end{tabular}         & Link blockages                                                                                 & \begin{tabular}[c]{@{}l@{}}A frozen variable algorithm is used to solve RIS-LED \\ association problem. A minorization-maximization \\ algorithm is used to solve the RIS-LED association/LED \\power control problem.\end{tabular}                                                             \\ \hline 
\cite{slysub}     & \begin{tabular}[c]{@{}l@{}}Data rate maximization for a VLC\\ system with an LC RIS receiver\end{tabular}             & \begin{tabular}[c]{@{}l@{}}
Beam steering\\ Beam amplification\end{tabular} & \begin{tabular}[c]{@{}l@{}}A sine-cosine based metaheuristic is used to configure an \\ optimal refractive index of the LC RIS.\end{tabular}

    \\ \hline
\cite{9756553}    & \begin{tabular}[c]{@{}l@{}}Secrecy rate maximization\\  for RIS-aided VLC systems\end{tabular}                & Security                                                                                       & \begin{tabular}[c]{@{}l@{}}An iterative Kuhn-Munkres algorithm solution is used to\\ match RIS elements to LEDs.\end{tabular}                                                                                                                                                      \\ \hline
\cite{9784887}   & \begin{tabular}[c]{@{}l@{}}Secrecy capacity maximization\\  for RIS-assisted VLC systems\end{tabular}        & Security                                                                                       & \begin{tabular}[c]{@{}l@{}}A deep reinforcement learning solution to adjust LEDs' \\ beamforming weights and RIS mirror orientations.\end{tabular}                                                                                                                 \\ \hline
\cite{9799770}    & \begin{tabular}[c]{@{}l@{}}Illumination and data rate \\ enhancement in RIS VLC networks\end{tabular} & Illumination                                                                                           & \begin{tabular}[c]{@{}l@{}}A two-stage heuristic approach is used to optimize RIS \\ mirror placement, LEDs' transmit powers, and LEDs-user \\ assignment.\end{tabular}                                                                             \\ \hline
\end{tabular}
}}
\end{table*}

\subsection{RIS for VLC Transmitter Performance Enhancement}
The received power signal from diffused wall reflections is always small for a high SNR system. An RIS is a good solution to this dilemma since it has multiple elements, and each element can provide different LoS signal towards the receiver. An RIS is made of materials with less loss at the reflection point, implying a good transition matrix, leading to a better SNR at the receiver when compared to the walls. 
In the outdoor OWC and VLC environments, moving objects such as vehicles and fixed object such a buildings can always disrupt the transmitted LoS signal. When a vehicle crosses the LoS VLC transmitted signal, it is likely to disrupt data transmission for that lap of time. Similarly, when both the transmitter and receiver are fix over two different buildings, it is likely that a new building can arise and disrupt the transmitted light signal. An RIS can be exploited in the above situation to re-establish the signal at the receiver by creating a cascaded system of two LoS systems.

\subsection{RIS for VLC Receiver Enhancement}
It is worth noticing that in the above subsection A, an RIS is used to reflect the incident light beam. When located inside the VLC receiver, the RIS can be exploited as a refractive device. Traditional VLC receivers use convex, spherical, or compound parabolic concentrators to steer the detected light, which provide limited FoV with high attenuation. These convex lenses generate about 10\% power loss for waves in the visible spectrum. For example, at 0$^{\circ}$ incidence of the incoming light rays, most glasses with a refractive index around 1.5 have a decrease of about 2 to 4\% and 5 to 10\% of the incident light power for clear and prismatic glasses, respectively. These values can reach 30\% for heat-absorbing glasses and represent an intensity loss due to reflection at the lens’s upper surface. However, the FoV is traded-off against higher received power. The impacts of these lenses become significant as detected signal is already considerably attenuated. Placed inside the receiver, the RIS can provide wider FoV, less light attenuation, digital beam steering, and higher SNR at the PD. {{\color{black}{In \cite{slysub}, the authors considered an LC RIS-based VLC system where the RIS is deployed at the receiver. For this system model, the authors proposed a channel model for the LC RIS and a sine-cosine based algorithm to optimize the configuration of the LC RIS such that the achievable data rate is maximized. Simulation results revealed that up to 700$\%$ improvement in the data rate can be achieved when compared to VLC receivers without LC RISs.}}}

\subsection{Summary and Lessons Learned}
{\color{black}{LoS blockages occur frequently in indoor VLC systems and can have significant detrimental effects on the data rate performance of VLC systems. Moreover, the orientation of users' devices can affect the quality of the LoS path between the transmitter and the receiver. This section has revealed the use of RIS technology in solving the VLC receiver orientation problem as well as the LoS blockage and the skip zone problems. The most recent works in this area of research have been summarized in Table~\ref{aug4}. While there have been many research activities on optimizing the configuration of IMA and IMR RISs in VLC systems (see Table~\ref{aug4}), few attempts have been made to investigate the RIS element to AP/user assignment design problem and the RIS array positioning design problem. Optimization problems involving the optimal configuration of RISs tend to be high-dimensional in nature and, as a result, result, it may be advantageous for future work in this area to leverage machine learning techniques to develop less computationally expensive algorithms. Moreover, most previous studies have assumed perfect knowledge of the CSI, which is impractical. As discussed, RISs can be deployed at the transmitter and receiver ends to  enhance the focusing capability of VLC transmitters and the quality of the received signals at receivers, respectively. However, there has been minimal attempt to examine deploying RISs at both the transmitter and the receiver.}}

\section{Integrating RIS with other Technologies in VLC}\label{int}
This section discusses the integration of RISs and other key enabling technologies for 5G and beyond networks, in the context of VLC systems. For each of these technologies, the fundamental motivation and the recent technical progresses are introduced. Then, the associated challenges and future research directions are provided. Although the integration of RISs and such other technologies have been investigated for RF communication systems (e.g., \cite{Zhang082020}), the techniques developed for RF channels cannot be directly applied to VLC channels due to the differences in their transmission protocols and modulation schemes.
\subsection{Angle Diversity Transmitters and RISs} Different techniques have been proposed in the literature to mitigate interference in indoor multi-cell VLC systems. Notable among them is ADTs. An ADT emlpoys multiple LED arrays with each array facing a direction specified by an elevation angle, to focus the optical beams and create multiple smaller cells for multi-user access \cite{Cheng122013,Aboagye062021}. This is different from traditional VLC transmitters that consist of LED arrays, typically modelled as a point source, or multiple, closely packed, LED arrays that are arranged facing vertically downwards. In comparison to  traditional VLC transmitter deployments, ADTs can provide diversity gain and offer a high degree of decorrelation between the optical channels of the  LED arrays with different elevation angles. There have been recent studies on the design, resource allocation optimization, and performance analysis of multi-cell VLC systems employing ADTs \cite{Aboagye102021,Cheng122013,Aboagye062021,Dixit122020, albraheem072018,Chen062017,Eroglu012018,Alsulami072019,Yin092015}. Simulation results from those work demonstrated that VLC systems with ADTs outperform those with traditional VLC transmitters in terms of interference mitigation and data rate performances. In addition, it was shown that choosing an appropriate elevation angle for the LED array can greatly affect the data rate performance of such VLC systems. However, in the above-mentioned studies, the elevation angle of the LED arrays are fixed and does not adapt to any changes in the VLC environment (e.g., number of users, position of users, and the occurrence of LoS blockages). As a result, the performance of the system can only be optimized for that particular ADT configuration. As pointed out in \cite{Abumarshoud042021}, there is the need to always adapt the spatial separation and width of the beams according to the number and position of users by adjusting the elevation angles of ADTs. Deploying RISs at the transmitter side can enable the realization of re-configurable ADTs to allow the creation of a varying number of smaller, interference-free cells to serve multiple users and improve the system performance. For such re-configurable ADTs, the LED arrays are deployed facing vertically downwards and the RIS can be utilized to dynamically change the beam direction (which is the elevation angle in regular ADTs) depending on the number and the position of users in the indoor environment. As illustrated in Fig.~\ref{figsNov25a} (C), the re-configurable ADT is able to transmit three highly directional optical beams to the PDs by controlling their width and divergence. In the event that the third user leaves the indoor environment, the re-configurable ADT can utilize the same number of LED arrays to create two directional beams for the remaining users to enhance the energy and spectral efficiency performances of the system.    

\begin{figure*}
	\centering
	\includegraphics[width=0.8\textwidth]{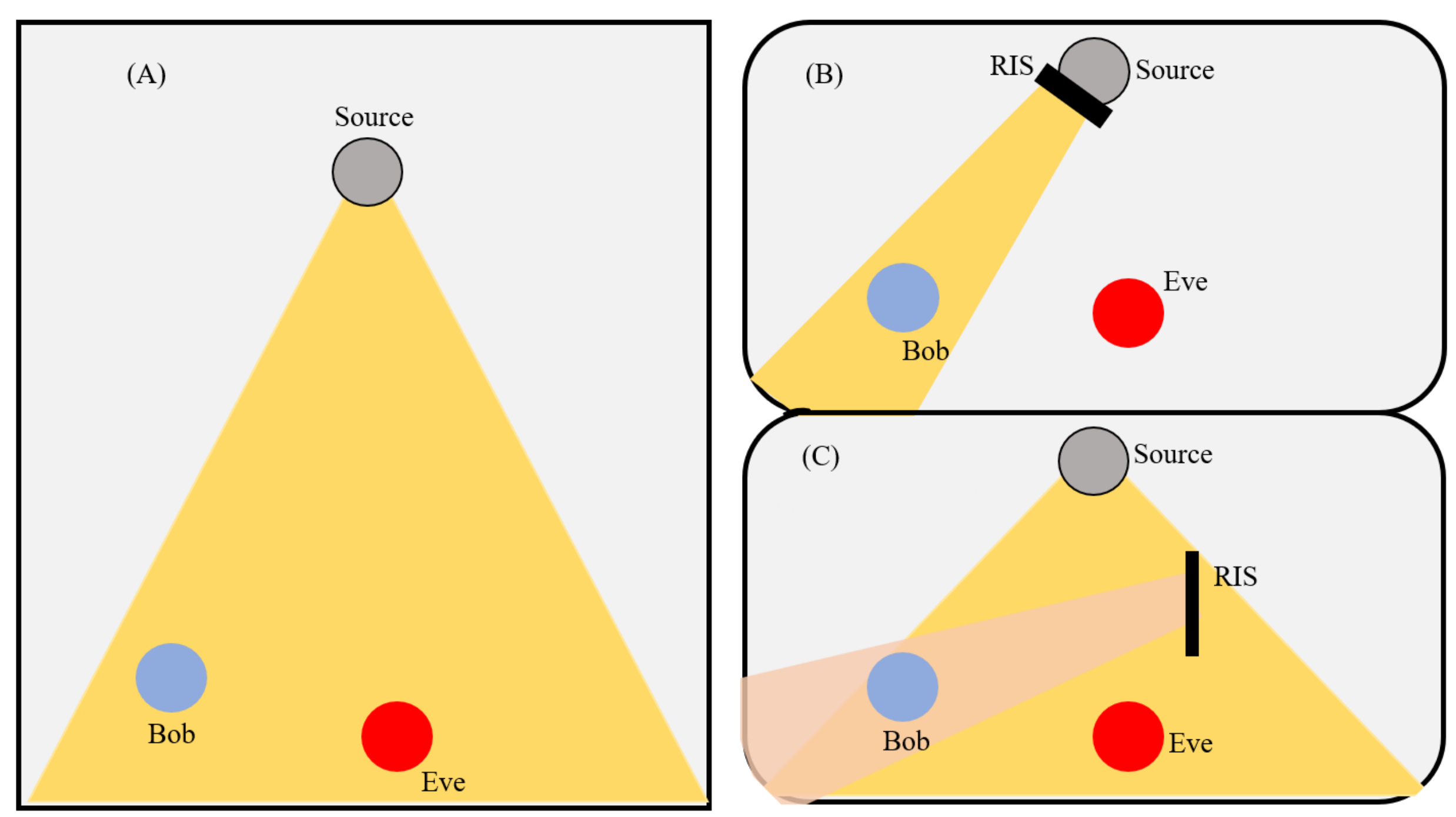}
	\caption{\small{Scenarios of RIS-assisted VLC for Secured Transmission and Reception: (A) VLC system without RIS; (B) VLC system with RIS at the transmitter side; (C) VLC system with RIS in the channel.}}
	\label{figrisv1}
\end{figure*}

\subsection{Physical Layer Security and RISs} 
Physical layer security (PLS) techniques typically exploit the characteristics of any wireless medium and its impairments including noise, fading, interference, dispersion, and diversity to ensure data transmission is realized in a most secured manner \cite{Hamamreh102019}. Specifically, the objective of any PLS technique is to enlarge the signal-to-interference-plus-noise ratio difference between that of intended user and the eavesdropper, and this can be realized by either reducing the received signal strength or increasing the noise (or interference) at the eavesdropper's end through the design of appropriate transmission schemes. In comparison to RF communications, VLC networks are known to offer a higher level of security for data transmission due to the requirement of the existence of LoS propagation between the transmitter and receiver, and the fact that light cannot penetrate walls. Nevertheless, the broadcast nature of the VLC channel makes it possible for the transmitted signal to be received by the intended user and intercepted by an eavesdropper located within the illumination area of the transmitter \cite{Mostafa072015,Chen112011,9869677}. Enhancing the security of VLC systems, especially in the multi-user scenario, has received significant research attention. On the other hand, PLS techniques have been a huge success in  improving the security of RF communication networks \cite{Mukherjee022014,Liu082017,Wang112019}, and has therefore recently been considered for VLC systems. An in-depth survey of earlier works on PLS-VLC can be found in \cite{Arfaoui044040,Hamamreh102019} and the most recent studies in \cite{Su042021,Ben062021,Abumarshoud082021,CHo122021,Duong122021,Peng062021,Qian062021,Pham012021,Cho112021}.

A basic system model for RIS-aided VLC system for secure transmission is presented in Fig.~\ref{figrisv1}. The objective of the security design in such a model will be to utilize the RIS module and its transmission/reflection/reception properties to make the system more robust to eavesdropping. As illustrated in Fig.~\ref{figrisv1}, instead of broadcasting the information intended for the legitimate user (i.e., Bob) as shown in Fig.~\ref{figrisv1} (A), an RIS can be utilized at the transmitter front end to focus all the emitted information carrying light signal towards Bob to guarantee the confidentiality of the transmitted signal as shown in Fig.~\ref{figrisv1} (B). Whenever the location of Bob changes, the direction of the emitted optical beam can be changed to the updated location by reconfiguring the properties of the RIS. As depicted in Fig.~\ref{figrisv1} (C), an RIS array can also be deployed on the wall to focus all the reflected signals at Bob to improve the channel gain and, as a result, the received signal strength. For information transmission and reception in Fig.~\ref{figrisv1} (A) and (B), the received signal of Bob and Eve can be given as 

\begin{equation}\label{dec14a}
\begin{array}{l}
y_{\rm Bob}=h^{\rm LoS}_{\rm Bob}s+n^{\rm Bob},\\\\
y_{\rm Eve}=h^{\rm LoS}_{\rm Eve}s+n^{\rm Eve},
\end{array}
\end{equation}
where $h^{\rm LoS}_{\rm Bob}$ and $h^{\rm LoS}_{\rm Eve}$ denote the channel gain of LoS links for Bob and Eve, respectively, $s$ is the transmitted signal, and $n^{\rm Bob}\sim {\mathcal N\left(0,\sigma^2\right)}$ and $n^{\rm Eve}\sim {\mathcal N\left(0,\sigma^2\right)}$ represent the noise at the receivers of Bob and Eve, respectively, which are modeled as additive white Gaussian noise with zero mean and variance $\sigma^2$. In Fig.~\ref{figrisv1}  (C), the received signal of Bob and Eve  can be expressed as

\begin{equation}\label{dec14b}
    \begin{array}{l}
    y_{\rm BoB}=\left(h^{\rm LoS}_{\rm Bob}+h^{\rm RIS}_{\rm Bob} \right)s+n^{\rm Bob},\\\\
    y_{\rm Eve}=\left(h^{\rm LoS}_{\rm Eve}+h^{\rm RIS}_{\rm Eve} \right)s+n^{\rm Eve},
    \end{array}
\end{equation}
where $h^{\rm RIS}_{\rm Bob}$ and $h^{\rm RIS}_{\rm Eve}$  are the channel gain of the links between Bob and the RIS, as well as Eve and the RIS, respectively, and their expressions can be obtained from \cite{Abdelhady122021}. PLS techniques for Fig.~\ref{figrisv1} (A) has been studied in \cite{Arfaoui044040,Hamamreh102019,Su042021,Ben062021,Abumarshoud082021,CHo122021,Duong122021,Peng062021,Pham012021,Cho112021} and, hence, is not considered in this survey. On the contrary, there has been no research on exploiting the model in Fig.~\ref{figrisv1} (B) and the works in \cite{Qian062021,9756553,9784887} are the only research on a deployment set-up analogous to that in Fig.~\ref{figrisv1} (C). 
Numerous research challenges remain to be solved to aid the practical implementation of the VLC-RIS set-ups in Figs.~\ref{figrisv1} (B) and (C) for secrecy performance improvements. Beginning with Fig.~\ref{figrisv1} (B), while the set-up looks promising, there has been no performance analysis on how deploying RIS (assuming a perfect RIS hardware exist) at the LED front-end to focus the emitted beam on the legitimate user can improve the channel gain while restricting the channel gain of the eavesdropper. In addition, it remains unclear which type of RIS material is suitable for deployment at the LED front end. Moreover, for multiple APs and user deployment scenarios for Fig.~\ref{figrisv1} (B), the RIS elements at the APs can be configured to create a dynamic interference-free cell-free VLC system in which a number of APs coordinate to focus all their emitted beams on the multiple legitimate users while no transmission is made to the eavesdroppers. However, novel schemes will be required to configure the RIS at the LEDs front-end to realize its beam focusing capabilities to ensure that the difference between the achievable rate of legitimate users and eavesdroppers is  maximized.

The following research issues need to be investigated for the set-up in Fig.~\ref{figrisv1} (C). In this figure, both Bob and Eve have direct LoS path with the transmitter. In VLC systems, the quality of the direct LoS path usually determines the system performance since any wall reflection components are considered negligible. Thus, Eve's LoS channel condition can be comparable to, or even better than, the channel quality of Bob, especially if Eve is closer to the transmitter than Bob. In this case, the question of ``can the reflected signal components from the RIS significantly increase the overall channel gain of Bob to realize any security improvements?'' arises. For such a scenario, optimizing only the RIS orientation parameter (as was done in \cite{Qian062021}) for the deployment in Fig.~\ref{figrisv1} (C) might not yield significant improvement in the security of the VLC system since any resultant performance gain from the RIS reflected signal components can be very minimal when compared with that of the LoS component. As a result, it is important to combine existing PLS schemes for VLC systems without RIS (e.g., artificial noise injection, secure beamforming/precoding, anti-eavesdropping signal design,  cooperation-based secure transmissions, power and resource allocation, etc.) and the exploitation of RISs to enhance the overall secrecy rate performance. Moreover, the reflected signals by the RIS can be added constructively with the direct LoS signals from the transmitter at Bob's side (i.e., signal enhancement), while being destructively added at Eve's side (i.e., signal cancellation). In the absence of a direct LoS transmission, the RIS can be used to improve the secrecy performance of the VLC system by reflecting all the incident beam towards Bob to improve the channel gain while making no signal available for Eve to eavesdrop. 

Another potential research area for the scenarios in Figs.~\ref{figrisv1} (B) and (C) is investigating appropriate algorithms (either centralized or distributed) to reliably obtain the channel state information (CSI) for the links between the RIS and the users since this information is needed by the RIS controller to properly configure its elements. This is crucial in the design of RIS-aided VLC systems since (i) RIS are low-powered structures that cannot initiate transmission to facilitate accurate channel estimation, and (ii) a large number of RIS elements and, as a consequence, a large number of extra channel coefficients  might be required to obtain a reasonable system performance  \cite{Zhang082020}.

\subsection{MIMO and RISs}
{{\color{black}{Typical indoor environments are characterized by the deployment of multiple light fixtures to ensure uniform illumination. These multiple light fixtures, when used for the dual purpose of communication and illumination, can be regarded as an  MIMO OWC system \cite{pathak4q2015}. Several studies have considered MIMO OWC systems, provided channel capacity bounds and proven the significant channel capacity improvements that MIMO brings in OWC systems \cite{Chaaban052018,7399676,9031328,Chaaban102018,7707381,8336902,4290022}. As in any OWC system, transmission in MIMO OWC systems may be hindered by an obstruction in the link between the transmitter and the receiver. Hence, the RIS technology could be used to connect users to the access point. Unlike RIS-based MIMO RF systems \cite{Yang082020, Shtaiwi062021, He092021, Wang032021, Khaleel092021, Perovic062021, 9392378, You122021}, and RIS-based cell-free MIMO RF systems \cite{Zhang062021, Bayan052021, AlNahhas062021, Trinh042021} which have been thoroughly studied, there has been minimal research on the application of RIS technology in MIMO OWC system.}}}

\begin{figure}
	\centering
	\includegraphics[width=0.49\textwidth]{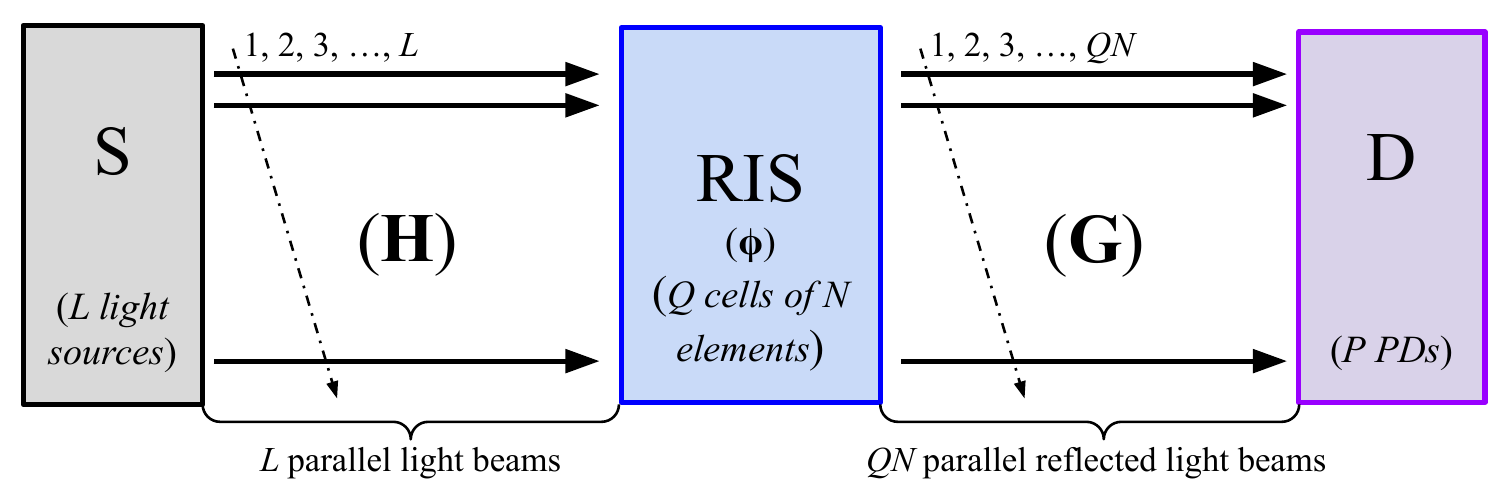}
	\caption{\small{\textcolor{black}{Model of an RIS-assisted MIMO OWC system \cite{ndjiongue112021}.}}}
	\label{Fig:00Model_MIMO_RIS}
\end{figure}

\begin{figure*}
	\centering
	\includegraphics[width=0.99\textwidth]{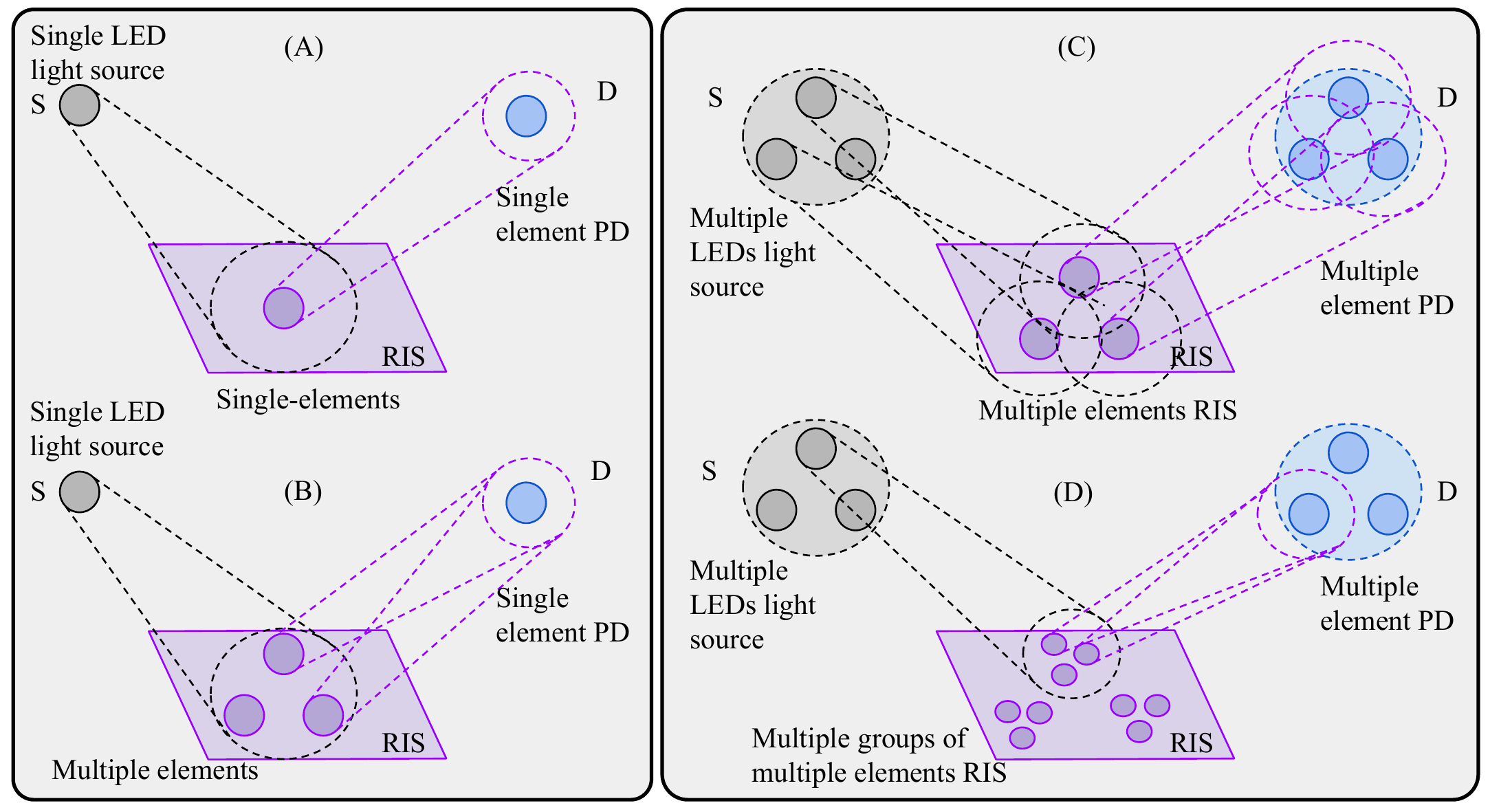}
	\caption{\small{\textcolor{black}{Scenarios of RIS-assisted MIMO OWC systems \cite{ndjiongue112021}.}}}
	\label{Fig:00P2P_RIS_OWC}
\end{figure*}

{{\color{black}{Figure~\ref{Fig:00Model_MIMO_RIS} depicts a generalized model of an RIS-aided MIMO OWC system \cite{ndjiongue112021}. In this figure, the data to be transmitted is distributed between the $L$ light sources constituting the transmitter. The transmitted light beams reflect off of the $QN$ RIS elements and are directed towards the $P$ PDs of the receiver. Several configurations of RIS-based MIMO OWC systems can be derived from Fig.~\ref{Fig:00Model_MIMO_RIS}. A few examples of such configurations are shown in Fig.~\ref{Fig:00P2P_RIS_OWC}.}}} They are all governed by the general transmission equation

\begin{equation}
\textbf{\textit{y}}(t) = \mathbb{H}\textbf{\textit{x}}(t) + \textbf{\textit{z}}(t),
\label{Eq:Genneral}
\end{equation}
where $ \textbf{\textit{x}}(t) = [x_1(t), \dots, x_{L}(t)]^T$ ($ x_i(t) \in \mathbb{R}_{+} $, $ i \in \{1, 2, \dots, M\}$) is the $ L $-dimensional real-valued S-RIS sub-channel input vectors (light intensity of the $ i^{th} $ transmitter at time instant $ t $). $ \textbf{\textit{y}}(t) = [y_1(t), \dots, y_{M}(t)]^T$ ($ y_k(t) \in \mathbb{R}_{+} $) denotes the $ QN $-dimensional real-valued RIS-D sub-channel output vectors, and $ \textbf{\textit{z}}(t) = [z_1(t), \dots, z_{QN}(t)]^T$ ($ z_k(t) \in \mathbb{R}_{+} $) is the $ P $-dimensional real-valued Gaussian noise vectors at D. The noise at the receiver, which is additive, independent, and identically distributed, is a combination of thermal and shot noises, which are well modeled as Gaussian. $ \textbf{\textit{y}}^{in}_{ris}$ and $ \textbf{\textit{y}}^{out}_{ris}$ are respectively given by $ \textbf{\textit{y}}^{in}_{ris}(t) = \textbf{H}\textbf{\textit{x}}(t) + \textbf{\textit{z}}^{in}_{ris}(t)$ and $ \textbf{\textit{y}}^{out}_{ris}(t) = \Phi\textbf{H}\textbf{\textit{x}}(t) + \textbf{\textit{z}}^{out}_{ris}(t)$, where $ \textbf{\textit{y}}^{in}_{ris} $ and $ \textbf{\textit{y}}^{out}_{ris} $ are $ QN $-dimensional real-valued signal measured at the input and output of the RIS module, respectively, and $\Phi$ denotes the phase shift matrix.  $ \textbf{\textit{z}}^{in}_{ris} $ and $ \textbf{\textit{z}}^{out}_{ris} $ represent the $ QN $-dimensional real-valued noise vectors before and after the RIS module, and the transmission from RIS to D is governed by $\textbf{\textit{y}}(t) = \textbf{G}\textbf{\textit{y}}^{out}_{ris}(t) + \textbf{\textit{z}}(t)$. In the matrix form, \eqref{Eq:Genneral}, defined in the $\mathbb{R}_+^{M\times M}$ space, can be given by
\begin{equation}
\mathbb{H} = \textbf{G}^{\text{T}}\Phi \textbf{H}.
\label{Eq:h}
\end{equation}
The entries of the matrices $\textbf{G}$ and $\textbf{H}$ strongly depend on the physical arrangements of the light sources ($\textbf{H}$) and photo detectors ($\textbf{G}$) \cite{ndjiongue112021}. 

{$x(t)$ in \eqref{Eq:Genneral} must always satisfy the non-negative and peak-intensity constraint, resulting in the generated light upper bounded by a total average value \cite{Chaaban052018, Chaaban102018}.
\begin{equation}
0 \leq x_i(t) \leq \mathbb{X} \:\: \text{and} \:\: \sum_{i = 1}^{M} \textbf{P}_i \leq p_o,
\label{Eq:constr}
\end{equation}
where $ \mathbb{X} $ represents the peak-input-intensity and $\textbf{P}_i$ = E$_{x_i}[x_i]$, with E$_{x}[\cdot] $ denoting the expectation with respect to the distribution of $ x $. The average optical power constraint can be written as $\|\textbf{P}\|_1$ = $\|\textbf{P}_1, \dots, \textbf{P}_M\|_1 \leq p_o$, where $\|\cdot\|_1 $ is the $ \ell_1 $-norm operator, and $ p_o $ is the transmitter's total optical power.}

With these two constraints, the channel capacity, $c(\mathbb{H}, p_o, \mathbb{X})$, of RIS-based MIMO OWC systems can be given by \cite{Chaaban052018}

\begin{equation}
c(\mathbb{H}, p_o, \mathbb{X}) = \underset{f(\textbf{\textit{x}})}{\mathrm{max}} I(\textbf{\textit{x}}:\mathbb{H}\textbf{\textit{x}}+\textbf{\textit{z}}).
\label{Eq:capa0}
\end{equation}
Figure~\ref{Fig:00P2P_RIS_OWC} illustrates four main scenarios for indoor RIS-assisted VLC systems. {Figure~\ref{Fig:00P2P_RIS_OWC}-(A) is a typical RIS-assisted SISO OWC system in which the transmitter, RIS, and receiver all consist of a single element. A system of this type is subject to a number of issues, including large geometric losses at the destination \cite{ndjiongue112021}. These losses are mitigated by using a multiple elements RIS (i.e., MIMO RIS), where the RIS elements converge their beam towards a single element receiver, as shown in Fig.~\ref{Fig:00P2P_RIS_OWC}-(B). Figures~\ref{Fig:00P2P_RIS_OWC}-(C) and -(D) shows the use of an RIS  to mitigate geometric losses in an OWC system where the transmitter, the RIS, and the receiver have multiple active components. The capacity bounds of all these channels can be evaluated based on $p_o/\mathbb{X} \geq L/2$ and $p_o/\mathbb{X} < L/2$, where $L/2$ represents a break in the channel capacity.}

\begin{figure}
	\centering
	\hspace*{-5mm}
	\includegraphics[width=0.55\textwidth]{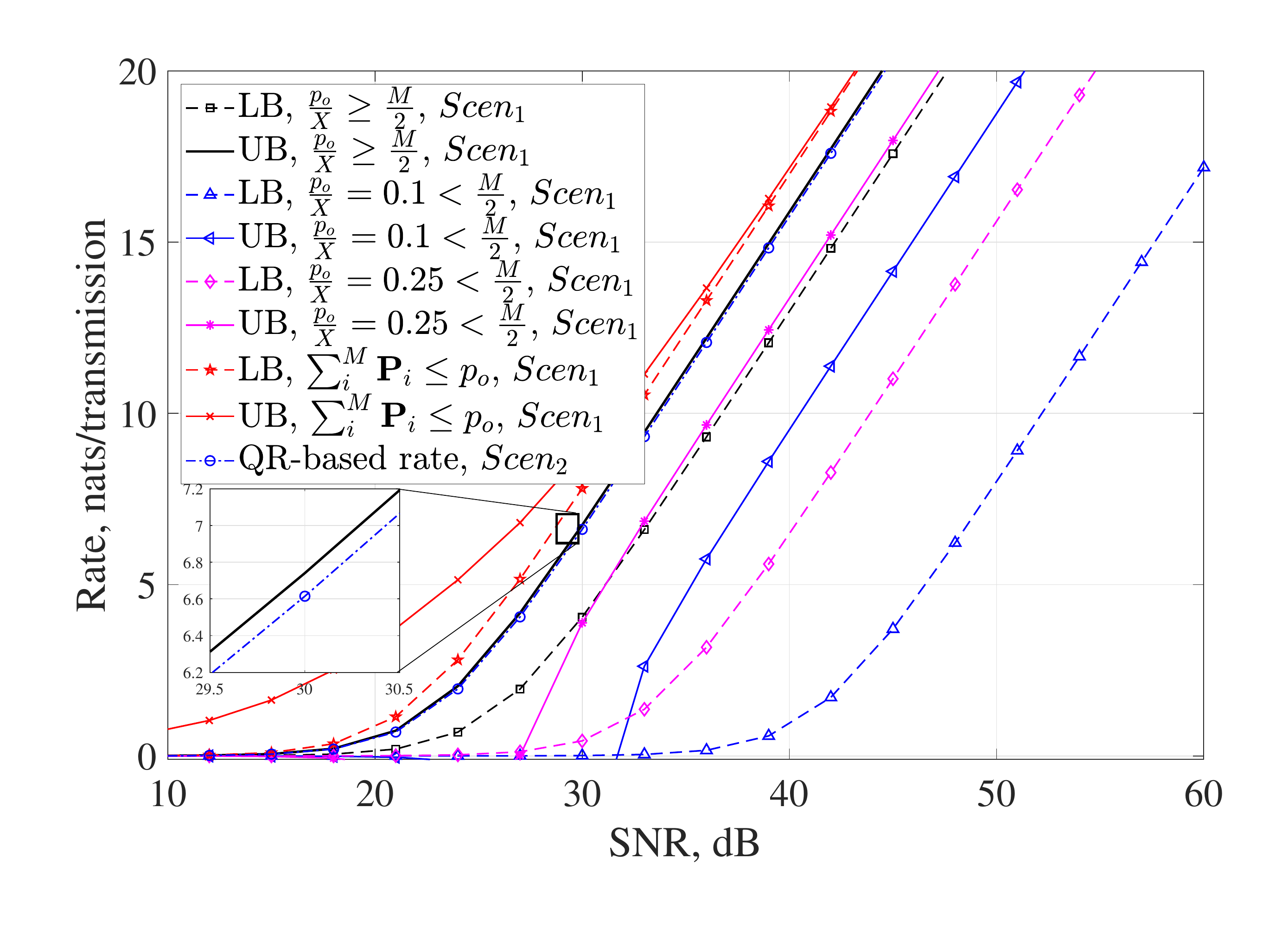}
	\caption{\textcolor{black}{Capacity bounds for an RIS-assisted MIMO OWC system. Example of an RIS-based MIMO FSO system \cite{ndjiongue112021}.}}
	\label{Fig:OWCMIMORIS}
	\vspace*{-3mm}
\end{figure}
As seen from Fig.~\ref{Fig:00P2P_RIS_OWC}, multiple RIS elements are required in MIMO OWC systems. Therefore, it is interesting to examine systems in which the RIS consists of multiple components. This type of system was proposed in \cite{ndjiongue112021}, where the authors utilized the $\it{QR}$-decomposition algorithm to determine the channel capacity of an RIS-based OWC system. This capacity relates to the disposition and orientation of the light sources, PDs, and RIS elements. In \cite{ndjiongue112021}, the authors considered two scenarios: One in which there is overlap between light beams at the receiver and a second one where there is no overlap. To maximize the channel capacity in the former, each channel must be optimized individually. The total maximized channel capacity is equal to the sum of all individual maximized channels. This is modeled in \cite{ndjiongue112021} as

\begin{equation}
c(\mathbb{H}^{N}_m, p_o, \mathbb{X}) =  \sum_{i=1}^{M} \mathrm{max} \left[g(\mathbb{H}_m, p_o, \mathbb{X})\right],
\label{Eq:para}
\end{equation}
where the function $ g(\mathbb{H}_m, p_o, \mathbb{X}) $ is the SISO channel capacity, which can be expressed using the considered transmission power and peak intensity constraints. When the light beams overlap at the receiver, the channel capacity may be determined by using the QR-decomposition algorithm. The $\it{QR}$-decomposition, in contrast to the single value decomposition extensively used in the analysis of MIMO RF systems, does not generate negative values and is thus very  suitable for the VLC channel. Such analysis is performed in \cite{ndjiongue112021} for OWC channels, with a focus on the situation where $p_o/\mathbb{X} \geq L/2$. The channel capacity of the overall channel can be expressed as 

\begin{equation}
c(\mathbb{H}_m, p_o, \mathbb{X}) = \underset{\mathbb{P}}{\mathrm{max}}  \sum_{i=1}^{M} \frac{1}{2} \log \left(1+\frac{2(\textbf{\textit{h}}_{m(i,i)})^2\mathbb{X}^2}{\pi 2e\sigma^2}\right),
\label{capa1}
\end{equation}  
where $\textbf{\textit{h}}_{m(i,i)}$ are diagonal entries of the overall channel matrix. Results of this analysis is proposed in Fig.~\ref{Fig:OWCMIMORIS}. The figure depicts upper bound (UB) and lower bound (LB) of the system with no overlaping ($Scen_1$) and overlaping ($Scen_2$), considering the constraint $p_o/\mathbb{X}$ with regard to the common number of light sources and PDs, and using the $\it{QR}$-decomposition. Open research challenges include determining the optimal position of the RISs as well as capacity optimization.    

\subsection{NOMA and RISs}
Non-orthogonal multiple access (NOMA) is among the key enabling techniques for massive connectivity in future wireless networks. In comparison to conventional orthogonal multiple access techniques such as OFDMA and TDMA, that serve a single user in each orthogonal resource block, NOMA is able to serve multiple users on the same resource block (which can be in time, frequency, or code domain) and, as result achieve high spectral efficiency gains \cite{Ding102017}. There have been many studies on the design and performance analysis of the two categories of NOMA (i.e., code and power domain NOMA) in RF communication networks (e.g., see \cite{Ding102017,Dai052018,Maraqa082020,Islam102016} and references therein). The success of NOMA in RF communication networks motivates its adoption for VLC systems. Specifically, NOMA is suited for VLC systems due to the following reasons: (\textit{i}) the modulation bandwidth of a typical VLC transmitter (e.g., LEDs) is very limited and that affects the achievable data rates; (\textit{ii}) NOMA is efficient in multiplexing a small number of users and in most indoor VLC systems each AP  typically have few number of users in its coverage area; (\textit{iii}) accurate CSI, which is crucial in NOMA-based systems (e.g., for superposition coding), can be obtained in VLC systems since the channel remains constant most of the time (due to the low mobility of users in VLC systems); (\textit{iv}) NOMA performs better in high SNR scenarios  \cite{Marshoud012016}, and VLC systems are able to provide very high SNR (when a LoS path exits) due to the short separation distance between the transmitter and receivers; and {\color{black}(\textit{v}) VLC systems offer a number of ways (e.g., optimizing the angles of irradiance and incidence through the use of angle diversity transmitters and receivers, tuning the  FoV of the transmitter and the receiver, etc.) to enhance the channel gain differences among users, which is  important in NOMA, although not a necessary requirement.\footnote{\color{black}{The performance of power-domain NOMA is highly dependent on having significant channel gain difference among users.}}} NOMA has been considered for VLC systems in few works, with \cite{Obeed032019} being the only survey paper on power domain NOMA-based VLC systems. In VLC, the use of power domain NOMA has received more attention as compared to code domain domain for the following reasons \cite{Marshoud042018}. Firstly, in a similar way that VLC systems typically serve few users, power domain NOMA is used to multiplex a small number of users. Secondly, power allocation in power domain NOMA largely depends on the availability of accurate CSI at the transmitter. In VLC, channel estimation is significantly less error-prone when compared to RF communication networks since the VLC channel is time-invariant and only changes when the user moves to another location. Finally, code domain NOMA are associated with high complexity multiuser detection algorithms \cite{Jamali062021}. 

Considering the downlink of a single cell indoor VLC system with ${\mathcal K}$ users, who are assumed to be static or quasi-static such that their CSI remains not outdated, the received signal at the $k$-th user according to the NOMA principle can be given by 

\begin{equation}\label{dec22}
    y_k=a_k{\sqrt{P}}h_k s_k + \sum_{j=1}^{{\mathcal K}-1} a_j{\sqrt{P}}h_k s_j + \sum_{i=k+1}^{\mathcal K} a_i{\sqrt{P}}h_k s_i  + n_k,
\end{equation}
where $a_k$ represents the power allocation coefficient $(0<~a_k<~1)$, $P$ refers to the total electrical power, $h_k$ is the channel power gain which can be calculated as in \cite{Aboagye102021}, $s_k$ denotes the desired information signal, and $n_k$ refers to the additive white Gaussian noise. In (\ref{dec22}), the first, second, and third terms on the right side of the equation represent the desired signal, the interference components cancelled by SIC, and the residual interference following SIC, respectively. Without loss of generality, an assumption is made that the $\mathcal K$ users have been sorted in ascending order according to their channel qualities

\begin{equation}
    h_1\le \cdots \le h_k \le \cdots \le h_{\mathcal K}.
\end{equation}
According to the NOMA principle, users with better channel conditions are allocated less power, meaning that $a_1 \ge \cdots \ge \cdots \ge a_{\mathcal K}$ and since the available transmit power is limited, the power allocation coefficients must satisfy the constraint

\begin{equation}
    \sum_{k=1}^{\mathcal K} a_k^2=1.
\end{equation}
It can be observed from (\ref{dec22}) that as the number of user users increases, the complexity of decoding the signal also increases since a total of ${\mathcal K}-1$ users (i.e., with the exception of the 1-st user) must perform SIC. Moreover, any residual interference resulting from inaccurate estimation of CSI increases with the number of users. In the sequel, most recent research on power doamin NOMA-VLC works are first summarized, and  then, the possible ways RIS can be utilized to enhance the performance of such communication systems are provided. 

\begin{figure*}
	\centering
	\includegraphics[width=0.8\textwidth]{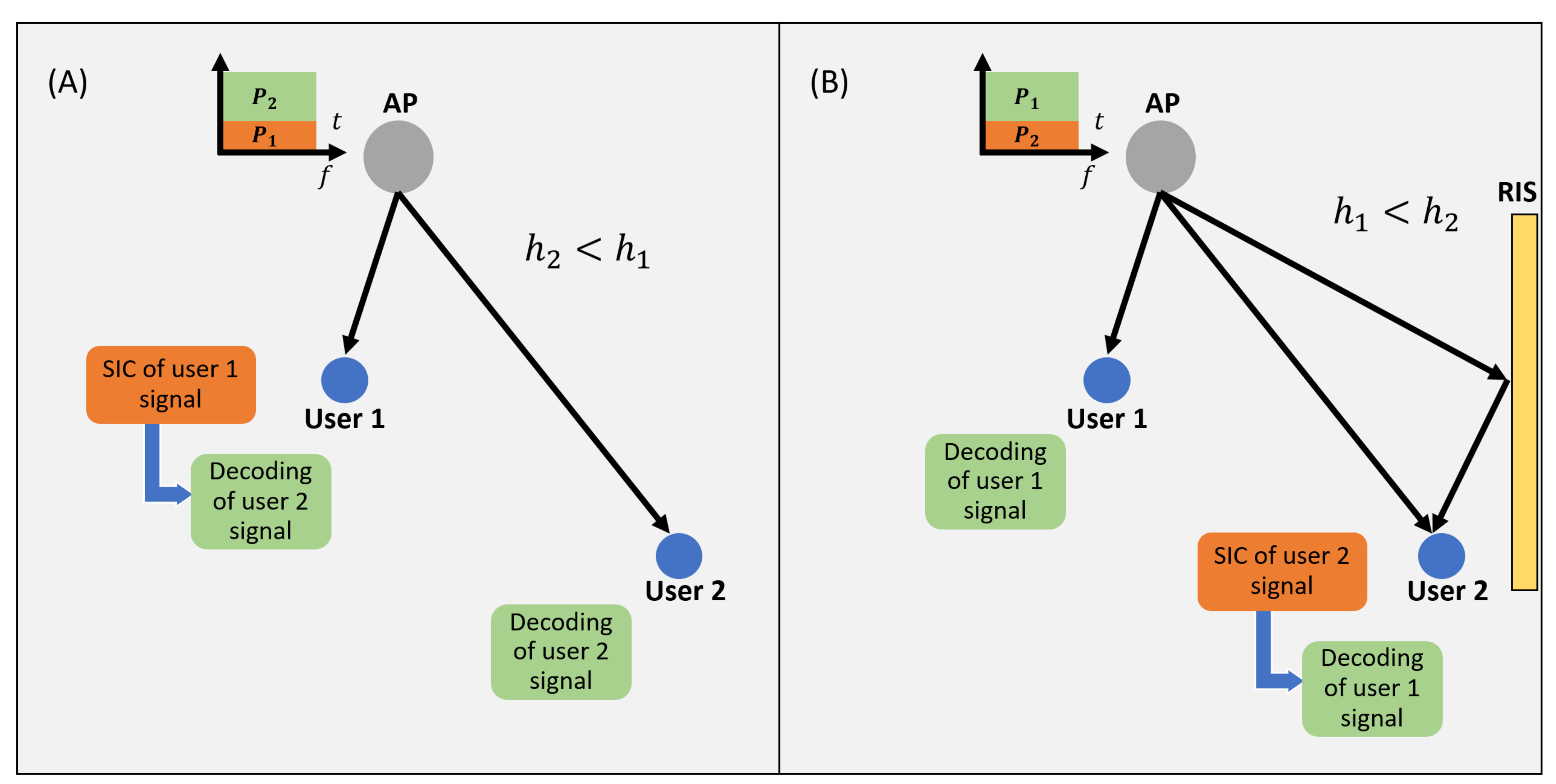}
	\caption{\small{Application of RISs in NOMA-based VLC systems: (A) NOMA principle in NOMA-VLC systems; and (B) Potential achievements of RIS-aided NOMA-VLC systems.}}
	\label{Fig:risnoma}
\end{figure*}
The authors in \cite{Kizilirmak092015,Lin112017,Yin122016,Yapici082019,Marshoud102017,Tran122021} demonstrated the superior sum rate and BER gains of NOMA-based VLC systems over OFDMA-based VLC systems. In \cite{Hammadi012021}, the authors proposed a general framework for energy efficiency analysis of a NOMA-based hybrid VLC and RF wireless system. Focusing on maximizing the sum-rate and the minimum user rate under transmit power constraints and QoS requirements, the authors in \cite{Shen122017} proposed a semi-closed form optimal power allocation scheme for a NOMA-based VLC system. The simulation results demonstrated that the proposed NOMA schemes outperformed OMA schemes. In \cite{Marshoud012016}, a novel gain ratio power allocation scheme was proposed, and the simulation results showed significant bit error rate (BER) and sum rate performance gain when compared to the static power allocation scheme. In \cite{Yang022017,Yang032021}, the authors considered a sum rate maximization problem under peak power and non-negativity constraints, and proposed a power allocation scheme based on the Lagrange dual method. In \cite{Zhang042017}, the authors proposed  gradient projection based algorithms for the sum rate and minimum user rate optimization problems while considering QoS requirements. In \cite{Ma032019}, the authors proposed bounds on the achievable rates of NOMA-based VLC networks. In that same paper, the authors proposed power allocation schemes for the static and mobile users scenarios. The joint optimization of user pairing, link selection, and power allocation to maximize the sum-rate of a hybrid VLC and RF communication system was investigated in \cite{Obeed032021,Obeed022021}. NOMA-based modulation schemes for single cell and multi-cell VLC systems where users located at the edge of the cells are served by multiple APs were examined in \cite{Uday102021} and \cite{Uday032021}, respectively. A user pairing scheme for a 3D VLC-NOMA systems was proposed in \cite{Janjua062020}. In \cite{Tahira082019}, a cuckoo search algorithm was proposed to optimize the transmit power, LED's semi-angle at half power, the FoV of the lens, and the gain of the optical filter such that the sum rate performance is maximized. In \cite{Eltokhey092021}, the optimization of power allocation in a NOMA-based VLC system using particle swarm optimization was investigated. Under MIMO settings, a normalized gain difference power allocation method was proposed in \cite{Chen022018} for a NOMA-based VLC system, and was shown to outperform the gain ratio power allocation method. An experimental demonstration of a MIMO NOMA-based VLC with single carrier mode of transmission and frequency domain successive interference cancellation was performed in \cite{Lin112017}.  

\begin{figure*}
	\centering
	\includegraphics[width=0.99\textwidth]{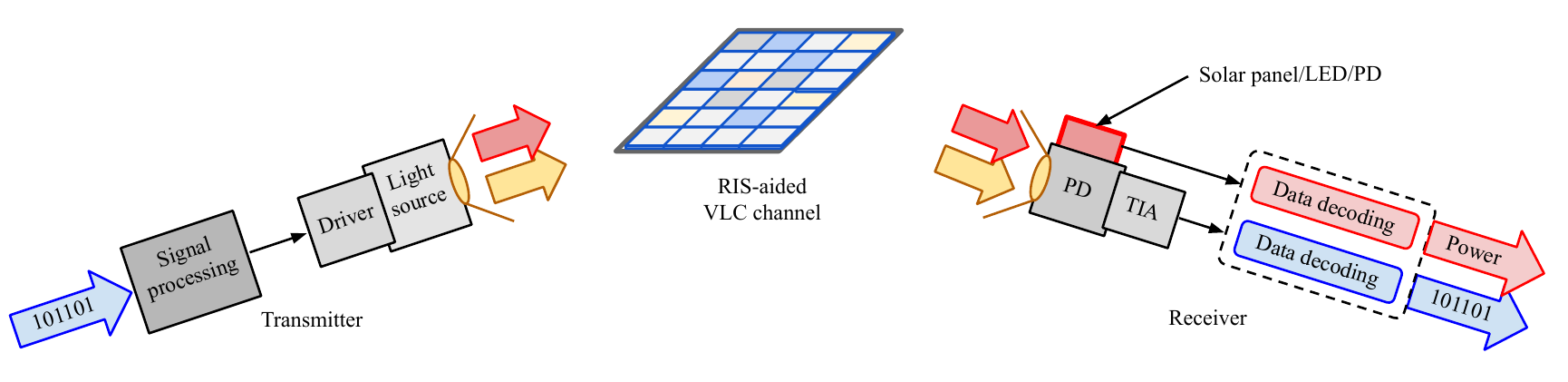}
	\caption{\small{Principles of simultaneous optical wireless data transmission and energy transfer over RIS-assisted VLC channels.}}
	\label{Fig:OwetRIS}
\end{figure*}

Many studies have exploited RISs to improve the performance gains in NOMA-based RF systems (e.g., see \cite{Zeng012021,Zengarxiv2021,Wanmingarxiv2022,Zuo112020,9437234,Wu092021} and references therein). However, this survey focuses on the use of RISs in NOMA-based VLC systems. RISs can be combined with NOMA in VLC systems for attractive performance gains. As previously mentioned, the NOMA principle requires the ordering of users' channel gains in ascending or descending order to enable them to employ SIC to recover the transmitted message. Thus, the performance of NOMA-based communication systems is highly dependent on the users' channel conditions. Moreover, it has been shown in \cite{Zhu022021} that NOMA based communication systems in general can attain superior performance gains than OMA-based systems only if the channel gains of different users are significantly distinct. Hence, the main functionality of an RIS in NOMA-based VLC systems is to modify the channel conditions of the users to yield additional performance improvements. Figure~\ref{Fig:risnoma} illustrates a potential application in a two-user RIS-aided NOMA-based VLC system. From this figure, it is demonstrated how RISs can further improve NOMA-VLC systems. Firstly, an RIS can be utilized in NOMA enabled VLC systems to control the channel gain ordering of users as well as to decide which users achieve higher data rate based on their QoS requirements. In VLC systems, the channel power gain is dependent on the angle of incidence, angle of irradiance, the LED's semi-angle at half power, and the distance between the AP and the user. All but the distance parameter are typically fixed. Hence, the closest users to the VLC AP will always have higher channel power gain and, as a result, the channel ordering of the users will always follow the location of users as shown in Fig.~\ref{Fig:risnoma} (A). However, the use of an RIS allows the VLC network to control the channel gains of the users and as result modify the ordering of the users' channel conditions as well as the potential achievable data rates. Specifically, the RIS elements can be optimized to enhance the channel conditions of users farther away from the AP and, as result, change the original order of the users' channel gain as illustrated in Fig.~\ref{Fig:risnoma} (B). Since the users with higher channel gain typically get higher data rate, the ability to modify the original order of the users' channel gain also allows the VLC network to sort its users based on their particular data rate requirements rather than on the uncontrollable random environment. For instance, users that are far away from the AP and have high data rate requirements in a conventional NOMA-VLC system might be in outage since irrespective of the allocated power, the weak users' rate will always be limited by interference from the strong user and the unfavorable channel conditions. However, the appropriate use of an RIS can allow the originally weak users with high data rate requirements to become strong users such that their required rates can be met. Secondly, the deployment of RISs in a NOMA-based VLC system can be used to enable massive connectivity by extending the coverage area of the VLC AP as depicted in Fig.~\ref{bkge1}. In conventional NOMA-based VLC systems and VLC systems in general, it can be difficult to guarantee uniform coverage in the indoor environment. This is because the VLC channel is easily affected by blockages, random device orientation as well as the limited FoV of the receivers. All these factors result in severe performance degradation. However, RISs can be deployed and optimized to provide coverage and better channel conditions for users outside the illumination area of the VLC AP. Finally, RISs can be used to create more distinct channel conditions for users in a NOMA-based VLC systems. As mentioned in \cite{Ding082016,Yin122016}, the performance of NOMA communication systems is very dependent on which users are paired together and could be enhanced by grouping users with more distinctive channel conditions. However, the channel conditions of VLC links are highly dependent on the location of the users. When the users are too close, that can negatively affect the VLC system performance when NOMA is applied. An RIS can be used in such a scenario to realize more distinct channel conditions for users, even when they are closely located. 

\subsection{Simultaneous Lightwave and Power Transfer and RISs}
In VLC technologies, the receiver may be self-powered. Thus, the integration of RISs into the VLC channels should enable energy forwarding (i.e., simultaneous lightwave and power transfer (SLPT)). The energy contained in the incident wave should also be reflected with the same orientation as the light.
Although there has been some work done on simultaneous wireless data and energy transfer using RIS-assisted RF systems \cite{Gao072021, Bhowal14232021, Mohjazi102021, Fernandez11122021}, this is not the case for OWC in general and VLC in particular. As a result, it is evident that there is still much to be done in the field of RIS-assisted SLPT optical systems. 

An RIS-supported SLPT OWC system can facilitate simultaneous wireless energy data transfer. This may be achieved by using a PD in conjunction with or without a TIA. The transmitter of such a system illuminates the environment as it broadcasts information, and the receiver harvests the energy from the light as it decodes the information. For example, an RIS-assisted SLPT VLC system may include three different technologies: lighting, data transmission, and energy harvesting. An example of such a system is shown in Fig.~\ref{Fig:OwetRIS}. The diagram in this figure illustrates the integration of an energy harvesting module. The transmitter is the same as the usual VLC transmitter, which includes a signal processing unit, an LED, and an LED driver. To detect light, three different types of devices can be used, namely PDs, LEDs, and photocells. In contrast to PDs and LEDs, photocells utilize photovoltaic processes to accomplish this. These devices are all capable of simultaneously reconstructing the message waveform and proceeding to data signal forwarding, along with offering energy harvesting capabilities. It is reported in \cite{Bhatti082016} that an indoor light can produce approximately 180 $ \mu $W of power harvested using an amorphous crystalline in a system without RIS. LEDs provide photon-to-current conversion, allowing up to 5V to be reached with a great current, which might be enough to charge a smart device's battery.

\subsection{Summary and Lessons Learned}
This section focused on the integration of RISs and other emerging technologies into VLC. Although such technologies have already been studied for VLC systems, we strongly believe their potentials have not been fully harnessed. In this section, we presented the various ways RISs can revolutionize the application of such new technologies in VLC systems to yield significant performance improvements. The key lessons are summarized as follows. Firstly, RISs can be utilized to design re-configurable ADTs that allow the creation of varying number of smaller cells depending on factors such as the number and the distribution of users. Secondly, RISs can be exploited to enhance the channel gain difference of the legitimate user and the eavesdropper to improve security performance. Thirdly, the various scenarios in which RISs can be deployed in MIMO networks and the corresponding channel structures, capacity, and achievable rates were discussed. Moreover, adopting RISs in NOMA-enabled VLC systems can help realize more distinct channel conditions and allow the control of the  channel gain ordering of users. RISs also enable coverage extension in NOMA-based VLC systems. Finally, the integration of RISs with simultaneous lightwave and power transfer was introduced.   

\section{Future Research Directions}
{\color{black}{Section~\ref{int} presented promising research directions on the integration of RISs and other technologies in VLC. This section further highlights other potential research problems in RIS-enabled VLC systems that need to be investigated in future work.}}
\subsubsection{Learning approach for blockage prediction in RIS-based VLC systems} {\color{black}{Current studies on RIS-aided VLC systems assume the availability of accurate CSI of the channel paths (i.e., between the APs, the RISs, and the users) for all users. The RISs' controller uses this information to optimize the configuration of the RISs' elements. However, acquisition of users' position information or the full real time CSI can be computationally expensive, and any errors in these estimates can severely impact the performance of the RIS-aided VLC system. This is due to the fact that the RIS can consist of a large number of elements and, as a result, the length of pilots required can be significant.  The idea of communication systems having the ability to predict the VLC environment (e.g., LoS link blockages), without knowledge of the CSI, such that the transmitter and the RIS can anticipate the blockage and act intelligently is an interesting and innovative concept that requires further investigation. As mentioned in \cite{Dileepa122021}, predicting blockage occurrences requires sensing the communication environment and processing the data using deep learning and ray casting techniques to allow for other functionalities such as users' and non-users' positioning. The lack of mathematical models for blockage predictions has motivated researchers to investigate predictors based on machine learning models (e.g., see \cite{Kalor122021} and references therein). Hence, the development of machine learning aided prediction techniques for RIS-enabled VLC systems is another promising research direction. Such designs should consider the unique characteristics of RIS-enabled VLC systems such as the limited range of the VLC APs, the structure of the RIS devices,  and the features of VLC signals. The concept of meta-learning \cite{9083856} should also be leveraged to reduce any training overheard involved.}}  

\subsubsection{Practical implementation and analysis} 
{\color{black}{The existing studies on RISs-enabled VLC systems reveal theoretical findings and simulation results but nothing on prototyping, practical implementation and analysis. Moreover, most studies make the assumption of perfect RIS hardware which will not generally be the case in practice. Thus, RIS technology for OWC systems in general is still in its infancy as there have not been any prototype designs and experimental results reported. The first step to the practical implementation of RISs in OWC is to develop a proof-of-concept platform to evaluate and illustrate the performance of RIS-enabled VLC systems. Such a platform will allow researchers and industry players to collaborate, evaluate, and compare  different RIS optimization algorithms, performance metrics (e.g., data rate, computational complexity, production cost, overhead cost, reliability, etc.) and system-level requirements for IMR and IMA RISs in VLC systems under more practical channel conditions and non-perfect RIS hardware scenarios that consider the impact of temperature on reflection/transmission characteristics, phase/yaw/roll responses with errors, etc. The results and analysis from such a  platform would validate theoretical and simulation findings, and accelerate the development of RISs for OWC in a low-cost and timely manner.}}  

\subsubsection{Transmitter-channel-receiver (TCaR)-RIS-assisted VLC systems} {\color{black}{Existing studies on RIS-enabled VLC systems have considered deploying the RIS either inside the receiver or in the channel. However, the RIS module could be deployed in the transmitter, the receiver, and the channel to realize transmitter-channel-receiver (TCaR)-RIS assisted VLC systems.  Located within the transmitter, the RIS could be used as a refractive element that assists in beam generation. Within the receiver, the RIS can enable dynamic FoV and perform incident light amplification and selective interference rejection. When deployed in the medium between the transmitter and the receiver (i.e., the channel), the RIS can achieve signal coverage and illumination expansion, and enhance security and signal power transmission. To date there have been no studies on TCaR-RIS-assisted VLC systems, and thus it will be interesting to explore how RISs can be deployed at the transmitter, the receiver, and in the channel to achieve performance improvements. Potential research areas include investigating key performance metrics to characterize such a novel system model, proposing channel models, and developing low-complexity optimization algorithms to efficiently optimize the performances of the RISs deployed in the various components of a TCaR-RIS-assisted VLC system. Thus, it is an important and challenging  research direction to investigate the performance trade-offs in TCaR-RIS-assisted VLC systems.}}

\subsubsection{Advances in optical RISs materials} {\color{black}{Although the application of reflect-array antennas in communication related applications has been well studied and developed \cite{9101000}, the same cannot be said about optical RISs. The materials (i.e., metasurfaces, mirror arrays, LCs, and their tuning mechanisms) used in the construction of optical RISs have not yet been well studied. For instance, the deployment of LC-based RISs at the receiver side has been demonstrated to provide beam steering and light amplification to enhance the data rate performance in VLC systems \cite{slysub}. Moreover, the concept of intelligent omni-surfaces capable of transmitting and reflecting signals simultaneously in OWC systems has been introduced in \cite{ndjiongue122021digital,ndjiongue122022omniDRIS}. 
However, several research challenges remain with regards to the RIS material type, channel modeling, optimum operating conditions (e.g., optimum refractive index, temperature, required voltage, etc.) and noise modeling in such novel applications. Furthermore, the absorption, reflection, and transmission characteristics of the various metasurface materials, mirror arrays, and LCs are not yet well understood as they depend on optical properties such as the optical wavelength range, refractive index, birefrigence, and temperature range. The performance of the various tuning mechanisms and the expected lifespan (e.g., for the actuators used in the IMA) need to analyzed.  Finally, it is important to factor in the availability of the various RIS materials and their cost.}}  

\subsubsection{RIS-enabled hybrid RF/VLC and standalone VLC systems} {\color{black}{VLC is typically deployed in the downlink of an OWC system while other technologies (e.g., RF or IR communications) are used for the uplink, forming a hybrid RF/VLC system with a decoupled uplink and downlink  \cite{7593453,9378787}. In such a hybrid design, RISs can be deployed for system performance enhancements when RF-based RIS   and VLC-based RIS techniques are utilized in the uplink and downlink, respectively. Moreover, RISs can be deployed in hybrid systems that use RF and VLC for the downlink and only RF for the uplink to improve the system performance. For instance, a single RIS with several elements (assuming an RIS that can reflect and transmit both RF and VLC signals exists) can be shared on  time-sharing basis or via dynamic clustering to aid uplink and downlink transmissions in a single user hybrid RF/VLC system. Furthermore, RISs can be deployed in several other standalone VLC networks as described in \cite{Ndjiongue012020}, that use VLC for both uplink and downlink paths. However, there have been no studies on RIS-enabled hybrid RF/VLC systems and RIS-aided standalone VLC systems that use a single/multiple RIS(s) to simultaneously enhance uplinks and downlinks. As a result, several technical challenges relating to the duplexing techniques, RIS control and resource optimization algorithms, distribution of the RISs, suitable RIS material for RF and VLC signals, as well as the associated overhead cost and implementation complexity remain to be tackled.}}

\section{Conclusion}
This paper has presented a comprehensive review of the design, application, and performance analysis of RISs in OWC in general and VLC in particular. Firstly, an overview of metasurfaces and mirrors and how they obtain their reconfigurability have been presented from the physics point of view. Then, the various RIS technologies suited for optical communications and their corresponding deployment scenarios have been explored. Afterward, the distinction between optical RISs and relays and their typical functionalities, benefits, and drawbacks have been presented. Furthermore, important issues such as RIS reflection design for solving LoS blockages, RISs for VLC transmitter and receiver performance enhancements have been thoroughly studied. More particularly, the standard RIS technologies for such roles have been presented, and design guidelines and performance optimization techniques (e.g., optimization problem structure, optimization variables, and promising solution techniques) have been provided. Simulation results have demonstrated how properly selecting an RIS technology for a VLC system and optimizing its performance can reduce the operational complexity of the RIS and yield significant data rate performance improvements. Finally, the potential role of RISs in other emerging wireless technologies, important challenges that should be expected, interesting open issues, and future research directions have been provided. It is hoped that this paper will provide helpful guidance for future research in this emerging and promising area of integrating RISs and VLC systems. This is crucial in harnessing the full potential of the RIS technology to enable the design of cost-effective, highly reliable, and sustainable future wireless networks.
\bibliographystyle{IEEEtran}
\bibliography{IEEEfull}

\begin{thebibliography}{100}
\providecommand{\url}[1]{#1}
\csname url@samestyle\endcsname
\providecommand{\newblock}{\relax}
\providecommand{\bibinfo}[2]{#2}
\providecommand{\BIBentrySTDinterwordspacing}{\spaceskip=0pt\relax}
\providecommand{\BIBentryALTinterwordstretchfactor}{4}
\providecommand{\BIBentryALTinterwordspacing}{\spaceskip=\fontdimen2\font plus
\BIBentryALTinterwordstretchfactor\fontdimen3\font minus
  \fontdimen4\font\relax}
\providecommand{\BIBforeignlanguage}[2]{{%
\expandafter\ifx\csname l@#1\endcsname\relax
\typeout{** WARNING: IEEEtran.bst: No hyphenation pattern has been}%
\typeout{** loaded for the language `#1'. Using the pattern for}%
\typeout{** the default language instead.}%
\else
\language=\csname l@#1\endcsname
\fi
#2}}
\providecommand{\BIBdecl}{\relax}
\BIBdecl

\bibitem{Bariah102020}
L.~Bariah, L.~Mohjazi, S.~Muhaidat, P.~C. Sofotasios, G.~K. Kurt,
  H.~Yanikomeroglu, and O.~A. Dobre, ``A prospective look: Key enabling
  technologies, applications and open research topics in 6g networks,''
  \emph{IEEE Access}, vol.~8, pp. 174\,792--174\,820, Aug. 2020.

\bibitem{Islam102016}
S.~M.~R. Islam, N.~Avazov, O.~A. Dobre, and K.-S. Kwak, ``{Power-domain
  non-orthogonal multiple access (NOMA) in 5G systems: Potentials and
  challenges},'' \emph{IEEE Commun. Surveys Tuts.}, vol.~19, no.~2, pp.
  721--742, Second Quarter 2017.

\bibitem{Ding102017}
Z.~Ding, X.~Lei, G.~K. Karagiannidis, R.~Schober, J.~Yuan, and V.~K. Bhargava,
  ``A survey on non-orthogonal multiple access for 5g networks: Research
  challenges and future trends,'' \emph{IEEE J. Sel. Areas Commun.}, vol.~35,
  no.~10, pp. 2181--2195, Oct. 2017.

\bibitem{Dai052018}
L.~Dai, B.~Wang, Z.~Ding, Z.~Wang, S.~Chen, and L.~Hanzo, ``{A survey of
  non-orthogonal multiple access for 5G},'' \emph{IEEE Commun. Surveys Tuts.},
  vol.~20, no.~3, pp. 2294--2323, Third Quarter 2018.

\bibitem{Yin122016}
L.~Yin, W.~O. Popoola, X.~Wu, and H.~Haas, ``{Performance evaluation of
  non-orthogonal multiple access in visible light communication},'' \emph{IEEE
  Trans. Commun.}, vol.~64, no.~12, pp. 5162--5175, Dec. 2016.

\bibitem{Bhardwaj102021}
P.~Bhardwaj and S.~M. Zafaruddin, ``Performance of dual-hop relaying for thz-rf
  wireless link over asymmetrical $\alpha$-$\mu$ fading,'' \emph{IEEE Trans.
  Veh. Technol.}, vol.~70, no.~10, pp. 10\,031--10\,047, Oct. 2021.

\bibitem{Mamaghani022022}
M.~Tatar~Mamaghani and Y.~Hong, ``Terahertz meets untrusted uav-relaying:
  Minimum secrecy energy efficiency maximization via trajectory and
  communication co-design,'' \emph{IEEE Trans. Veh. Technol.}, Feb. 2022, Early
  Access.

\bibitem{Xia122021}
Q.~Xia and J.~M. Jornet, ``Multi-hop relaying distribution strategies for
  terahertz-band communication networks: A cross-layer analysis,'' \emph{IEEE
  Trans. Wireless Commun.}, Dec. 2021, Early Access.

\bibitem{Chen012021}
Z.~Chen and D.~Smith, ``Mmwave m2m networks: Improving delay performance of
  relaying,'' \emph{IEEE Trans. Wireless Commun.}, vol.~20, no.~1, pp.
  577--589, Jan. 2021.

\bibitem{Ruiz022021}
C.~G. Ruiz, A.~Pascual-Iserte, and O.~Muñoz, ``Analysis of blocking in mmwave
  cellular systems: Application to relay positioning,'' \emph{IEEE Trans.
  Commun.}, vol.~69, no.~2, pp. 1329--1342, Feb. 2021.

\bibitem{Yalcin062020}
A.~Z. Yalçın and Y.~Yapıcı, ``Multiuser precoding for sum-rate maximization
  in relay-aided mmwave communications,'' \emph{IEEE Trans. Veh. Technol.},
  vol.~69, no.~6, pp. 6808--6812, Jun. 2020.

\bibitem{Kim062020}
H.~Kim, K.~Granström, L.~Gao, G.~Battistelli, S.~Kim, and H.~Wymeersch, ``5g
  mmwave cooperative positioning and mapping using multi-model phd filter and
  map fusion,'' \emph{IEEE Trans. Wireless Commun.}, vol.~19, no.~6, pp.
  3782--3795, Jun. 2020.

\bibitem{Zhang022021}
Y.~Zhang, J.~Mu, and J.~Xiaojun, ``Performance of multi-cell mmwave noma
  networks with base station cooperation,'' \emph{IEEE Commun. Lett.}, vol.~25,
  no.~2, pp. 442--445, Feb. 2021.

\bibitem{Bayaki122012}
E.~Bayaki, D.~S. Michalopoulos, and R.~Schober, ``Edfa-based all-optical
  relaying in free-space optical systems,'' \emph{IEEE Trans. Commun.},
  vol.~60, no.~12, pp. 3797--3807, Dec. 2012.

\bibitem{Liu102020}
W.~Liu, J.~Ding, J.~Zheng, X.~Chen, and C.-L. I, ``Relay-assisted technology in
  optical wireless communications: A survey,'' \emph{IEEE Access}, vol.~8, pp.
  194\,384--194\,409, Oct. 2020.

\bibitem{Najafi072017}
M.~Najafi, V.~Jamali, and R.~Schober, ``Optimal relay selection for the
  parallel hybrid rf/fso relay channel: Non-buffer-aided and buffer-aided
  designs,'' \emph{IEEE Trans. Commun.}, vol.~65, no.~7, pp. 2794--2810, Jul.
  2017.

\bibitem{Kamga052021}
G.~N.~Kamga, S.~Aïssa, T.~R. Rasethuntsa, and M.-S. Alouini, ``Mixed rf/fso
  communications with outdated-csi-based relay selection under double
  generalized gamma turbulence, generalized pointing errors, and nakagami-m
  fading,'' \emph{IEEE Trans. Wireless Commun.}, vol.~20, no.~5, pp.
  2761--2775, May 2021.

\bibitem{Wang072019}
Z.-Y. Wang, H.-Y. Yu, D.-M. Wang, and Y.-Y. Zhang, ``Optimized cooperative
  finite-alphabet two-way relaying strategy for indoor visible light
  communication systems,'' \emph{IEEE Trans. Wireless Commun.}, vol.~18,
  no.~10, pp. 4886--4901, Oct. 2019.

\bibitem{Kizilirmak102015}
R.~C. Kizilirmak, O.~Narmanlioglu, and M.~Uysal, ``Relay-assisted ofdm-based
  visible light communications,'' \emph{IEEE Trans. Commun.}, vol.~63, no.~10,
  pp. 3765--3778, Oct. 2015.

\bibitem{Aboagye102021}
S.~Aboagye, T.~M.~N. Ngatched, and O.~A. Dobre, ``{Subchannel and power
  allocation in downlink VLC under different system configurations},''
  \emph{IEEE Trans. Wireless Commun.}, Early Access, 2021.

\bibitem{Pan112020}
G.~Pan, J.~Ye, C.~Zhang, J.~An, H.~Lei, Z.~Ding, and M.~S. Alouini, ``Secure
  cooperative hybrid vlc-rf systems,'' \emph{IEEE Trans. Wireless Commun.},
  vol.~19, no.~11, pp. 7097--7107, Nov. 2020.

\bibitem{Guzman112015}
B.~G. Guzman, A.~L. Serrano, and V.~P. Gil~Jimenez, ``Cooperative optical
  wireless transmission for improving performance in indoor scenarios for
  visible light communications,'' \emph{IEEE Trans. Consum. Electron.},
  vol.~61, no.~4, pp. 393--401, Apr. 2015.

\bibitem{Guzman072020}
B.~G. Guzmán, A.~A. Dowhuszko, V.~P.~G. Jiménez, and A.~I. Pérez-Neira,
  ``Resource allocation for cooperative transmission in optical wireless
  cellular networks with illumination requirements,'' \emph{IEEE Trans.
  Commun.}, vol.~68, no.~10, pp. 6440--6455, Oct. 2020.

\bibitem{Aboagye062021}
S.~Aboagye, T.~M.~N. Ngatched, O.~A. Dobre, and A.~G. Armada, ``Energy
  efficient subchannel and power allocation in cooperative vlc systems,''
  \emph{IEEE Commun. Lett.}, vol.~25, no.~6, pp. 1935--1939, Jun. 2021.

\bibitem{basar022020}
E.~Basar, ``{Reconfigurable intelligent surface-based index modulation: A new
  beyond MIMO paradigm for 6G},'' \emph{IEEE Trans. Commun.}, vol.~68, no.~5,
  pp. 3187--3196, May 2020.

\bibitem{wu012020}
{Q. {Wu} and R. {Zhang}}, ``{Towards smart and reconfigurable environment:
  Intelligent reflecting surface aided wireless network},'' \emph{IEEE Commun.
  Mag.}, vol.~58, no.~1, pp. 106--112, Jan. 2020.

\bibitem{liu112020}
H.~Liu, X.~Yuan, and Y.-J.~A. Zhang, ``{Matrix-calibration-based cascaded
  channel estimation for reconfigurable intelligent surface assisted multiuser
  MIMO},'' \emph{IEEE J. Sel. Areas Commun.}, vol.~38, no.~11, pp. 2621--2636,
  Nov. 2020.

\bibitem{Yang082020}
X.~Yang, C.-K. Wen, and S.~Jin, ``{MIMO detection for reconfigurable
  intelligent surface-assisted millimeter wave systems},'' \emph{IEEE J. Sel.
  Areas Commun.}, vol.~38, no.~8, pp. 1777--1792, Aug. 2020.

\bibitem{Shtaiwi062021}
E.~Shtaiwi, H.~Zhang, S.~Vishwanath, M.~Youssef, A.~Abdelhadi, and Z.~Han,
  ``{Channel estimation approach for RIS assisted MIMO systems},'' \emph{IEEE
  Trans. Cogn. Commun. Netw}, vol.~7, no.~2, pp. 452--465, Jun. 2021.

\bibitem{He092021}
J.~He, H.~Wymeersch, and M.~Juntti, ``{Channel estimation for RIS-aided mmWave
  MIMO systems via atomic norm ninimization},'' \emph{IEEE Trans. Wireless
  Commun.}, vol.~20, no.~9, pp. 5786--5797, Sep. 2021.

\bibitem{Abumarshoud042021}
H.~Abumarshoud, L.~Mohjazi, O.~A. Dobre, M.~Di~Renzo, M.~A. Imran, and H.~Haas,
  ``{LiFi through reconfigurable intelligent surfaces: A new frontier for
  6G?}'' \emph{IEEE Veh. Technol Mag.}, vol.~17, no.~1, pp. 37--46, Mar. 2022.

\bibitem{ndjiongue112021}
A.~R. Ndjiongue, T.~M.~N. Ngatched, and O.~A. Dobre, ``{On the capacity of
  RIS-assisted intensity-modulation optical channels},'' \emph{IEEE Commun.
  Lett.}, vol.~26, no.~2, pp. 389--393, 2022.

\bibitem{Gao072021}
Z.~Gao, Y.~Xu, Q.~Wang, Q.~Wu, and D.~Li, ``{Outage-constrained energy
  efficiency maximization for RIS-assisted WPCNs},'' \emph{IEEE Commun. Lett.},
  vol.~25, no.~10, pp. 3370--3374, Jul. 2021.

\bibitem{Gong062020}
S.~Gong, X.~Lu, D.~T. Hoang, D.~Niyato, L.~Shu, D.~I. Kim, and Y.-C. Liang,
  ``Toward smart wireless communications via intelligent reflecting surfaces: A
  contemporary survey,'' \emph{IEEE Commun. Surveys Tuts.,}, vol.~22, no.~4,
  pp. 2283--2314, Fourth Quarter 2020.

\bibitem{Ning022021}
B.~Ning, Z.~Chen, W.~Chen, Y.~Du, and J.~Fang, ``Terahertz multi-user massive
  mimo with intelligent reflecting surface: Beam training and hybrid
  beamforming,'' \emph{IEEE Trans. Veh. Technol.}, vol.~70, no.~2, pp.
  1376--1393, Feb. 2021.

\bibitem{Pei122021}
X.~Pei, H.~Yin, L.~Tan, L.~Cao, Z.~Li, K.~Wang, K.~Zhang, and E.~Björnson,
  ``Ris-aided wireless communications: Prototyping, adaptive beamforming, and
  indoor/outdoor field trials,'' \emph{IEEE Trans. Commun.}, vol.~69, no.~12,
  pp. 8627--8640, Dec. 2021.

\bibitem{Zhang082020}
J.~Zhang, E.~Björnson, M.~Matthaiou, D.~W.~K. Ng, H.~Yang, and D.~J. Love,
  ``{Prospective multiple antenna technologies for beyond 5G},'' \emph{IEEE J.
  Sel. Areas Commun.}, vol.~38, no.~8, pp. 1637--1660, Aug. 2020.

\bibitem{Zeng062021}
M.~Zeng, E.~Bedeer, O.~A. Dobre, P.~Fortier, Q.-V. Pham, and W.~Hao,
  ``Energy-efficient resource allocation for irs-assisted multi-antenna uplink
  systems,'' \emph{IEEE Wireless Commun. Lett.}, vol.~10, no.~6, pp.
  1261--1265, Jun. 2021.

\bibitem{Hao082021}
W.~Hao, G.~Sun, M.~Zeng, Z.~Chu, Z.~Zhu, O.~A. Dobre, and P.~Xiao, ``Robust
  design for intelligent reflecting surface-assisted mimo-ofdma terahertz iot
  networks,'' \emph{IEEE Internet Things J.}, vol.~8, no.~16, pp.
  13\,052--13\,064, Aug. 2021.

\bibitem{GLYBOVSKI052016}
S.~B. Glybovski, S.~A. Tretyakov, P.~A. Belov, Y.~S. Kivshar, and C.~R.
  Simovski, ``Metasurfaces: From microwaves to visible,'' \emph{Phys. Reports},
  vol. 634, pp. 1--72, May 2016.

\bibitem{ndjiongue052021}
A.~R. Ndjiongue, T.~M.~N. Ngatched, O.~A. Dobre, and H.~Haas,
  ``{Re-configurable intelligent surface-based VLC receivers using tunable
  liquid-crystals: The concept},'' \emph{IEEE/OSA J. Lightw. Technol.},
  vol.~39, no.~10, pp. 3193--3200, May 2021.

\bibitem{Aboagye122021}
S.~Aboagye, T.~M.~N. Ngatched, O.~A. Dobre, and A.~R. Ndjiongue, ``{Intelligent
  reflecting surface-aided indoor visible light communication systems},''
  \emph{IEEE Commun. Lett.}, vol.~25, no.~12, pp. 3913--3917, Dec. 2021.

\bibitem{Cao022020}
B.~Cao, M.~Chen, Z.~Yang, M.~Zhang, J.~Zhao, and M.~Chen, ``{Reflecting the
  light: Energy efficient visible light communication with reconfigurable
  intelligent surface},'' in \emph{Proc. 92nd 2020 VTC Conf. (VTC2020-Fall),
  Victoria, BC, Canada}, Nov. 2020, pp. 1--5.

\bibitem{Qian062021}
L.~Qian, X.~Chi, L.~Zhao, and A.~Chaaban, ``{Secure visible light
  communications via intelligent reflecting surfaces},'' in \emph{Proc. ICC
  Conf., Montreal, QC, Canada}, Jun. 2021, pp. 1--6.

\bibitem{Abdelhady142022}
A.~M. Abdelhady, O.~Amin, A.~K. Sultan, M.-S. Alouini, and B.~Shihada,
  ``Channel characterization of irs-based visible light communication
  systems,'' \emph{IEEE Trans. Commun.}, Jan. 2022, Early Access.

\bibitem{Sun112021}
S.~Sun, F.~Yang, and J.~Song, ``{Sum rate maximization for intelligent
  reflecting surface-aided visible light communications},'' \emph{IEEE Commun.
  Lett.}, vol.~25, no.~11, pp. 3619--3623, Nov. 2021.

\bibitem{Ssun2022}
S.~Sun, F.~Yang, and Z.~Han, ``{Joint resource management for intelligent
  reflecting surface-aided visible light communications},'' \emph{IEEE Trans.
  Wireless Commun.}, vol.~21, no.~8, pp. 6508--6522, Aug. 2022.

\bibitem{9437234}
J.~Xu, Y.~Liu, X.~Mu, and O.~A. Dobre, ``{STAR-RISs}: Simultaneous transmitting
  and reflecting reconfigurable intelligent surfaces,'' \emph{IEEE Commun.
  Lett.}, vol.~25, no.~9, pp. 3134--3138, May 2021\color{black}.

\bibitem{svilen}
S.~Dimitrov and H.~Haas, \emph{Principle of {LED} light communications: Towards
  networked {Li-Fi}}.\hskip 1em plus 0.5em minus 0.4em\relax Cambridge
  university press, 2015.

\bibitem{slysub}
S.~Aboagye, A.~R. Ndjiongue, T.~M.~N. Ngatched, and O.~A. Dobre, ``Design and
  optimization of liquid crystal {RIS}-based visible light communication
  receivers,'' \emph{IEEE Photon. J.}, vol.~14, no.~6, pp. 1--7, Dec. 2022.

\bibitem{9609592}
A.~Krohn, S.~Pachnicke, and P.~A. Hoeher, ``Genetic optimization of liquid
  crystal matrix based interference suppression for {VLC} {MIMO}
  transmissions,'' \emph{IEEE Photon. J.}, vol.~14, no.~1, pp. 1--5, Feb. 2022.

\bibitem{9860058}
A.~Krohn, A.~Harlakin, S.~Arms, S.~Pachnicke, and P.~A. Hoeher, ``Impact of
  liquid crystal based interference mitigation and precoding on the multiuser
  performance of {VLC} massive {MIMO} arrays,'' \emph{IEEE Photon. J.},
  vol.~14, no.~5, pp. 1--12, 2022.

\bibitem{8839844}
A.~Krohn, G.~J.~M. Forkel, P.~A. Hoeher, and S.~Pachnicke, ``{LCD}-based
  optical filtering suitable for non-imaging channel decorrelation in {VLC}
  applications,'' \emph{IEEE/OSA J. Lightw. Technol.}, vol.~37, no.~23, pp.
  5892--5898, Dec. 2019\color{black}{}.

\bibitem{Obeed032019}
M.~Obeed, A.~M. Salhab, M.-S. Alouini, and S.~A. Zummo, ``{On optimizing VLC
  networks for downlink multi-user transmission: A survey},'' \emph{IEEE
  Commun. Surveys Tuts.}, vol.~21, no.~3, pp. 2947--2976, Third Quarter
  2019\color{black}{}.

\bibitem{9241073}
A.~Memedi and F.~Dressler, ``Vehicular visible light communications: A
  survey,'' \emph{IEEE Commun. Surveys Tuts.}, vol.~23, no.~1, pp. 161--181,
  First Quarter 2021.

\bibitem{7932857}
A.-M. Cailean and M.~Dimian, ``Current challenges for visible light
  communications usage in vehicle applications: A survey,'' \emph{IEEE Commun.
  Surveys Tuts.}, vol.~19, no.~4, pp. 2681--2703, Fourth Quarter 2017.

\bibitem{9351549}
X.~Wu, M.~D. Soltani, L.~Zhou, M.~Safari, and H.~Haas, ``Hybrid lifi and wifi
  networks: A survey,'' \emph{IEEE Commun. Surveys Tuts.}, vol.~23, no.~2, pp.
  1398--1420, Second Quarter 2021.

\bibitem{8015106}
J.~Luo, L.~Fan, and H.~Li, ``Indoor positioning systems based on visible light
  communication: State of the art,'' \emph{IEEE Commun. Surveys Tuts.},
  vol.~19, no.~4, pp. 2871--2893, Fourth Quarter 2017.

\bibitem{6497926}
A.~Sevincer, A.~Bhattarai, M.~Bilgi, M.~Yuksel, and N.~Pala, ``Lightnets: Smart
  lighting and mobile optical wireless networks - a survey,'' \emph{IEEE
  Commun. Surveys Tuts.}, vol.~15, no.~4, pp. 1620--1641, Fourth Quarter 2013.

\bibitem{8308722}
X.~Li, R.~Zhang, and L.~Hanzo, ``Optimization of visible-light optical wireless
  systems: Network-centric versus user-centric designs,'' \emph{IEEE Commun.
  Surveys Tuts.}, vol.~20, no.~3, pp. 1878--1904, Third Quarter 2018.

\bibitem{8698841}
L.~E.~M. Matheus, A.~B. Vieira, L.~F.~M. Vieira, M.~A.~M. Vieira, and
  O.~Gnawali, ``Visible light communication: Concepts, applications and
  challenges,'' \emph{IEEE Commun. Surveys Tuts.}, vol.~21, no.~4, pp.
  3204--3237, Fourth Quarter 2019.

\bibitem{7072557}
D.~Karunatilaka, F.~Zafar, V.~Kalavally, and R.~Parthiban, ``Led based indoor
  visible light communications: State of the art,'' \emph{IEEE Commun. Surveys
  Tuts.}, vol.~17, no.~3, pp. 1649--1678, Third Quarter 2015.

\bibitem{kumarapr2010}
N.~Kumar and N.~R. Lourenco, ``Led-based visible light communication system: A
  brief survey and investigation,'' \emph{J. Eng. Appl. Sci.}, vol.~5, no.~4,
  pp. 296--307, Apr. 2010.

\bibitem{tsonevfeb2014}
L.~Qian, X.~Chi, L.~Zhao, and A.~Chaaban, ``Light fidelity (li-fi): Towards
  all-optical networking,'' in \emph{Proc. SPIE OPTO, San Francisco, CA, USA},
  2014, pp. 1--10.

\bibitem{wunovdec2014}
S.~Wu, H.~Wang, and C.-H. Youn, ``Visible light communications for 5g wireless
  networking systems: From fixed to mobile communications,'' \emph{IEEE Netw.},
  Jun. 2014.

\bibitem{pathak4q2015}
P.~H. Pathak, X.~Feng, P.~Hu, and P.~Mohapatra, ``Visible light communication,
  networking, and sensing: A survey, potential and challenges,'' \emph{IEEE
  Commun. Surveys Tuts.}, vol.~17, no.~4, pp. 2047--2077, Fourth Quarter 2015.

\bibitem{sahaoct2015}
N.~Saha, M.~S. Ifthekhar, N.~T. Le, and Y.~M. Jang, ``Survey on optical camera
  communications: Challenges and opportunities,'' \emph{IET Optoelectron.},
  vol.~9, no.~5, pp. 172--183, Oct. 2015.

\bibitem{qiuoct2016}
Y.~Qiu, H.~Chen, W.~Meng, and Y.~M. Jang, ``Channel modeling for visible light
  communications—a survey,'' \emph{Wireless Commun. Mobile Comput.}, vol.~16,
  no.~14, pp. 2016--2034, Oct. 2016.

\bibitem{sindhubala042016}
K.~Sindhubala and B.~Vijayalakshmi, ``Survey on noise sources and restrain
  techniques in visible-light communication,'' \emph{Light Eng.}, vol.~24,
  no.~2, pp. 107--117, Apr. 2016.

\bibitem{do052016}
T.~Do and M.~Yoo, ``An in-depth survey of visible light communication based
  positioning systems,'' \emph{Sensors}, vol.~16, no.~5, pp. 1--40, May 2016.

\bibitem{7096279}
R.~Zhang, J.~Wang, Z.~Wang, Z.~Xu, C.~Zhao, and L.~Hanzo, ``Visible light
  communications in heterogeneous networks: Paving the way for user-centric
  design,'' \emph{IEEE Wireless Commun.}, vol.~22, no.~2, pp. 8--16, Apr. 2015.

\bibitem{komine5012004}
T.~Komine and M.~Nakagawa, ``Fundamental analysis for visible-light
  communication system using led lights,'' \emph{IEEE Trans. Consum.
  Electron.}, vol.~50, no.~1, pp. 100--107, Feb. 2004\color{black}{}.

\bibitem{Tang022021}
T.~Tang, T.~Shang, and Q.~Li, ``{Impact of multiple shadows on visible light
  communication channel},'' \emph{IEEE Commun. Lett.}, vol.~25, no.~2, pp.
  513--517, Feb. 2021.

\bibitem{Soltani032019}
M.~D. Soltani, A.~A. Purwita, Z.~Zeng, H.~Haas, and M.~Safari, ``{Modeling the
  random orientation of mobile devices: Measurement, analysis and liFi use
  case},'' \emph{IEEE Trans. Commun.}, vol.~67, no.~3, pp. 2157--2172, Mar.
  2019.

\bibitem{Ndjiongue012020}
{A. R. {Ndjiongue}, T. M. N. {Ngatched}, O. A. {Dobre}, and A. G. {Armada}},
  ``{{VLC}-based networking: Feasibility and challenges},'' \emph{IEEE Netw.},
  vol.~34, no.~4, pp. 158--165, Jan. 2020.

\bibitem{hu022019}
W.-W. Hu, ``{PAPR reduction in DCO-OFDM visible light communication systems
  using optimized odd and even sequences combination},'' \emph{IEEE Photon.
  J.}, vol.~11, no.~1, pp. 1--15, Feb. 2019.

\bibitem{na032018}
Z.~Na, Y.~Wang, M.~Xiong, X.~Liu, and J.~Xia, ``{Modeling and throughput
  analysis of an ADO-OFDM based relay-assisted VLC system for 5G networks},''
  \emph{IEEE Access}, vol.~6, pp. 17\,586--17\,594, Mar. 2018.

\bibitem{wang012019}
T.~Q. Wang, H.~Li, and X.~Huang, ``{Analysis and mitigation of clipping noise
  in layered ACO-OFDM based visible light communication systems},'' \emph{IEEE
  Trans. Commun.}, vol.~67, no.~1, pp. 564--577, Jan. 2019.

\bibitem{Ndjiongue062018}
A.~R. Ndjiongue and H.~C. Ferreira, ``{An overview of outdoor visible light
  communications},'' \emph{Wiley Trans. Emerg. Telecommun. Technol.}, vol.~29,
  no.~7, p. e3448, Jun. 2018.

\bibitem{Haas032016}
{H. {Haas} {\textit{et al.}}}, ``{What is LiFi?}'' \emph{IEEE/OSA J. Lightw.
  Technol.}, vol.~34, no.~6, pp. 1533--1544, Mar. 2016.

\bibitem{Alshaer092018}
H.~{Alshaer} and H.~{Haas}, ``{Bidirectional LiFi attocell access point slicing
  scheme},'' \emph{IEEE Trans. Netw. Service Manag.}, vol.~15, no.~3, pp.
  909--922, Sep. 2018.

\bibitem{figueiredo102017}
{M. {Figueiredo} {\textit{et al.}}}, ``{Lighting the wireless world: The
  promise and challenges of visible light communication},'' \emph{IEEE Consum.
  Electron. Mag.}, vol.~6, no.~4, pp. 28--37, Oct. 2017.

\bibitem{papanikolaou032018}
{V. K. {Papanikolaou} {\textit{et al.}}}, ``{Li-Fi and Wi-Fi with common
  backhaul: Coordination and resource allocation},'' in \emph{Proc. IEEE
  Wireless Commun. Netw. Conf., Barcelona, Spain}, 15-18 Apr. 2018, pp. 1--6.

\bibitem{albraheem072018}
{L. I. {Albraheem} {\textit{et al.}}}, ``{Toward designing a Li-Fi-based
  hierarchical IoT architecture},'' \emph{IEEE Access}, vol.~6, pp.
  40\,811--40\,825, Jul. 2018.

\bibitem{wu122017}
{X. {Wu} {\textit{et al.}}}, ``{Access point selection for hybrid Li-Fi and
  Wi-Fi networks},'' \emph{IEEE Trans. Commun.}, vol.~65, no.~12, pp.
  5375--5385, Dec. 2017.

\bibitem{soltani032017}
{M. D. {Soltani} {\textit{et al.}}}, ``{Handover modeling for indoor Li-Fi
  cellular networks: The effects of receiver mobility and rotation},'' in
  \emph{Proc. IEEE Wireless Commun. Netw. Conf., San Francisco, CA, USA}, Mar.
  2017, pp. 1--6.

\bibitem{surampudi082018}
A.~{Surampudi} and R.~K. {Ganti}, ``{Interference characterization in downlink
  Li-Fi optical attocell networks},'' \emph{IEEE/OSA J. Lightw. Technol.},
  vol.~36, no.~16, pp. 3211--3228, Aug. 2018.

\bibitem{pittolo042016}
{A. {Pittolo} {\textit{et al.}}}, ``{In-vehicle power line communication:
  Differences and similarities among the in-car and the in-ship scenarios},''
  \emph{IEEE Veh. Technol. Mag.}, vol.~11, no.~2, pp. 43--51, Apr. 2016.

\bibitem{artale032018}
{G. {Artale} {\textit{et al.}}}, ``{A new low cost power line communication
  solution for smart grid monitoring and management},'' \emph{IEEE Instrum.
  Meas. Mag.}, vol.~21, no.~2, pp. 29--33, Mar. 2018.

\bibitem{prasad112020}
{G. {Prasad} and L. {Lampe}}, ``{Full-duplex power line communications: Design
  and applications from multimedia to smart grid},'' \emph{IEEE Commun. Mag.},
  vol.~58, no.~2, pp. 106--112, Nov. 2020.

\bibitem{belhassen032020}
{H. {Belhassen} and E. {Verney}}, ``{Proof of concept of vehicle to
  infrastructure power line communication link for tramway CCTV},'' \emph{IEEE
  Intell. Transp. Syst. Mag.}, vol.~13, no.~3, pp. 89--98, Mar. 2020.

\bibitem{lin042017}
{B. {Lin} {\textit{et al.}}}, ``{Experimental demonstration of an indoor VLC
  positioning system based on OFDMA},'' \emph{IEEE Photon. J.}, vol.~9, no.~2,
  pp. 1--9, Apr. 2017.

\bibitem{li072018}
{Y. {Li} {\textit{et al.}}}, ``{A VLC smartphone camera based indoor
  positioning system},'' \emph{IEEE Photon. Technol. Lett.}, vol.~30, no.~13,
  pp. 1171--1174, Jul. 2018.

\bibitem{zhu012018}
{B. {Zhu} {\textit{et al.}}}, ``{Three-dimensional VLC positioning based on
  angle difference of arrival with arbitrary tilting angle of receiver},''
  \emph{IEEE J. Sel. Areas Commun.}, vol.~36, no.~1, pp. 8--22, Jan. 2018.

\bibitem{park042017}
{J. K. {Park} {\textit{et al.}}}, ``{Hadamard matrix design for a low-cost
  indoor positioning system in visible light communication},'' \emph{IEEE
  Photon. J.}, vol.~9, no.~2, pp. 1--10, Apr. 2017.

\bibitem{gu052016}
{W. {Gu} {\textit{et al.}}}, ``{Impact of multipath reflections on the
  performance of indoor visible light positioning systems},'' \emph{IEEE/OSA J.
  Lightw. Technol.}, vol.~34, no.~10, pp. 2578--2587, May 2016.

\bibitem{alam072019}
{F. {Alam} {\textit{et al.}}}, ``{Indoor visible light positioning using
  spring-relaxation technique in real-world setting},'' \emph{IEEE Access},
  vol.~7, pp. 91\,347--91\,359, Jul. 2019.

\bibitem{feng052016}
{S. {Feng} {\textit{et al.}}}, ``{Hybrid positioning aided amorphous-cell
  assisted user-centric visible light downlink techniques},'' \emph{IEEE
  Access}, vol.~4, pp. 2705--2713, May 2016.

\bibitem{hou122016}
{Y. {Hou} {\textit{et al.}}}, ``{Single LED beacon-based 3-D indoor positioning
  using off-the-shelf devices},'' \emph{IEEE Photon. J.}, vol.~8, no.~6, pp.
  1--11, Dec. 2016.

\bibitem{qin012019}
C.~{Qin} and X.~{Zhan}, ``{VLIP: Tightly coupled visible-light/inertial
  positioning system to cope with intermittent outage},'' \emph{IEEE Photon.
  Technol. Lett.}, vol.~31, no.~2, pp. 129--132, Jan. 2019.

\bibitem{du082018}
{P. {Du} {\textit{et al.}}}, ``{Demonstration of a low-complexity indoor
  visible light positioning system using an enhanced TDOA scheme},'' \emph{IEEE
  Photon. J.}, vol.~10, no.~4, pp. 1--10, Aug. 2018.

\bibitem{guo012019}
{X. {Guo} {\textit{et al.}}}, ``{Indoor localization using visible light via
  two-layer fusion network},'' \emph{IEEE Access}, vol.~7, pp.
  16\,421--16\,430, Jan. 2019.

\bibitem{yu122018}
{X. {Yu} {\textit{et al.}}}, ``{Single LED-based indoor positioning system
  using multiple photodetectors},'' \emph{IEEE Photon. J.}, vol.~10, no.~6, pp.
  1--8, Dec.\color{black}{} 2018.

\bibitem{9664274}
L.~Bariah, M.~Elamassie, S.~Muhaidat, P.~C. Sofotasios, and M.~Uysal,
  ``Non-orthogonal multiple access-based underwater vlc systems in the presence
  of turbulence,'' \emph{IEEE Photon. J.}, vol.~14, no.~1, pp. 1--7, Feb. 2022.

\bibitem{7593257}
Z.~Zeng, S.~Fu, H.~Zhang, Y.~Dong, and J.~Cheng, ``A survey of underwater
  optical wireless communications,'' \emph{IEEE Commun. Surveys Tuts.},
  vol.~19, no.~1, pp. 204--238, First Quarter 2017.

\bibitem{8449320}
M.~Elamassie, F.~Miramirkhani, and M.~Uysal, ``Performance characterization of
  underwater visible light communication,'' \emph{IEEE Trans. Commun.},
  vol.~67, no.~1, pp. 543--552, Jan. 2019.

\bibitem{8370053}
M.~V. Jamali, A.~Mirani, A.~Parsay, B.~Abolhassani, P.~Nabavi, A.~Chizari,
  P.~Khorramshahi, S.~Abdollahramezani, and J.~A. Salehi, ``Statistical studies
  of fading in underwater wireless optical channels in the presence of air
  bubble, temperature, and salinity random variations,'' \emph{IEEE Trans.
  Commun.}, vol.~66, no.~10, pp. 4706--4723, Oct. 2018.

\bibitem{9140399}
M.~Elamassie and M.~Uysal, ``Vertical underwater visible light communication
  links: Channel modeling and performance analysis,'' \emph{IEEE Trans.
  Wireless Commun.}, vol.~19, no.~10, pp. 6948--6959, Oct. 2020.

\bibitem{9585115}
M.~Elamassie, L.~Bariah, M.~Uysal, S.~Muhaidat, and P.~C. Sofotasios,
  ``Capacity analysis of noma-enabled underwater vlc networks,'' \emph{IEEE
  Access}, vol.~9, pp. 153\,305--153\,315, Oct. 2021.

\bibitem{9590553}
R.~Hamagami, T.~Ebihara, N.~Wakatsuki, and K.~Mizutani, ``Optimal modulation
  technique for underwater visible light communication using rolling-shutter
  sensor,'' \emph{IEEE Access}, vol.~9, pp. 146\,422--146\,436, Oct. 2021.

\bibitem{9324916}
W.~Niu, H.~Chen, J.~Zhang, Y.~Ha, P.~Zou, and N.~Chi, ``Nonlinearity mitigation
  based on modulus pruned look-up table for multi-bit delta-sigma 32-cap
  modulation in underwater visible light communication system,'' \emph{IEEE
  Photon. J.}, vol.~13, no.~1, pp. 1--12, Feb. 2021.

\bibitem{8610103}
Y.-W. Ji, G.-F. Wu, and C.~Wang, ``Generalized likelihood block detection for
  spad-based underwater vlc system,'' \emph{IEEE Photon. J.}, vol.~12, no.~1,
  pp. 1--10, Feb. 2020.

\bibitem{8935430}
M.~Chen, P.~Zou, L.~Zhang, and N.~Chi, ``Demonstration of a 2.34 gbit/s
  real-time single silicon-substrate blue led-based underwater vlc system,''
  \emph{IEEE Photon. J.}, vol.~12, no.~1, pp. 1--11, Feb. 2020.

\bibitem{9760160}
Q.~Hu, C.~Gong, T.~Lin, J.~Luo, and Z.~Xu, ``Secrecy performance analysis for
  water-to-air visible light communication,'' \emph{IEEE/OSA J. Light.
  Technol.}, vol.~40, no.~14, pp. 4607--4620, Jul. 2022\color{black}{}.

\bibitem{di042020}
{B. {Di} {\textit{et al.}}}, ``{Practical hybrid beamforming with
  limited-resolution phase shifters for reconfigurable intelligent surface
  based multi-user communications},'' \emph{IEEE Trans. Veh. Technol.},
  vol.~69, no.~4, pp. 4565--4570, Apr. 2020.

\bibitem{hu032018}
{S. {Hu} {\textit{et al.}}}, ``{Beyond massive MIMO: The potential of data
  transmission with large intelligent surfaces},'' \emph{IEEE Trans. Signal
  Process.}, vol.~66, no.~10, pp. 2746--2758, Mar. 2018.

\bibitem{liang062019}
{Y. C. Liang {\textit{et al.}}}, ``{Large intelligent surface/antennas (LISA):
  Making reflective radios smart},'' \emph{IEEE/OSA J. Commun. Inf. Netw.},
  vol.~4, no.~2, pp. 40--50, Jun. 2019.

\bibitem{basar091019}
{E. Basar {\textit{et al.}}}, ``{Wireless communications through reconfigurable
  intelligent surfaces},'' \emph{IEEE Access}, vol.~7, pp. 116\,753--116\,773,
  Aug. 2019.

\bibitem{han062019}
{Y. {Han} {\textit{et al.}}}, ``{Large intelligent surface-assisted wireless
  communication exploiting statistical CSI},'' \emph{IEEE Trans. Veh.
  Technol.}, vol.~68, no.~8, pp. 8238--8242, Jun. 2019.

\bibitem{huang062019}
{C. Huang {\textit{et al.}}}, ``{Reconfigurable intelligent surfaces for energy
  efficiency in wireless communication},'' \emph{IEEE Trans. Wireless Commun.},
  vol.~18, pp. 4157--4170, Jun. 2019.

\bibitem{IEEE092011}
``{IEEE standard for local and metropolitan area networks-part 15.7:
  Short-range wireless optical communication using visible light},'' \emph{IEEE
  Std 802.15.7-2011}, pp. 1--309, Sep. 2011.

\bibitem{IEEE072018}
``{IEEE draft standard for local and metropolitan area networks - part 15.7:
  Short-range optical wireless communications},'' \emph{IEEE P802.15.7/D2a,
  Jun. 2018}, pp. 1--428, Jul. 2018.

\bibitem{IEEE122018}
``{IEEE approved draft standard for local and metropolitan area networks - part
  15.7: Short-range optical wireless communications},'' \emph{IEEE
  P802.15.7/D3a, Aug. 2018}, pp. 1--428, Dec. 2018.

\bibitem{IEEE042019}
``{IEEE standard for local and metropolitan area networks-part 15.7:
  Short-range optical wireless communications},'' \emph{IEEE Std 802.15.7-2018
  (Revision of IEEE Std 802.15.7-2011)}, pp. 1--407, Apr. 2019.

\bibitem{Renzo052019}
M.~D. Renzo, M.~Debbah, D.-T. Phan-Huy, A.~Zappone, M.-S. Alouini, C.~Yuen,
  V.~Sciancalepore, G.~C. Alexandropoulos, J.~Hoydis, H.~Gacanin, J.~d. Rosny,
  A.~Bounceur, G.~Lerosey, and M.~Fink, ``{Smart radio environments empowered
  by reconfigurable AI meta-surfaces: An idea whose time has come},''
  \emph{EURASIP J. Wireless Commun. Netw.}, vol. 2019, no.~1, p. 129, May 2019.

\bibitem{Luo082021}
S.~Luo, J.~Hao, F.~Ye, J.~Li, Y.~Ruan, H.~Cui, W.~Liu, and L.~Chen,
  ``{Evolution of the electromagnetic manipulation: From tunable to
  programmable and intelligent metasurfaces},'' \emph{Micromachines}, vol.~12,
  no.~8, pp. 1--25, Jul. 2021.

\bibitem{Ma102019}
Q.~Ma, G.~D. Bai, H.~B. Jing, C.~Yang, L.~Li, and T.~J. Cui, ``Smart
  metasurface with self-adaptively reprogrammable functions,'' \emph{Light: Sc.
  App.}, vol.~8, no.~1, p.~98, Oct. 2019.

\bibitem{Hum012014}
S.~V. Hum and J.~Perruisseau-Carrier, ``{Reconfigurable reflectarrays and array
  lenses for dynamic antenna beam control: A review},'' \emph{IEEE Trans.
  Antennas Propag.}, vol.~62, no.~1, pp. 183--198, Oct. 2014.

\bibitem{Saman012016}
J.~Saman and J.~Zubin, ``All-dielectric metamaterials,'' \emph{Nat.
  Nanotechnol.}, vol.~11, pp. 23--36, Jan. 2016.

\bibitem{nicolas072015}
N.~Bonod, ``Large-scale dielectric metasurfaces,'' \emph{Nat. Material},
  vol.~14, pp. 664--665, Jul. 2015.

\bibitem{Boltasseva012011}
A.~Boltasseva and H.~A. Atwater, ``Low-loss plasmonic metamaterials,''
  \emph{Science}, vol. 331, no. 6015, pp. 290--291, Jan. 2011.

\bibitem{Li062018}
A.~Li, S.~Singh, and D.~Sievenpiper, ``Metasurfaces and their applications:,''
  \emph{Nanophoton.}, vol.~7, no.~6, pp. 989--1011, Jun. 2018.

\bibitem{liu052018}
F.~Liu, A.~Pitilakis, M.~S. Mirmoosa, O.~Tsilipakos, X.~Wang, A.~C.
  Tasolamprou, S.~Abadal, A.~Cabellos-Aparicio, E.~Alarcón, C.~Liaskos, N.~V.
  Kantartzis, M.~Kafesaki, E.~N. Economou, C.~M. Soukoulis, and S.~Tretyakov,
  ``{Programmable metasurfaces: State of the art and prospects},'' in
  \emph{Proc. IEEE Int. Symp. Circuits and Sys. (ISCAS), Florence, Italy}, May
  2018, pp. 1--5.

\bibitem{chang072018}
S.~Chang, X.~Guo, and X.~Ni, ``Optical metasurfaces: Progress and
  applications,'' \emph{Annual Review of Materials Research}, vol.~48, no.~1,
  pp. 279--302, Jul. 2018.

\bibitem{Lininger082020}
A.~Lininger, A.~Y. Zhu, J.-S. Park, G.~Palermo, S.~Chatterjee, J.~Boyd,
  F.~Capasso, and G.~Strangi, ``{Optical properties of metasurfaces infiltrated
  with liquid crystals},'' vol. 117, no.~34, pp. 20\,390--20\,396, Aug. 2020.

\bibitem{Sun062019}
M.~Sun, X.~Xu, X.~W. Sun, X.~Liang, V.~Valuckas, Y.~Zheng,
  R.~Paniagua-Dom{\'i}nguez, and A.~I. Kuznetsov, ``Efficient visible light
  modulation based on electrically tunable all dielectric metasurfaces embedded
  in thin-layer nematic liquid crystals,'' \emph{Scientific Reports}, vol.~9,
  no.~1, p. 8673, Jun. 2019.

\bibitem{Wang122017}
X.~chen Wang and S.~Tretyakov, ``{Tunable perfect absorption in continuous
  graphene sheets on metasurface substrates},'' \emph{arXiv: Opt.}, 2017.

\bibitem{Aldrigo102014}
M.~Aldrigo, M.~Dragoman, A.~Costanzo, and D.~Masotti, ``{Exploitation of
  graphene as HIS and RIS for devices in the MW and THz frequency ranges},'' in
  \emph{Proc. 44th European Microw. Conf., Rome, Italy}, Oct. 2014, pp.
  355--358.

\bibitem{Lee092012}
S.~H. Lee, M.~Choi, T.-T. Kim, S.~Lee, M.~Liu, X.~Yin, H.~K. Choi, S.~S. Lee,
  C.-G. Choi, S.-Y. Choi, X.~Zhang, and B.~Min, ``{Switching terahertz waves
  with gate-controlled active graphene metamaterials},'' \emph{Nat. Materials},
  vol.~11, no.~11, pp. 936--941, Nov. 2012.

\bibitem{Giddens042018}
H.~Giddens, L.~Yang, J.~Tian, and Y.~Hao, ``{Mid-infrared reflect-array antenna
  with beam switching enabled by continuous graphene layer},'' \emph{IEEE
  Photon. Technol. Lett.}, vol.~30, no.~8, pp. 748--751, Apr. 2018.

\bibitem{Jiang092021}
X.-Q. Jiang, W.-H. Fan, C.~Song, X.~Chen, and Q.~Wu, ``Terahertz
  photoconductive antenna based on antireflection dielectric metasurfaces with
  embedded plasmonic nanodisks,'' \emph{Appl. Opt.}, vol.~60, no.~26, pp.
  7921--7928, Sep. 2021.

\bibitem{Bhattacharya102019}
A.~Bhattacharya, D.~Ghindani, and S.~S. Prabhu, ``Enhanced terahertz emission
  bandwidth from photoconductive antenna by manipulating carrier dynamics of
  semiconducting substrate with embedded plasmonic metasurface,'' \emph{Opt.
  Express}, vol.~27, no.~21, pp. 30\,272--30\,279, Oct. 2019.

\bibitem{Shen012011}
N.-H. Shen, M.~Massaouti, M.~Gokkavas, J.-M. Manceau, E.~Ozbay, M.~Kafesaki,
  T.~Koschny, S.~Tzortzakis, and C.~M. Soukoulis, ``{Optically implemented
  broadband blueshift switch in the terahertz regime},'' \emph{Phys. Rev.
  Lett.}, vol. 106, pp. 1--4, Jan. 2011.

\bibitem{MAJDAZDANCEWICZ201892}
E.~Majda-Zdancewicz, M.~Suproniuk, M.~Pawłowski, and M.~Wierzbowski, ``Current
  state of photoconductive semiconductor switch engineering,''
  \emph{Opto-Electron. Review}, vol.~26, no.~2, pp. 92--102, May 2018.

\bibitem{Shen042009}
N.-H. Shen, M.~Kafesaki, T.~Koschny, L.~Zhang, E.~N. Economou, and C.~M.
  Soukoulis, ``Broadband blueshift tunable metamaterials and dual-band
  switches,'' \emph{Phys. Rev. B}, vol.~79, p. 161102, Apr. 2009.

\bibitem{9491943}
S.~Zhang, H.~Zhang, B.~Di, Y.~Tan, M.~Di~Renzo, Z.~Han, H.~Vincent~Poor, and
  L.~Song, ``Intelligent omni-surfaces: Ubiquitous wireless transmission by
  reflective-refractive metasurfaces,'' \emph{IEEE Trans. Wireless Commun.},
  vol.~21, no.~1, pp. 219--233, Jan. 2022.

\bibitem{Chen122004}
J.~Chen, W.~Weingartner, A.~Azarov, and R.~Giles, ``Tilt-angle stabilization of
  electrostatically actuated micromechanical mirrors beyond the pull-in
  point,'' \emph{J. Microelectromechanical Sys.}, vol.~13, no.~6, pp. 988--997,
  Dec. 2004.

\bibitem{Lei122010}
L.~Wu, S.~Dooley, E.~A. Watson, P.~F. McManamon, and H.~Xie, ``{A
  tip-tilt-piston micromirror array for optical phased array applications},''
  \emph{J. Microelectromechanical Sys.}, vol.~19, no.~6, pp. 1450--1461, May
  2010.

\bibitem{David082016}
D.~Torres, T.~Wang, J.~Zhang, X.~Zhang, S.~Dooley, X.~Tan, H.~Xie, and
  N.~Sepúlveda, ``{VO$_2$-Based MEMS Mirrors},'' \emph{J.
  Microelectromechanical Sys.}, vol.~25, no.~4, pp. 780--787, Aug. 2016.

\bibitem{Hillmer072018}
H.~Hillmer, B.~Al-Qargholi, M.~M. Khan, N.~Worapattrakul, H.~Wilke, C.~Woidt,
  and A.~Tatzel, ``Optical {MEMS}-based micromirror arrays for active light
  steering in smart windows,'' vol.~57, no. 8S2, pp. 1--13, Jul. 2018.

\bibitem{ndjiongue062021}
A.~R. Ndjiongue, T.~M.~N. Ngatched, O.~A. Dobre, and H.~Haas, ``{Toward the use
  of re-configurable intelligent surfaces in VLC systems: Beam steering},''
  \emph{IEEE Wireless Commun. Mag.}, vol.~28, no.~3, pp. 156--162, Jun. 2021.

\bibitem{9662064}
S.~Sun, T.~Wang, F.~Yang, J.~Song, and Z.~Han, ``Intelligent reflecting
  surface-aided visible light communications: Potentials and challenges,''
  \emph{IEEE Veh. Tech. Mag.}, vol.~17, no.~1, pp. 47--56, Mar. 2022.

\bibitem{lee032010}
I.-H. Lee and D.~Kim, ``{Achieving maximum spatial diversity with
  decouple-and-forward relaying in dual-hop OSTBC transmissions},'' \emph{IEEE
  Trans. Wireless Commun.}, vol.~9, no.~3, pp. 921--925, Mar. 2010.

\bibitem{gu062020}
P.~Gu, C.~Hua, W.~Xu, R.~Khatoun, Y.~Wu, and A.~Serhrouchni, ``{Control channel
  anti-jamming in vehicular networks via cooperative relay beamforming},''
  \emph{IEEE Internet Things J.}, vol.~7, no.~6, pp. 5064--5077, Jun. 2020.

\bibitem{mahboobi082015}
B.~Mahboobi, S.~Mehrizi, and M.~Ardebilipour, ``Multicast relay beamforming in
  cdma networks: Nonregenerative approach,'' \emph{IEEE Commun. Lett.},
  vol.~19, no.~8, pp. 1418--1421, Aug. 2015.

\bibitem{louie062019}
R.~H. Louie, Y.~Li, H.~A. Suraweera, and B.~Vucetic, ``{Performance analysis of
  beamforming in two hop amplify and forward relay networks with antenna
  correlation},'' \emph{IEEE Trans. Wireless Commun.}, vol.~8, no.~6, pp.
  3132--3141, Jun. 2009.

\bibitem{nakai022019}
R.~Nakai and S.~Sugiura, ``{Physical layer security in buffer-state-based
  max-ratio relay selection exploiting broadcasting with cooperative
  beamforming and jamming},'' \emph{IEEE Trans. Inf. Forensics Security},
  vol.~14, no.~2, pp. 431--444, Feb. 2019.

\bibitem{liu052021}
J.~Liu, W.~Chen, Z.~Cao, and Y.~J.~A. Zhang, ``{Cooperative beamforming for
  cognitive radio networks: A cross-layer design},'' \emph{IEEE Trans.
  Commun.}, vol.~60, no.~5, pp. 1420--1431, May 2012.

\bibitem{liu082012}
J.~Liu, W.~Chen, Z.~Cao, and Y.~J. Zhang, ``{Delay optimal scheduling for
  cognitive radios with cooperative beamforming: A structured matrix-geometric
  method},'' \emph{IEEE Trans. Mobile Comput.}, vol.~11, no.~8, pp. 1412--1423,
  Aug. 2012.

\bibitem{oliveira062021}
R.~Oliveira, R.~N. Nogueira, and M.~V. Drummond, ``{A photonic beamformer based
  on complex-valued filtering of wavelength-division multiplexed signals},'' in
  \emph{Proc. SPIE 11852, Int. Conf. Space Opt.}, vol. 11852, Jun. 2021, pp.
  1--10.

\bibitem{liu08092010}
J.~Liu, W.~Noonpakdee, H.~Takano, and S.~Shimamoto, ``{A cross layer design for
  optical relay system employing RF subcarrier beamforming},'' in \emph{Proc.
  IEEE Int. Conf. Wireless Inf. Technol. Sys., Honolulu, HI, USA}, Aug. 2010,
  pp. 1--4.

\bibitem{liu052011}
J.~Liu, W.~Noonpakdee, H.~Takano, and S.~Shimamoto, ``{A novel RF signal
  beamforming scheme over optical wireless communications},'' in \emph{Proc.
  Int. Conf. Space Opt. Sys. App. (ICSOS), Santa Monica, CA, USA}, May 2011,
  pp. 346--350.

\bibitem{alain122021}
A.~R. Ndjiongue, T.~M.~N. Ngatched, O.~A. Dobre, and H.~Haas, ``{Design of a
  power amplifying-RIS for free-space optical communication systems},''
  \emph{IEEE Wireless Commun. Mag.}, vol.~28, no.~6, pp. 152--159, Dec.
  2021\color{black}{}.

\bibitem{9756553}
S.~Sun, F.~Yang, J.~Song, and Z.~Han, ``Optimization on multiuser physical
  layer security of intelligent reflecting surface-aided vlc,'' \emph{IEEE
  Wireless Commun. Let.}, vol.~11, no.~7, pp. 1344--1348, Jul. 2022.

\bibitem{9784887}
D.~Saifaldeen, B.~S. Ciftler, M.~Abdallah, and K.~Qaraqe, ``{DRL}-based
  {IRS}-assisted secure visible light communications,'' \emph{IEEE Photon. J.},
  pp. 1--9, Early Access, 2022.

\bibitem{9799770}
S.~I. Mushfique, A.~Alsharoa, and M.~Yuksel, ``Mirror{VLC}: Optimal mirror
  placement for multi-element vlc networks,'' \emph{IEEE Trans. Wireless
  Commun.}, pp. 1--15, Early Access, 2022.

\bibitem{Demir032021}
M.~S. Demir and M.~Uysal, ``{A cross-layer design for dynamic resource
  management of VLC networks},'' \emph{IEEE Trans. Commun.}, vol.~69, no.~3,
  pp. 1858--1867, Mar. 2021\color{black}{}.

\bibitem{Aboagye102021a}
S.~Aboagye, T.~M.~N. Ngatched, O.~A. Dobre, and A.~Ibrahim, ``{Joint access
  point assignment and power allocation in multi-tier hybrid RF/VLC hetNets},''
  \emph{IEEE Trans. Wireless Commun.}, vol.~20, no.~10, pp. 6329--6342, Oct.
  2021.

\bibitem{Guo102021}
Y.~Guo, K.~Xiong, Y.~Lu, D.~Wang, P.~Fan, and K.~B. Letaief, ``{Achievable
  information rate in hybrid VLC-RF networks with lighting energy
  harvesting},'' \emph{IEEE Trans. Commun.}, vol.~69, no.~10, pp. 6852--6864,
  Oct. 2021.

\bibitem{Obeed032021}
M.~Obeed, H.~Dahrouj, A.~M. Salhab, S.~A. Zummo, and M.-S. Alouini, ``{User
  pairing, link selection, and power allocation for cooperative NOMA hybrid
  VLC/RF systems},'' \emph{IEEE Trans. Wireless Commun.}, vol.~20, no.~3, pp.
  1785--1800, Mar. 2021.

\bibitem{Eroglu022019}
Y.~S. Eroglu, Y.~Yapici, and Ismail.~Guvenc, ``{Impact of random receiver
  orientation on visible light communications channel},'' \emph{IEEE Trans.
  Commun.}, vol.~67, no.~2, pp. 1313--1325, Feb. 2019.

\bibitem{Dehghani032019}
M.~Dehghani~Soltani, A.~A. Purwita, I.~Tavakkolnia, H.~Haas, and M.~Safari,
  ``{Impact of device orientation on error performance of LiFi systems},''
  \emph{IEEE Access}, vol.~7, pp. 41\,690--41\,701, Mar. 2019.

\bibitem{Zeng082018}
Z.~Zeng, M.~D. Soltani, H.~Haas, and M.~Safari, ``{Orientation model of mobile
  device for indoor VLC and millimetre wave systems},'' in \emph{Proc. 88th
  IEEE VTC Conf. (VTC2018-Fall), Chicago, IL, USA}, 27-30 Aug. 2018, pp. 1--6.

\bibitem{Purwita082019}
A.~A. Purwita, M.~D. Soltani, M.~Safari, and H.~Haas, ``{Terminal orientation
  in OFDM-based LiFi systems},'' \emph{IEEE Trans. Wireless Commun.}, vol.~18,
  no.~8, pp. 4003--4016, Aug. 2019.

\bibitem{Zeng052020}
Z.~Zeng, M.~Dehghani~Soltani, Y.~Wang, X.~Wu, and H.~Haas, ``{Realistic indoor
  hybrid WiFi and OFDMA-based LiFi networks},'' \emph{IEEE Trans. Commun.},
  vol.~68, no.~5, pp. 2978--2991, May 2020.

\bibitem{Abdelhady122021}
A.~M. Abdelhady, A.~K.~S. Salem, O.~Amin, B.~Shihada, and M.-S. Alouini,
  ``{Visible light communications via intelligent reflecting surfaces:
  Metasurfaces vs. mirror arrays},'' \emph{IEEE Open J. Commun. Soc.}, vol.~2,
  pp. 1--20, Dec. 2021.

\bibitem{boyd2004}
S.~Boyd and L.~Vandenberghe, \emph{Convex optimization}.\hskip 1em plus 0.5em
  minus 0.4em\relax Cambridge university press, 2004.

\bibitem{Cheng122013}
C.~Chen, D.~Tsonev, and H.~Haas, ``Joint transmission in indoor visible light
  communication downlink cellular networks,'' in \emph{Proc. IEEE Globecom
  Workshops (GC Wkshps), Atlanta, GA, USA}, 2013, pp. 1127--1132.

\bibitem{Dixit122020}
V.~Dixit and A.~Kumar, ``{Performance analysis of indoor visible light
  communication system with angle diversity transmitter},'' in \emph{Proc. 4th
  IEEE Conf. Inf. Commun. Technol. (CICT), Chennai, India}, 3-5 Dec. 2020, pp.
  1--5.

\bibitem{Chen062017}
Z.~Chen, D.~A. Basnayaka, and H.~Haas, ``{Space division multiple access for
  optical attocell network using angle diversity transmitters},''
  \emph{IEEE/OSA J. Lightw. Technol.}, vol.~35, no.~11, pp. 2118--2131, Jun.
  2017.

\bibitem{Eroglu012018}
Y.~S. Eroglu, Ismail~Guvenc, A.~Sahin, Y.~Yapici, N.~Pala, and M.~Yuksel,
  ``{Multi-element VLC networks: LED assignment, power control, and optimum
  combining},'' \emph{IEEE J. Sel. Areas Commun.}, vol.~36, no.~1, pp.
  121--135, Jan. 2018.

\bibitem{Alsulami072019}
O.~Z. Alsulami, M.~T. Alresheedi, and J.~M.~H. Elmirghani, ``{Transmitter
  diversity with beam steering},'' in \emph{Proc. 21st Int. Conf. Transparent
  Opt. Netw. (ICTON), Angers, France}, Jul. 2019, pp. 1--5.

\bibitem{Yin092015}
L.~Yin, X.~Wu, and H.~Haas, ``{Indoor visible light positioning with angle
  diversity transmitter},'' in \emph{Proc. 82nd IEEE VTC Conf. (VTC2015-Fall),
  Boston, MA, USA}, Sep. 2015, pp. 1--5.

\bibitem{Hamamreh102019}
J.~M. Hamamreh, H.~M. Furqan, and H.~Arslan, ``{Classifications and
  applications of physical layer security techniques for confidentiality: A
  comprehensive survey},'' \emph{IEEE Commun. Surveys Tuts}, vol.~21, no.~2,
  pp. 1773--1828, Second Quarter 2019.

\bibitem{Mostafa072015}
A.~Mostafa and L.~Lampe, ``{Enhancing the security of VLC links: Physical-layer
  approaches},'' in \emph{Proc. IEEE Summer Topicals Meeting Series (SUM)},
  2015.

\bibitem{Chen112011}
L.~Chen, ``Physical layer security for cooperative relaying in broadcast
  networks,'' in \emph{Proc. IEEE MILCOM Conf., Baltimore, MD, USA}, Nov. 2011,
  pp. 91--96.

\bibitem{9869677}
G.~Shi, S.~Aboagye, T.~M.~N. Ngatched, O.~A. Dobre, Y.~Li, and W.~Cheng,
  ``Secure transmission in {NOMA}-aided multi-user visible light communication
  broadcasting network with cooperative precoding design,'' \emph{IEEE Trans.
  Inf. Forensics Security}, vol.~17, pp. 3123--3138, 2022.

\bibitem{Mukherjee022014}
A.~Mukherjee, S.~A.~A. Fakoorian, J.~Huang, and A.~L. Swindlehurst,
  ``{Principles of physical layer security in multiuser wireless networks: A
  survey},'' \emph{IEEE Commun. Surveys Tuts.}, vol.~16, no.~3, pp. 1550--1573,
  Third Quarter 2014.

\bibitem{Liu082017}
Y.~Liu, H.-H. Chen, and L.~Wang, ``{Physical layer security for next generation
  wireless networks: Theories, technologies, and challenges},'' \emph{IEEE
  Commun. Surveys Tuts.}, vol.~19, no.~1, pp. 347--376, First Quarter 2017.

\bibitem{Wang112019}
D.~Wang, B.~Bai, W.~Zhao, and Z.~Han, ``{A survey of optimization approaches
  for wireless physical layer security},'' \emph{IEEE Commun. Surveys Tuts.},
  vol.~21, no.~2, pp. 1878--1911, Second Quarter 2019.

\bibitem{Arfaoui044040}
M.~A. Arfaoui, M.~D. Soltani, I.~Tavakkolnia, A.~Ghrayeb, M.~Safari, C.~M.
  Assi, and H.~Haas, ``{Physical layer security for visible light communication
  systems: A survey},'' \emph{IEEE Commun. Surveys Tuts.}, vol.~22, no.~3, pp.
  1887--1908, Third Quarter 2020.

\bibitem{Su042021}
N.~Su, E.~Panayirci, M.~Koca, A.~Yesilkaya, H.~V. Poor, and H.~Haas,
  ``{Physical layer security for multi-user MIMO visible light communication
  systems with generalized space shift keying},'' \emph{IEEE Trans. Commun.},
  vol.~69, no.~4, pp. 2585--2598, Apr. 2021.

\bibitem{Ben062021}
Y.~Ben, M.~Chen, B.~Cao, Z.~Yang, Z.~Li, Y.~Cang, and Z.~Xu, ``{On secrecy
  sum-rate of artificial-noise-aided multi-user visible light communication
  systems},'' in \emph{Proc. IEEE ICC Workshops, Montreal, QC, Canada}, 14-23
  Jun. 2021, pp. 1--6.

\bibitem{Abumarshoud082021}
H.~Abumarshoud, M.~D. Soltani, M.~Safari, and H.~Haas, ``{Realistic secrecy
  performance analysis for LiFi systems},'' \emph{IEEE Access}, vol.~9, pp.
  120\,675--120\,688, Aug. 2021.

\bibitem{CHo122021}
S.~Cho, G.~Chen, and J.~P. Coon, ``{Cooperative beamforming and jamming for
  secure VLC system in the presence of active and passive eavesdroppers},''
  \emph{IEEE Trans. Green Commun. Netw.}, vol.~5, no.~4, pp. 1988--1998, Dec.
  2021.

\bibitem{Duong122021}
S.~T. Duong, T.~V. Pham, C.~T. Nguyen, and A.~T. Pham, ``{Energy-efficient
  precoding designs for multi-user visible light communication systems with
  confidential messages},'' \emph{IEEE Trans. Green Commun. Netw.}, vol.~5,
  no.~4, pp. 1974--1987, Dec. 2021.

\bibitem{Peng062021}
H.~Peng, Z.~Wang, S.~Han, and Y.~Jiang, ``{Physical layer security for MISO
  NOMA VLC system under eavesdropper collusion},'' \emph{IEEE Trans. Veh.
  Technol.}, vol.~70, no.~6, pp. 6249--6254, Jun. 2021.

\bibitem{Pham012021}
T.~V. Pham and A.~T. Pham, ``{Energy efficient artificial noise-aided precoding
  designs for secured visible light communication systems},'' \emph{IEEE Trans.
  Wireless Commun.}, vol.~20, no.~1, pp. 653--666, Jan. 2021.

\bibitem{Cho112021}
S.~Cho, G.~Chen, and J.~P. Coon, ``{Zero-forcing beamforming for active and
  passive eavesdropper mitigation in visible light communication systems},''
  \emph{IEEE Trans. Inf. Forensics Security}, vol.~16, pp. 1495--1505, Nov.
  2021.

\bibitem{Chaaban052018}
A.~Chaaban, Z.~Rezki, and M.-S. Alouini, ``{Capacity bounds and high-SNR
  capacity of MIMO intensity-modulation optical channels},'' \emph{IEEE Trans.
  Wireless Commun.}, vol.~17, no.~5, pp. 3003--3017, May 2018\color{black}{}.

\bibitem{7399676}
M.~R. Bhatnagar and Z.~Ghassemlooy, ``Performance analysis of gamma–gamma
  fading {FSO MIMO} links with pointing errors,'' \emph{IEEE/OSA J. Lightw.
  Technol.}, vol.~34, no.~9, pp. 2158--2169, Sep. 2016.

\bibitem{9031328}
L.~Li, S.~M. Moser, L.~Wang, and M.~Wigger, ``On the capacity of {MIMO} optical
  wireless channels,'' \emph{IEEE Trans. Info. Theory}, vol.~66, no.~9, pp.
  5660--5682, Sep. 2020\color{black}{}.

\bibitem{Chaaban102018}
A.~Chaaban, Z.~Rezki, and M.-S. Alouini, ``{Low-SNR asymptotic capacity of MIMO
  optical intensity channels with peak and average constraints},'' \emph{IEEE
  Trans. Commun.}, vol.~66, no.~10, pp. 4694--4705, Oct. 2018\color{black}{}.

\bibitem{7707381}
A.~Chaaban, Z.~Rezki, and M.-S. Alouini, ``Fundamental limits of parallel
  optical wireless channels: Capacity results and outage formulation,''
  \emph{IEEE Tran. Commun.}, vol.~65, no.~1, pp. 296--311, Jan. 2017.

\bibitem{8336902}
S.~M. Moser, L.~Wang, and M.~Wigger, ``Capacity results on multiple-input
  single-output wireless optical channels,'' \emph{IEEE Trans. Inf. Theory},
  vol.~64, no.~11, pp. 6954--6966, Nov. 2018.

\bibitem{4290022}
S.~M. Navidpour, M.~Uysal, and M.~Kavehrad, ``{BER} performance of free-space
  optical transmission with spatial diversity,'' \emph{IEEE Trans. Wireless
  Commun.}, vol.~6, no.~8, pp. 2813--2819, Aug. 2007\color{black}{}.

\bibitem{Wang032021}
X.~Wang, Z.~Fei, J.~Guo, Z.~Zheng, and B.~Li, ``{RIS-assisted spectrum sharing
  between MIMO radar and MU-MISO communication systems},'' \emph{IEEE Wireless
  Commun. Lett.}, vol.~10, no.~3, pp. 594--598, Mar. 2021\color{black}{}.

\bibitem{Khaleel092021}
A.~Khaleel and E.~Basar, ``{Reconfigurable intelligent surface-empowered MIMO
  systems},'' \emph{IEEE Sys. J.}, vol.~15, no.~3, pp. 4358--4366, Sep.
  2021\color{black}{}.

\bibitem{Perovic062021}
N.~S. Perović, L.-N. Tran, M.~Di~Renzo, and M.~F. Flanagan, ``{Achievable rate
  optimization for MIMO systems with reconfigurable intelligent surfaces},''
  \emph{IEEE Trans. Wireless Commun.}, vol.~20, no.~6, pp. 3865--3882, Jun.
  2021.

\bibitem{9392378}
J.~Zhang, J.~Liu, S.~Ma, C.-K. Wen, and S.~Jin, ``{Large system achievable rate
  analysis of RIS-assisted MIMO wireless communication with statistical
  CSIT},'' \emph{IEEE Trans. Wireless Commun.}, vol.~20, no.~9, pp. 5572--5585,
  Sep. 2021.

\bibitem{You122021}
L.~You, J.~Xiong, D.~W.~K. Ng, C.~Yuen, W.~Wang, and X.~Gao, ``{Energy
  efficiency and spectral efficiency tradeoff in RIS-aided multiuser MIMO
  uplink transmission},'' \emph{IEEE Trans. Signal Process.}, vol.~69, pp.
  1407--1421, Dec. 2021.

\bibitem{Zhang062021}
Y.~Zhang, B.~Di, H.~Zhang, J.~Lin, C.~Xu, D.~Zhang, Y.~Li, and L.~Song,
  ``{Beyond cell-free MIMO: Energy efficient reconfigurable intelligent surface
  aided cell-free MIMO communications},'' \emph{IEEE Trans. Cogn. Commun.
  Netw.}, vol.~7, no.~2, pp. 412--426, Jun. 2021.

\bibitem{Bayan052021}
B.~Al-Nahhas, M.~Obeed, A.~Chaaban, and M.~J. Hossain, ``{RIS-aided cell-free
  massive MIMO: Performance analysis and competitiveness},'' \emph{arXiv}, May
  2021.

\bibitem{AlNahhas062021}
B.~Al-Nahhas, M.~Obeed, A.~Chaaban, and M.~J. Hossain, ``{RIS-aided cell-free
  massive MIMO: Performance analysis and competitiveness},'' in \emph{Proc.
  IEEE ICC Workshops, Montreal, QC, Canada}, 14-23 Jun. 2021, pp. 1--6.

\bibitem{Trinh042021}
T.~V. Chien, H.~Q. Ngo, S.~Chatzinotas, M.~D. Renzo, and B.~Ottersten,
  ``{Reconfigurable intelligent surface-assisted cell-free massive MIMO systems
  over spatially-correlated channels},'' \emph{arXiv}, Apr. 2021.

\bibitem{Maraqa082020}
O.~Maraqa, A.~S. Rajasekaran, S.~Al-Ahmadi, H.~Yanikomeroglu, and S.~M. Sait,
  ``{A survey of rate-optimal power domain NOMA with enabling technologies of
  future wireless networks},'' \emph{IEEE Commun. Surveys Tuts.}, vol.~22,
  no.~4, pp. 2192--2235, Fourth Quarter 2020.

\bibitem{Marshoud012016}
H.~Marshoud, V.~M. Kapinas, G.~K. Karagiannidis, and S.~Muhaidat,
  ``{Non-orthogonal multiple access for visible light communications},''
  \emph{IEEE Photon. Technol. Lett.}, vol.~28, no.~1, pp. 51--54, Jan. 2016.

\bibitem{Marshoud042018}
H.~Marshoud, S.~Muhaidat, P.~C. Sofotasios, S.~Hussain, M.~A. Imran, and B.~S.
  Sharif, ``{Optical non-orthogonal multiple access for visible light
  communication},'' \emph{IEEE Wireless Commun. Mag.}, vol.~25, no.~2, pp.
  82--88, Apr. 2018.

\bibitem{Jamali062021}
M.~V. Jamali and H.~Mahdavifar, ``{Massive coded-NOMA for low-capacity
  channels: A low-complexity recursive approach},'' \emph{IEEE Trans. Commun.},
  vol.~69, no.~6, pp. 3664--3681, Jun. 2021.

\bibitem{Kizilirmak092015}
R.~C. Kizilirmak, C.~R. Rowell, and M.~Uysal, ``{Non-orthogonal multiple access
  (NOMA) for indoor visible light communications},'' in \emph{Proc. 4th Int.
  Worksh. Opt. Wireless Commun. (IWOW), Istanbul, Turkey}, Sep. 2015, pp.
  98--101.

\bibitem{Lin112017}
B.~Lin, X.~Tang, Z.~Ghassemlooy, C.~Lin, M.~Zhang, Z.~Zhou, Y.~Wu, and H.~Li,
  ``{A NOMA scheme for visible light communications using a single carrier
  transmission},'' in \emph{Proc. 1st South American Colloquium on Visible
  Light Commun. (SACVLC), Santiago, Chile}, Nov. 2017, pp. 1--4.

\bibitem{Yapici082019}
Y.~Yapici and Ismail.~Guvenc, ``{NOMA for VLC downlink transmission with random
  receiver orientation},'' \emph{IEEE Trans. Commun.}, vol.~67, no.~8, pp.
  5558--5573, Aug. 2019.

\bibitem{Marshoud102017}
H.~Marshoud, P.~C. Sofotasios, S.~Muhaidat, G.~K. Karagiannidis, and B.~S.
  Sharif, ``{On the performance of visible light communication systems with
  non-orthogonal multiple access},'' \emph{IEEE Trans. Wireless Commun.},
  vol.~16, no.~10, pp. 6350--6364, Oct. 2017.

\bibitem{Tran122021}
M.~Le-Tran, T.-H. Vu, and S.~Kim, ``{Performance analysis of optical backhauled
  cooperative NOMA visible light communication},'' \emph{IEEE Trans. Veh.
  Technol.}, vol.~70, no.~12, pp. 12\,932--12\,945, Dec. 2021.

\bibitem{Hammadi012021}
A.~Al~Hammadi, P.~C. Sofotasios, S.~Muhaidat, M.~Al-Qutayri, and H.~Elgala,
  ``{Non-orthogonal multiple access for hybrid VLC-RF networks with imperfect
  channel state information},'' \emph{IEEE Trans. Veh. Technol.}, vol.~70,
  no.~1, pp. 398--411, Jan. 2021.

\bibitem{Shen122017}
H.~Shen, Y.~Wu, W.~Xu, and C.~Zhao, ``{Optimal power allocation for downlink
  two-user non-orthogonal multiple access in visible light communication},''
  \emph{IEEE J. Commun. Inf. Netw.}, vol.~2, no.~4, pp. 57--64, Dec. 2017.

\bibitem{Yang022017}
Z.~Yang, W.~Xu, and Y.~Li, ``{Fair non-orthogonal multiple access for visible
  light communication downlinks},'' \emph{IEEE Wireless Commun. Lett.}, vol.~6,
  no.~1, pp. 66--69, Feb. 2017.

\bibitem{Yang032021}
F.~Yang, X.~Ji, X.~Liu, and M.~Peng, ``{Power allocation optimization for NOMA
  based visible light communications},'' in \emph{Proc. IEEE WCNC Conf.,
  Nanjing, China}, Mar. 2021, pp. 1--6.

\bibitem{Zhang042017}
X.~Zhang, Q.~Gao, C.~Gong, and Z.~Xu, ``{User grouping and power allocation for
  NOMA visible light communication multi-cell networks},'' \emph{IEEE Commun.
  Lett.}, vol.~21, no.~4, pp. 777--780, Apr. 2017.

\bibitem{Ma032019}
S.~Ma, Y.~He, H.~Li, S.~Lu, F.~Zhang, and S.~Li, ``{Optimal power allocation
  for mobile users in non-orthogonal multiple access visible light
  communication networks},'' \emph{IEEE Trans. Commun.}, vol.~67, no.~3, pp.
  2233--2244, Mar. 2019.

\bibitem{Obeed022021}
M.~Obeed, H.~Dahrouj, A.~M. Salhab, A.~Chaaban, S.~A. Zummo, and M.-S. Alouini,
  ``{Power allocation and link selection for multicell cooperative NOMA hybrid
  VLC/RF systems},'' \emph{IEEE Commun. Lett.}, vol.~25, no.~2, pp. 560--564,
  Feb. 2021.

\bibitem{Uday102021}
T.~Uday, A.~Kumar, and L.~Natarajan, ``{Joint NOMA for improved SER of
  cell-edge users in multi-cell indoor VLC},'' \emph{IEEE Wireless Commun.
  Lett.}, vol.~11, no.~1, pp. 13--17, Jan. 2021.

\bibitem{Uday032021}
T.~Uday, A.~Kumar, and L.~Natarajan, ``{NOMA for multiple access channel and
  broadcast channel in indoor VLC},'' \emph{IEEE Wireless Commun. Lett.},
  vol.~10, no.~3, pp. 609--613, Mar. 2021.

\bibitem{Janjua062020}
M.~B. Janjua, D.~B. da~Costa, and H.~Arslan, ``{User pairing and power
  allocation strategies for 3D VLC-NOMA systems},'' \emph{IEEE Wireless Commun.
  Lett.}, vol.~9, no.~6, pp. 866--870, Jun. 2020.

\bibitem{Tahira082019}
Z.~Tahira, H.~M. Asif, A.~A. Khan, S.~Baig, S.~Mumtaz, and S.~Al-Rubaye,
  ``{Optimization of non-orthogonal multiple access based visible light
  communication systems},'' \emph{IEEE Commun. Lett.}, vol.~23, no.~8, pp.
  1365--1368, Aug. 2019.

\bibitem{Eltokhey092021}
M.~Wafik~Eltokhey, M.-A. Khalighi, and Z.~Ghassemlooy, ``{Power allocation
  optimization in NOMA-based multi-cell VLC networks},'' in \emph{Proc. 17th
  Int. Symp. Wireless Commun. Sys. (ISWCS), Berlin, Germany}, Sep. 2021, pp.
  1--5.

\bibitem{Chen022018}
C.~Chen, W.-D. Zhong, H.~Yang, and P.~Du, ``{On the performance of
  MIMO-NOMA-based visible light communication systems},'' \emph{IEEE Photon.
  Technol. Lett.}, vol.~30, no.~4, pp. 307--310, Feb. 2018.

\bibitem{Zeng012021}
M.~Zeng, X.~Li, G.~Li, W.~Hao, and O.~A. Dobre, ``Sum rate maximization for
  {IRS}-assisted uplink {NOMA},'' \emph{IEEE Commun. Lett.}, vol.~25, no.~1,
  pp. 234--238, Jan. 2021.

\bibitem{Zengarxiv2021}
\BIBentryALTinterwordspacing
M.~Zeng, E.~Bedeer, X.~Li, Q.-V. Pham, O.~A. Dobre, P.~Fortier, and L.~A.
  Rusch, ``{IRS}-empowered wireless communications: State-of-the-art, key
  techniques, and open issues,'' 2021. [Online]. Available:
  \url{https://arxiv.org/abs/2101.07394}
\BIBentrySTDinterwordspacing

\bibitem{Wanmingarxiv2022}
\BIBentryALTinterwordspacing
W.~Hao, F.~Zhou, M.~Zeng, O.~A. Dobre, and N.~Al-Dhahir, ``Ultra wide band
  {THz} {IRS} communications: Applications, challenges, key techniques, and
  research opportunities,'' 2022. [Online]. Available:
  \url{https://arxiv.org/abs/2202.07137}
\BIBentrySTDinterwordspacing

\bibitem{Zuo112020}
J.~Zuo, Y.~Liu, E.~Basar, and O.~A. Dobre, ``Intelligent reflecting surface
  enhanced millimeter-wave {NOMA} systems,'' \emph{IEEE Commun. Lett.},
  vol.~24, no.~11, pp. 2632--2636, Nov. 2020.

\bibitem{Wu092021}
C.~Wu, Y.~Liu, X.~Mu, X.~Gu, and O.~A. Dobre, ``Coverage characterization of
  {STAR-RIS} networks: {NOMA} and {OMA},'' \emph{IEEE Commun. Lett.}, vol.~25,
  no.~9, pp. 3036--3040, Sep. 2021.

\bibitem{Zhu022021}
J.~Zhu, Y.~Huang, J.~Wang, K.~Navaie, and Z.~Ding, ``{Power efficient
  IRS-assisted NOMA},'' \emph{IEEE Trans. Commun.}, vol.~69, no.~2, pp.
  900--913, Feb. 2021.

\bibitem{Ding082016}
Z.~Ding, P.~Fan, and H.~V. Poor, ``{Impact of user pairing on 5G nonorthogonal
  multiple-access downlink transmissions},'' \emph{IEEE Trans. Veh. Technol},
  vol.~65, no.~8, pp. 6010--6023, Aug. 2016.

\bibitem{Bhowal14232021}
A.~Bhowal, S.~Aïssa, and R.~Singh~Kshetrimayum, ``{RIS-assisted spatial
  modulation and space shift keying for ambient backscattering
  communications},'' in \emph{IEEE ICC Conf., Montreal, QC, Canada}, Jun. 2021,
  pp. 1--6.

\bibitem{Mohjazi102021}
L.~Mohjazi, S.~Muhaidat, Q.~H. Abbasi, M.~A. Imran, O.~A. Dobre, and
  M.~Di~Renzo, ``Battery recharging time models for reconfigurable intelligent
  surfaces-assisted wireless power transfer systems,'' \emph{IEEE Trans. Green
  Commun. Netw.}, Oct. Early Access, 2021.

\bibitem{Fernandez11122021}
S.~Fernández, F.~Gregorio, B.~K. Chalise, and J.~Cousseau, ``{Wireless
  information and power transfer assisted by reconfigurable intelligent
  surfaces: Invited paper},'' in \emph{Proc. Argentine Conf. Electron. (CAE),
  Bahia Blanca, Argentina}, Mar. 2021, pp. 73--77.

\bibitem{Bhatti082016}
{Bhatti Naveed Anwar {\textit{et al.}}}, ``{Energy harvesting and wireless
  transfer in sensor network applications: Concepts and experiences},''
  \emph{ACM Trans. Sensors Netw.}, vol.~12, no.~3, pp. 1--40, Aug. 2016.

\bibitem{Dileepa122021}
D.~Marasinghe, N.~Rajatheva, and M.~Latva-aho, ``Lidar aided human blockage
  prediction for 6g,'' in \emph{2021 IEEE Globecom Workshops (GC Wkshps),
  Madrid, Spain}, 2021, pp. 1--6.

\bibitem{Kalor122021}
A.~E. Kalør, O.~Simeone, and P.~Popovski, ``Prediction of {mmWave/THz} link
  blockages through meta-learning and recurrent neural networks,'' \emph{IEEE
  Wireless Commun. Lett.}, vol.~10, no.~12, pp. 2815--2819, Dec.
  2021\color{black}{}.

\bibitem{9083856}
O.~Simeone, S.~Park, and J.~Kang, ``From learning to meta-learning: Reduced
  training overhead and complexity for communication systems,'' in \emph{Proc.
  2nd {6G} Wireless Summit ({6G} SUMMIT), Levi, Finland}, 2020, pp. 1--5.

\bibitem{9101000}
M.~R. Chaharmir, J.~Ethier, and J.~Shaker, \emph{Reflectarray Antennas:
  Analysis, Design, Fabrication, and Measurement}.\hskip 1em plus 0.5em minus
  0.4em\relax Artech House, 2013\color{black}{}.

\bibitem{ndjiongue122021digital}
A.~R. Ndjiongue, T.~M.~N. Ngatched, O.~A. Dobre, and H.~Haas, ``{Digital RIS
  (DRIS): The future of digital beam management in RIS-assisted OWC systems},''
  \emph{IEEE/OSA J. Lightw. Technol.}, vol.~40, no.~16, pp. 5597--5604, Aug.
  2022.

\bibitem{ndjiongue122022omniDRIS}
A.~R. Ndjiongue, T.~M.~N. Ngatched, O.~A. Dobre, and H.~Haas, ``{Double-sided
  beamforming in OWC systems using omni-digital reconfigurable intelligent
  surfaces},'' \emph{ArXiv}, Mar. 2022\color{black}{}.

\bibitem{7593453}
L.~Feng, R.~Q. Hu, J.~Wang, P.~Xu, and Y.~Qian, ``Applying {VLC} in {5G}
  networks: Architectures and key technologies,'' \emph{IEEE Network}, vol.~30,
  no.~6, pp. 77--83, Nov./Dec. 2016.

\bibitem{9378787}
X.~Zhang, Z.~Babar, P.~Petropoulos, H.~Haas, and L.~Hanzo, ``The evolution of
  optical {OFDM},'' \emph{IEEE Commun. Surveys Tuts.}, vol.~23, no.~3, pp.
  1430--1457, Third Quarter 2021.

\end{thebibliography}

\end{document}